\documentclass{article}
\pdfoutput=1

\PassOptionsToPackage{table}{xcolor}

\usepackage{iclr2027_conference,times}
\newif\ifarxivpreprint
\arxivpreprinttrue        %
\ifarxivpreprint\iclrfinalcopy\fi

\usepackage[utf8]{inputenc}
\usepackage[T1]{fontenc}
\usepackage{hyperref}
\ificlrfinal
  \hypersetup{
    pdftitle={HALLMARK: How to Diagnose LLM Citation Verifiers and When to Deploy Them},
    pdfauthor={Kayoon Kim, Wieland Brendel, Patrik Reizinger},
  }
\else
  \hypersetup{
    pdftitle={HALLMARK: How to Diagnose LLM Citation Verifiers and When to Deploy Them},
    pdfauthor={Anonymous Author(s)},
  }
\fi
\usepackage{url}
\usepackage{xurl}   %
\usepackage{booktabs}
\usepackage{amsfonts}
\usepackage{amsmath}
\usepackage{nicefrac}
\usepackage{microtype}
\usepackage{xcolor}   %
\usepackage{graphicx}
\usepackage[nameinlink]{cleveref}
\usepackage{soul}
\usepackage[normalem]{ulem}
\let\st\sout
\crefname{section}{\S}{\S\S}
\Crefname{section}{\S}{\S\S}
\crefname{subsection}{\S}{\S\S}
\Crefname{subsection}{\S}{\S\S}
\crefname{subsubsection}{\S}{\S\S}
\Crefname{subsubsection}{\S}{\S\S}
\crefname{figure}{Fig.}{Figs.}
\Crefname{figure}{Fig.}{Figs.}
\crefname{table}{Tab.}{Tabs.}
\Crefname{table}{Tab.}{Tabs.}
\crefname{equation}{Eq.}{Eqs.}
\Crefname{equation}{Eq.}{Eqs.}
\crefname{appendix}{Appx.}{Appxs.}
\Crefname{appendix}{Appx.}{Appxs.}
\usepackage{multirow}
\usepackage{longtable}   %
\usepackage{enumitem}
\usepackage{subcaption}
\usepackage{pifont}
\usepackage{listings}
\usepackage{tikz}
\usetikzlibrary{arrows.meta, positioning, calc}
\usepackage[most]{tcolorbox}
\newtcolorbox{takeaway}{%
  enhanced,
  colback=blue!4, colframe=blue!50!black,
  leftrule=2pt, rightrule=0.4pt, toprule=0.4pt, bottomrule=0.4pt,
  boxsep=2pt, left=4pt, right=4pt, top=2pt, bottom=2pt,
  before skip=4pt, after skip=4pt,
  fontupper=\small
}
\newtcolorbox{keytakeaway}{%
  enhanced,
  colback=orange!6, colframe=orange!70!black,
  leftrule=4pt, rightrule=0.4pt, toprule=0.4pt, bottomrule=0.4pt,
  boxsep=2pt, left=4pt, right=4pt, top=2pt, bottom=2pt,
  before skip=4pt, after skip=4pt,
  fontupper=\small
}
\usepackage[toc,page,header]{appendix}
\usepackage{minitoc}

\definecolor{hallmark-blue}{HTML}{2563EB}
\definecolor{hallmark-red}{HTML}{DC2626}
\definecolor{hallmark-green}{HTML}{16A34A}
\colorlet{cellgood}{hallmark-blue!12}  %
\colorlet{cellbad}{hallmark-red!12}
\definecolor{tier1}{HTML}{009E73}  %
\definecolor{tier2}{HTML}{E69F00}  %
\definecolor{tier3}{HTML}{D55E00}  %
\definecolor{pipelineOverride}{HTML}{009E73}
\definecolor{pipelineAgg}{HTML}{E69F00}

\setlist{nosep,leftmargin=*}

\makeatletter
\renewcommand\paragraph{\@startsection{paragraph}{4}{\z@}%
  {1.2ex \@plus 0.4ex \@minus 0.2ex}%
  {-0.5em}%
  {\normalfont\normalsize\bfseries}}
\makeatother
\ificlrfinal
  \newcommand{\repourl}{https://github.com/rpatrik96/hallmark}
  \newcommand{\reponote}{}
  \newcommand{\btuurl}{https://github.com/rpatrik96/bibtexupdater}
\else
  \newcommand{\repourl}{https://anonymous.4open.science/r/hallmark/}
  \newcommand{\reponote}{ (anonymized for review)}
  \newcommand{\btuurl}{https://anonymous.4open.science/r/bibtexupdater/}
\fi

\newif\ifneuripschecklist \neuripschecklistfalse

\newcommand{\answerTODO}[1][]{\textcolor{red}{\bf [TODO]}}

\usepackage[acronym, automake, toc, nomain, nopostdot, style=tree, nonumberlist, numberedsection]{glossaries}

\makeglossaries

\newacronym{dr}{DR}{detection rate}
\newacronym{fpr}{FPR}{false positive rate}
\newacronym{f1}{F1}{F1 score}
\newacronym{twf1}{TW-F1}{tier-weighted F1}
\newacronym{mcc}{MCC}{Matthews Correlation Coefficient}
\newacronym{ece}{ECE}{Expected Calibration Error}
\newacronym{auroc}{AUROC}{Area Under the ROC Curve}
\newacronym{ppv}{PPV}{positive predictive value}
\newacronym{mde}{MDE}{Minimum Detectable Effect}

\newacronym{pf}{PF}{parse-failure rate}

\newacronym{llm}{LLM}{Large Language Model}
\newacronym{rlhf}{RLHF}{Reinforcement Learning from Human Feedback}
\newacronym{loo}{LOO}{leave-one-out}
\newacronym{iaa}{IAA}{Inter-Annotator Agreement}
\newacronym{gepa}{GEPA}{Genetic-Pareto}

\newacronym{doi}{DOI}{Digital Object Identifier}

\glsunset{f1}  %
\newif\ifdraftnotes
\draftnotesfalse          %
\ifdraftnotes
  \usepackage[textsize=footnotesize,colorinlistoftodos]{todonotes}
  \newcommand{\claude}[1]{\textcolor{teal}{[Claude: #1]}}
  \newcommand{\confirm}[1]{\textcolor{red}{\textbf{[confirm: #1]}}}
\else
  \usepackage[disable]{todonotes}

  \renewcommand{\st}[1]{}   %
  
  \newcommand{\claude}[1]{}  %
  \newcommand{\confirm}[1]{%
    \GenericError{}{%
      UNFILLED \string\confirm\space in the clean build: #1%
    }{}{%
      Supply the fact and delete the \string\confirm\space call, or set
      \string\draftnotestrue\space to keep building the draft.%
    }}
\fi

\newcommand{\fitwidth}[1]{%
  \resizebox{\ifdim\width>\linewidth \linewidth\else \width\fi}{!}{#1}}

\title{HALLMARK: How to Diagnose LLM Citation Verifiers and When to Deploy Them}

\ificlrfinal
  \author{%
    Kayoon Kim$^{1,2}$ \quad Wieland Brendel$^{2,3}$ \quad Patrik Reizinger$^{2,3}$ \\[2pt]
    $^1$University of T\"ubingen \quad $^2$ELLIS Institute T\"ubingen \\
    $^3$Max Planck Institute for Intelligent Systems \\
    \texttt{kyoonkm@gmail.com}
  }
\fi

\begin{document}
\doparttoc %
\faketableofcontents %

\maketitle
\ifarxivpreprint\lhead{Preprint.}\fi

\ificlrfinal
  {\let\thefootnote\relax
   \footnotetext{Benchmark and code: \url{\repourl}; verification tool (\texttt{bibtex-updater}): \url{\btuurl}.}}
\fi

\begin{abstract}
GPTZero found 53 NeurIPS~2025 accepted papers with hallucinated citations~\citep{gptzero2025,neurips2025incident}, and venues now screen submissions with automated reference checkers.
Rule- and \gls{llm}-based verifiers are emerging, but many are released without a shared evaluation and each flags a different kind of error, so it is unclear which to trust or when to deploy them.
We address this with \textsc{Hallmark} (\textbf{Hall}ucination bench\textbf{mark}): 2{,}526 BibTeX entries spanning 14 hallucination types, with per-type and diagnostic sub-test labels and further temporal, cross-domain, and real-world evaluation splits.
We evaluate zero-shot and tool-augmented \glspl{llm}, a \acrshort{doi}-lookup baseline, and our own co-designed rule-based verifier, \texttt{bibtex-updater}.
Verifiers can miss hallucinations and wrongly flag correct papers, so every flag still needs human review. The \gls{fpr} determines the review effort and thus limits deployment.
The tested models exhibit an order-of-magnitude spread in \gls{fpr}; agentic lookups buy recall but inflate \gls{fpr}, and neither smaller models nor prompt optimization reach the frontier's operating point.
A two-stage cascade with \texttt{bibtex-updater} avoids that inflation, providing an upper bound, given the co-design.
These rankings, however, depend on what counts as a hallucination: re-scoring the types that describe a real work cited inaccurately changes which verifier ranks first, and venues have not agreed on a definition.
We call for a community definition and release per-type labels and a re-scoring protocol so that any agreed definition can be scored against every verifier.
\end{abstract}

\section{Introduction}
\label{sec:introduction}

Hallucinated citations---fabricated references generated by language models---have become a prevalent threat to scientific publishing, especially because they often \emph{seem} plausible.
Shortly after NeurIPS~2025, GPTZero audited the 4{,}841 accepted papers and found 53 with fabricated citations that had passed peer review~\citep{gptzero2025,neurips2025incident}, and independent audits report hundreds more across venues and a sharp rise since 2021~\citep{ghostcite2026,hallucitation2026,mysterious2026}.
Venues now screen submissions for them: ICLR~2026 ran an automated reference checker, desk-rejected papers with confirmed hallucinations, and reported a significant false-positive rate~\citep{iclr2026retrospective}.

Verifiers have multiplied, from \acrshort{doi} resolvers to multi-database cross-referencing systems and \gls{llm}-based checkers.
Many are released without a shared evaluation, and often focus on different error types, ranging from records that cannot be found to mismatched authors, venue or year. It is unclear which to trust, or how to deploy them.

To enable comparisons and to provide practical guidance on deployment for authors and venues, we release \textsc{Hallmark}, comprising 2{,}526 BibTeX entries spanning 14 hallucination types across three difficulty tiers, with taxonomy labels and six diagnostic sub-tests.
We evaluate a \acrshort{doi}-lookup baseline, frontier and small \glspl{llm} zero-shot, tool-augmented agents, and our own rule-based verifier \texttt{bibtex-updater}, which was co-designed with the taxonomy and is therefore reported as an upper bound rather than ranked.
We score every verifier under a single protocol (\cref{fig:overview}) and rely on both a held-out test set and four additional evaluation-only splits that add recent papers, a second domain, and authentic ChatGPT citations.

\begin{figure}[tbp]
\centering
\begin{tikzpicture}[
    scale=1.005, every node/.style={scale=1.005},
    node distance=0.35cm,
    box/.style={
        rectangle, rounded corners=3pt, draw=#1!60, fill=#1!8,
        minimum height=0.65cm, text width=1.75cm, align=center,
        font=\scriptsize\rmfamily, line width=0.6pt,
    },
    pipelinebox/.style={
        rectangle, rounded corners=3pt, draw=gray!70, fill=gray!5,
        minimum height=1.15cm, text width=2.25cm, align=center,
        font=\scriptsize\rmfamily, line width=0.6pt,
    },
    arrow/.style={-{Stealth[length=4pt]}, thick, gray!70},
    tierblock/.style={
        rectangle, rounded corners=2pt, draw=#1!50, fill=#1!12,
        minimum height=0.55cm, text width=4.8cm, align=left,
        font=\tiny\ttfamily, line width=0.5pt, inner sep=3pt,
    },
    tierlabel/.style={
        font=\tiny\sffamily\bfseries, black,
    },
    sublabel/.style={
        font=\tiny\rmfamily, black,
    },
]

\node[pipelinebox] (source) at (-4.09cm, 0) {%
    \textbf{Source Papers}\\[-1pt]
    {\tiny DBLP 2021--2023}\\[-1pt]
    {\tiny 1,036 valid entries}
};

\node[pipelinebox, right=0.25cm of source] (generation) {%
    \textbf{Generation}\\[-1pt]
    {\tiny Perturbation, LLM,}\\[-1pt]
    {\tiny real-world, adversarial}
};

\node[pipelinebox, right=0.25cm of generation] (dataset) {%
    \textbf{Benchmark}\\[-1pt]
    {\tiny 2,526 entries}\\[-1pt]
    {\tiny 6 sub-tests each}\\[-1pt]
    {\tiny + 4 eval-only splits}
};

\node[pipelinebox, right=0.25cm of dataset] (tools) {%
    \textbf{Tool Evaluation}\\[-1pt]
    {\tiny 13 tools: 1 DB,}\\[-1pt]
    {\tiny 12 LLMs}
};

\node[pipelinebox, right=0.25cm of tools] (metrics) {%
    \textbf{Metrics}\\[-1pt]
    {\tiny DR, FPR, TW-F1,}\\[-1pt]
    {\tiny ECE, MCC}
};

\draw[arrow] (source) -- (generation);
\draw[arrow] (generation) -- (dataset);
\draw[arrow] (dataset) -- (tools);
\draw[arrow] (tools) -- (metrics);

\coordinate (pipeheadY) at ([yshift=0.15cm]dataset.north);

\coordinate (taxstart) at ([yshift=-0.9cm, xshift=6.33cm]source.south west);

\node[tierblock=tier1, anchor=north west] (t1) at ([yshift=-0.12cm]taxstart) {%
    fabricated\_doi\quad nonexistent\_venue\quad placeholder\_authors\quad future\_date%
};
\node[tierlabel=tier1!80!black, anchor=east] at ([xshift=-0.15cm]t1.west) {%
    \rotatebox{90}{\tiny Tier 1}%
};
\node[sublabel, anchor=west] (t1sub) at (t1.east) {\hspace{2pt}4 types (easy)};

\node[tierblock=tier2, anchor=north west, below=1.5pt of t1.south west] (t2) {%
    chimeric\_title\quad wrong\_venue\quad swapped\_authors\quad preprint\_as\_pub.%
    \newline hybrid\_fabrication%
};
\node[tierlabel=tier2!80!black, anchor=east] at ([xshift=-0.15cm]t2.west) {%
    \rotatebox{90}{\tiny Tier 2}%
};
\node[sublabel, anchor=west] (t2sub) at (t2.east) {\hspace{2pt}5 types (medium)};

\node[tierblock=tier3, anchor=north west, below=1.5pt of t2.south west] (t3) {%
    near\_miss\_title\quad plausible\_fabrication%
};
\node[tierlabel=tier3!80!black, anchor=east] (t3lab) at ([xshift=-0.15cm]t3.west) {%
    \rotatebox{90}{\tiny Tier 3}%
};
\node[sublabel, anchor=west] at (t3.east) {\hspace{2pt}2 types (hard)};

\node[sublabel, anchor=north west] at ([yshift=-3pt]t3.south west) {%
    $\triangle$\,not shown: 3 stress-test types (2 in Tier~2, 1 in Tier~3), scored separately --- 14 in total%
};

\coordinate (difftail) at ([xshift=-0.30cm, yshift=0.34cm]$(t1.north west)!0.18!(t3.north west)$);
\coordinate (diffhead) at ([xshift=-0.30cm, yshift=0.34cm]t3.north west);
\draw[-{Stealth[length=6pt, width=4pt]}, thick, gray!55]
    (difftail) -- (diffhead);
\node[font=\tiny\rmfamily, black, fill=white, inner sep=1.5pt, anchor=center]
    at ($(difftail)!0.5!(diffhead)$) {Difficulty};

\coordinate (figmid) at ($(source.west)!0.5!(metrics.east)$);
\coordinate (taxmid) at ($(t3lab.west)!0.5!(t1sub.east)$);
\node[font=\footnotesize\rmfamily\bfseries, black, anchor=south]
    at (pipeheadY -| figmid) {Benchmark Pipeline};
\node[font=\footnotesize\rmfamily\bfseries, black, anchor=south]
    at ([yshift=0.08cm]t1.north -| taxmid) {Hallucination Taxonomy};

\end{tikzpicture}
\caption{%
\textbf{Overview of \textsc{Hallmark}.}
\textbf{Top:} The benchmark pipeline: real papers sourced from DBLP are transformed via perturbation, LLM generation, and real-world collection into 2,526 annotated BibTeX entries, each with six diagnostic sub-tests; four evaluation-only splits (1,349 entries) are scored separately (\cref{tab:stats}).
Thirteen full-coverage verification tools (1 citation-database, 12 zero-shot LLMs) are evaluated using tier-weighted metrics.
\textbf{Bottom:} The three-tier hallucination taxonomy with the 11 \emph{main} types, ordered by the verification effort required to detect them; 3 further stress-test types are evaluated separately and listed in \cref{tab:taxonomy}, for 14 in total.
}
\label{fig:overview}
\end{figure}

\paragraph{Scope.}
\textsc{Hallmark} targets citation \emph{metadata}---detecting fabricated, inconsistent, or nonexistent bibliographic records---not claim-level hallucination, which needs full-text analysis.

{Current citation verifiers are not perfect; thus, human review is required, and \emph{the false-positive rate, not recall, becomes the bottleneck for deployment}, as we show in our \gls{ppv} analysis in \mbox{\cref{sec:ppv}}, with \gls{fpr} spanning an order of magnitude ($0.043$ to $0.702$).}
We observe and analyze three failure modes (\cref{tab:failure-modes}): (i) agentic lookups buy recall but inflate \gls{fpr}; (ii) precision collapses at realistic prevalence; and (iii) most \glspl{llm} by default over-flag papers published after their training cutoff.
Some of these failure modes can be mitigated: a two-stage cascade with \texttt{bibtex-updater} avoids inflating \gls{fpr} by passing only uncertain entries to an \gls{llm}. It catches nearly all hallucinations at a low false-positive rate, which we treat as an upper bound given the co-design. A prompt that reminds the model of its training cutoff (i.e., cutoff-aware) reduces over-flagging of recent papers, but mostly through abstention; mitigating it selectively is left for future work. Prompt optimization~\citep{agrawal2026gepareflectivepromptevolution} also has limits, especially for smaller models, which maintain high \gls{fpr}.

Verifier performance and ranking depend on the---perhaps surprisingly ambiguous---definition of a hallucination. For example, a real work cited with the wrong venue is not fully correct, but is that a hallucination?
Existing verifiers address this differently in what they check, and venues have not (yet) reached consensus.
We therefore urge the community to have more discussion on definitions of hallucinated citations; meanwhile, we release per-type labels and a re-scoring protocol so that any agreed definition can be scored against every verifier.

Our \textbf{contributions} are:
\begin{enumerate}
    \item \textbf{A benchmark and taxonomy} for citation-hallucination detection, with per-type and sub-test labels and temporal, cross-domain, and real-world splits (\cref{sec:benchmark}).
    \item \textbf{A failure-mode analysis} of citation verifiers under one protocol, showing that the false-positive rate decides deployability (\cref{sec:analysis}).
    \item \textbf{A re-scoring protocol} that lets any definition of a hallucinated citation be applied to every verifier.
\end{enumerate}

\section{Related work}
\label{sec:related}

\paragraph{Citation hallucination in the LLM era.}
\Gls{llm} hallucination~\citep{ji2023survey,huang2023survey} manifests distinctly in citations~\citep{alkaissi2023,agrawal2024,walters2023fabrication}; venue-level audits document scale---GhostCite~\citep{ghostcite2026} on 56K papers, HalluCitation~\citep{hallucitation2026} on ACL/NAACL/EMNLP, the HPC-venue study of \citet{mysterious2026}, and CITADEL's audit of 2.5 million PubMed Central papers~\citep{topaz2026fabricated}---measuring \emph{prevalence}, not detection-tool performance.
Detection tools---HaRC~\citep{harc2024}, verify-citations~\citep{verifycitations2025}, RefChecker~\citep{refchecker2024}, CheckIfExist~\citep{checkifexist2026}---report incompatible metrics on disjoint test data, preventing direct comparison; CheckIfExist (a CrossRef/Semantic-Scholar/OpenAlex cascade verifier) ships without dataset or metrics, and RefChecker requires full-text manuscript context.
Concurrent work~\citep{rao2026bibtex,raowong2026urls,hallucitechecker2026} evaluates \gls{llm}-as-citation-generator accuracy or releases lightweight checkers. The closest work to \textsc{Hallmark} is CiteAudit~\citep{citeaudit}, which likewise benchmarks \glspl{llm} and commercial verifiers on a human-validated/annotated dataset via a multi-agent pipeline.
\textsc{Hallmark} differs by focusing on understanding failure modes by rich diagnostics. It provides a 14-type, tier-stratified error taxonomy and post-cutoff evaluation splits that expose training-cutoff effects, so it reports \emph{which} metadata failures each tool catches.
Adjacent concurrent detection work~\citep{raowong2026urls} targets citation-URL liveness on DRBench (the stale-vs-hallucinated distinction), a scope orthogonal to \textsc{Hallmark}'s BibTeX-metadata cross-consistency checks. \Cref{app:related} details boundary cases of audit-vs-benchmark categorization.

\paragraph{Hallucination detection benchmarks.}
General-purpose hallucination benchmarks~\citep{hallueval2023,halogen2024} target factual claims, not metadata; adjacent work---FactScore~\citep{min2023factscore}, RARR~\citep{gao2023rarr}, SciFact~\citep{wadden2020scifact}---does not evaluate BibTeX integrity (\gls{doi}/title/author/venue cross-consistency against external databases).
Concurrent generator-side benchmarks such as HalluHard~\citep{halluhard2026} measure whether \glspl{llm} produce \emph{claims} grounded by retrievable inline citations in multi-turn dialogs. None evaluates the verifiers that check citation metadata. \textsc{Hallmark}'s design borrows from established benchmarks: temporal segmentation, as in SWE-bench~\citep{swebench} and LiveCodeBench~\citep{livecodebench2024}, separates recall from verification, and multi-criteria sub-tests, as in HumanEval~\citep{humaneval}, attribute each failure to a field (\cref{app:related}).

\section{The \textsc{Hallmark} benchmark}
\label{sec:benchmark}

\subsection{Hallucination taxonomy}
\label{sec:taxonomy}

We define 14 citation hallucination types---11 empirically grounded and 3 stress-test types---organized into three difficulty tiers based on the verification effort required (\cref{tab:taxonomy}, in the appendix).
 \textbf{Tier~1 (Easy)} hallucinations are settled by one check, a single lookup (\texttt{fabricated\_doi}, \texttt{nonexistent\_venue}) or none at all (\texttt{placeholder\_authors}, \texttt{future\_date}).
\textbf{Tier~2 (Medium)} hallucinations require cross-referencing multiple metadata fields: e.g., \texttt{chimeric\_title} pairs real authors with a fabricated title, \texttt{wrong\_venue} assigns a paper to the wrong conference, and \texttt{hybrid\_fabrication} uses a real \gls{doi} whose resolved record does not match the BibTeX metadata.
\textbf{Tier~3 (Hard)} hallucinations require deep verification or semantic reasoning: \texttt{near\_miss\_title} entries differ by one or two words from a real paper; \texttt{plausible\_fabrication} entries are entirely invented but realistic; and \texttt{arxiv\_version\_mismatch} entries cite a preprint with wrong venue and shifted year. The tiers rank the effort a human verifier needs, not difficulty for an \gls{llm}.

The taxonomy is grounded in real incidents: we derived it from hallucinated citations in the NeurIPS~2025 incident and related audits~\citep{ghostcite2026,hallucitation2026}, then stress-tested it with adversarial brainstorming of failure modes existing tools might miss.
Of the 108 real-world hallucinated citations in the benchmark, the 72 we analyzed in detail map cleanly to a single type 79\% of the time, with the rest assigned to their highest applicable tier (\cref{app:real-world-mapping}).
Three types---\texttt{merged\_citation}, \texttt{partial\_author\_list}, \texttt{arxiv\_version\_mismatch}---we could not trace to a documented incident, so we treat them as stress-test types. They appear in every split (18.3\% of hallucinated entries on \texttt{dev\_public}, 20.2\% on \texttt{test\_public}); per-type labels let each type be scored separately.
Perturbation rules, generation prompts and the collection procedure are in \cref{app:dataset,app:prompt-template}.

\subsection{Dataset construction}
\label{sec:dataset}

The dataset contains two classes of entries: \emph{valid} references scraped from DBLP and \emph{hallucinated} references built with the four methods below.

\paragraph{Valid entries.}
We scraped BibTeX records from DBLP~\citep{dblp} for papers published at major ML venues (NeurIPS, ICML, ICLR, AAAI, CVPR) between 2021 and 2023, plus 35 arXiv preprints from February 2026 (22 \texttt{dev\_public}, 13 \texttt{test\_public}) as a recency probe.
Each entry was verified by confirming \gls{doi} resolution where a \gls{doi} is present ($46\%$ of valid entries), title existence in at least two databases, and author-venue consistency.
This retained 1,036 valid entries.

\paragraph{Hallucinated entries.}
We generated hallucinated entries using four methods:
\emph{Systematic perturbation}: modifying specific fields of valid entries to produce targeted hallucination types (e.g., replacing a \gls{doi} with a non-resolving one for \texttt{fabricated\_doi}, swapping author lists between papers for \texttt{swapped\_authors}).
\emph{\gls{llm} generation}: prompting language models to generate plausible but fictional references for types requiring coherent fabrication (\texttt{plausible\_fabrication}, \texttt{chimeric\_title}).
\emph{Adversarial crafting}: manually constructing entries designed to evade specific detection strategies.
\emph{Real-world collection}: harvesting actual hallucinated citations from published papers identified in audits.
An evaluation-only supplement of 341 authentic ChatGPT-generated citations (172 valid / 169 hallucinated), hand-verified by \citet{walters2023fabrication} across 42 multidisciplinary topics, adds real-world hallucinations (\cref{app:walters_wilder}).

\paragraph{Labeling.}
Labels are assigned deterministically by the generation pipeline; \gls{llm}-generated entries are also filtered against bibliographic databases, and real-world and adversarial entries receive manual review (\cref{app:dataset}).
\paragraph{Splits and release.}
Every hallucinated entry is built by modifying some real \emph{base paper}. Across the two public splits those base papers do not overlap: no hallucinated entry on one split is built from a paper that appears as a valid entry on the other. \texttt{test\_hidden} and \texttt{stress\_test} do not have this property: 109 of their 244 and 68 of their 121 hallucinated entries modify a paper that is a valid entry in public data.
Most types on \texttt{dev\_public} and \texttt{test\_public} carry $\geq 30$ instances, enough for meaningful per-type comparison; \texttt{test\_hidden} spreads 244 hallucinated entries over 14 types and does not clear that floor, so its per-type intervals are wider (statistical power in \cref{app:statistics}).

Every entry passes automated validation, the \texttt{url} field is blinded to avoid using URL resolution as a shortcut for verification, unique marker tokens in the metadata expose memorization if a model reproduces one, and types are roughly balanced to maximize per-type power (\cref{app:dataset}).
The core splits are \texttt{dev\_public} (1{,}119 entries), \texttt{test\_public} (831), \texttt{test\_hidden} (454), and \texttt{stress\_test} (122); per-split counts and the four evaluation-only splits are in \cref{tab:stats}.
\label{sec:temporal}
Entries carry a temporal segment tag (\emph{pre-2023}, \emph{2023--2024}, \emph{2025+}), used for the post-cutoff analysis in \cref{sec:temporal_robustness,app:temporal}.

\paragraph{Sub-test design.}
\label{sec:subtests}

To understand what is hallucinated and \emph{why}, each entry decomposes citation validity into six independently verifiable dimensions, each scored True/False/N/A: \textbf{\Gls{doi} resolves}, \textbf{Title exists} in bibliographic databases, \textbf{Authors match} the identified paper, \textbf{Venue real} and correctly attributed, \textbf{Fields complete}, \textbf{Cross-DB agreement}.

The six sub-tests also make an entry's defect \emph{count} observable: one type can span several failed sub-tests, and multi-defect entries are the norm in the real-world subset, whereas under perturbation they are rare. Detection by defect count is in \cref{app:multidefects}.

\section{Experiments}
\label{sec:experiments}

\subsection{Metrics}
\label{sec:evaluation}

\label{sec:metrics}

Our comparison focuses on two \emph{prevalence-independent} metrics, ensuring they remain comparable across splits with differing class ratios.
\textbf{\Gls{dr}} is recall on the hallucinated class; \textbf{\gls{fpr}} is the fraction of valid entries wrongly flagged.
\textbf{\Gls{twf1}} weights each hallucinated entry by its tier (1/2/3), rewarding the detection of harder hallucinations.
\textbf{\Gls{mcc}} reads all four confusion-matrix cells; it moves with prevalence by ${\le}0.03$ between our splits ($54.2\%$ vs.\ $62.5\%$ hallucinated).
\textbf{\Gls{ece}}~\citep{naeini2015} measures how well a tool's confidence tracks its accuracy; an \gls{ece} below $0.05$ is excellent and above $0.2$ unreliable.
Formal definitions are in \cref{app:metric_details}; per-tier and per-type breakdowns in \cref{sec:per_type,app:pertype}.

\subsection{Evaluated tools}
\label{sec:baselines}
We evaluate thirteen independent full-coverage tools (i.e., returning a verdict on every entry, neither co-designed with the benchmark nor given tool access) on \texttt{dev\_public}: a \gls{doi}-only lower bound and twelve zero-shot \glspl{llm} from OpenAI, Anthropic, Google, Meta, Mistral, DeepSeek and Qwen (\cref{tab:results}; model IDs in \cref{app:llm-setup}), including a later-cutoff GPT-5.4 control (Aug~2025 cutoff) for \cref{sec:temporal_robustness}.
All \glspl{llm} use the same zero-shot prompt at temperature $0$ and seed $42$ with one API call per entry; DeepSeek-R1's chain-of-thought raises its median latency to ${\sim}19$\,s against $1$--$10$\,s for the single-pass cohort (\cref{tab:cost_latency}).
On top of these, four \emph{agentic} variants (\cref{app:llm-setup}) allow up to five database lookups per entry; prompted to cross-reference databases, the model tends to flag an entry as soon as one lookup returns no match---the disposition behind the inflated false-positive rate we report in \cref{sec:analysis}.
HaRC~\citep{harc2024} and verify-citations~\citep{verifycitations2025} are excluded from \cref{tab:results}: neither returns its own verdicts at benchmark scale, so any row for them would report the shared pre-screening layer (\cref{app:harc_disclaimer}).
The \gls{doi}-only baseline only verifies each entry's \gls{doi} against the DOI.org resolver.
 The optional pre-screening layer (\gls{doi} format, year bounds, author heuristics) is applied to the database-lookup tools and the cascade, not to the \gls{llm} verifiers, so those rows include their verdicts (\cref{sec:prescreening,app:cascade}).
We label \texttt{bibtex-updater} \emph{co-designed} because its development overlapped the taxonomy: the typed sub-tests partly mirror the tool's verification stages, so it may score better here than on a novel hallucination distribution.
We read its row as an unranked \emph{upper-bound reference} (\cref{app:codesign}).

\subsection{Main results}
\label{sec:main_results}

\begin{table}[!t]
\caption{\textbf{Main results on \texttt{dev\_public}.} \textbf{Bold} marks the best \emph{point estimate} among independent full-coverage tools (\gls{doi}-only and the zero-shot \glspl{llm}, i.e., neither co-designed nor tool-augmented; see the \emph{Scoring conventions} paragraph, \cref{sec:main_results}). The shaded \emph{co-designed} block was developed alongside the benchmark taxonomy and is a reference \emph{upper bound} \emph{excluded from ranking}; no co-designed cell is bolded (\cref{app:codesign}). The zero-shot block is sorted by \gls{fpr} ascending. \textbf{Coverage} (Cov.) is the fraction of entries a tool decides about, scored under the \emph{Scoring conventions}. $\Delta$FPR is the cross-split shift \texttt{test\_public} $-$ \texttt{dev\_public}. FPR cells are shaded: blue marks the low-\gls{fpr} band (FPR ${\le}.13$), red marks over-flagging operating points (FPR ${\ge}.41$); the gray co-designed block is excluded from shading as from ranking. The three cascade rows count entries neither stage resolves as hallucinated (\cref{tab:cascade}); they share one Stage~1 (Stage~1: \texttt{bibtex-updater}; Stage~2: an \gls{llm}; \cref{sec:cascade}), and flagging rather than dropping the entries Stage~2 leaves undecided changes no cascade's \gls{fpr}. \texttt{bibtex-updater} is scored over the full split with abstentions resolved to \textsc{valid}; under the \gls{llm} scoring rule (abstentions excluded; \cref{app:coverage}) it reaches \gls{dr}~0.992 / \gls{fpr}~0.056 at coverage 0.86. \emph{Always-call}: \texttt{bibtex-updater} runs on every entry and its output is given to the \gls{llm} in the prompt.}
\label{tab:results}
\centering
\small
\setlength{\tabcolsep}{4pt}
\fitwidth{%
\begin{tabular}{lcccccccc}
\toprule
 & \multicolumn{5}{c}{\textbf{Performance}} & \textbf{Calibr.} & \textbf{Coverage} & \textbf{Robustness} \\
\cmidrule(lr){2-6}\cmidrule(lr){7-7}\cmidrule(lr){8-8}\cmidrule(lr){9-9}
\textbf{Tool} & \textbf{DR $\uparrow$} & \textbf{FPR $\downarrow$} & \textbf{F1 $\uparrow$} & \textbf{MCC $\uparrow$} & \textbf{TW-F1 $\uparrow$} & \textbf{ECE $\downarrow$} & \textbf{Cov. $\uparrow$} & \textbf{$\Delta$FPR $\downarrow$} \\
\midrule
\multicolumn{9}{l}{\emph{Citation-database tools}} \\
DOI-only                & 0.218          & \cellcolor{cellgood}\textbf{0.043} & 0.347          & 0.253          & 0.268          & 0.082 & 1.00 & $\phantom{+}-0.001$ \\
\midrule
\multicolumn{9}{l}{\emph{Zero-shot LLMs (sorted by FPR)}} \\
Gemini~2.5~Pro          & 0.476          & \cellcolor{cellgood}0.050          & 0.627          & 0.473          & 0.609          & 0.297 & 0.97 & $+0.009$ \\
Gemini~2.5~Flash        & 0.500          & \cellcolor{cellgood}0.100          & 0.631          & 0.429          & 0.628          & 0.265 & 0.99 & $+0.006$ \\
Llama~4~Maverick        & 0.614          & 0.146          & 0.707          & 0.476          & 0.709          & 0.176 & 1.00 & $+0.020$ \\
Claude Opus~4.7$^{\S}$         & 0.906 & 0.154 & 0.890 & 0.755 & 0.913 & 0.052 & 1.00 & $+0.051$ \\
Claude Sonnet~4.6$^{\S}$       & \textbf{0.916} & 0.161 & \textbf{0.893} & \textbf{0.760} & \textbf{0.917} & \textbf{0.041} & 0.98 & $+0.026$ \\
GPT-5.4 (zero-shot)     & 0.767          & 0.228          & 0.783          & 0.538          & 0.807          & 0.202 & 1.00 & $\phantom{+}-0.004$ \\
Mistral Large           & 0.716          & 0.250          & 0.742          & 0.465          & 0.765          & 0.229 & 0.99 & $+0.032$ \\
GPT-5.1 (zero-shot)     & 0.837          & \cellcolor{cellbad}0.411          & 0.766          & 0.442          & 0.822          & 0.190 & 1.00 & $+0.069$ \\
Qwen3-235B              & 0.860          & \cellcolor{cellbad}0.533          & 0.744          & 0.358          & 0.821          & 0.279 & 1.00 & $+0.082$ \\
Qwen3-VL-235B           & 0.860          & \cellcolor{cellbad}0.551          & 0.740          & 0.342          & 0.818          & 0.286 & 1.00 & $+0.077$ \\
DeepSeek-R1             & 0.896          & \cellcolor{cellbad}0.623          & 0.739          & 0.324          & 0.825          & 0.238 & 0.98 & $\phantom{+}-0.303$$^{\P}$ \\
DeepSeek-V3.2           & 0.911 & \cellcolor{cellbad}0.702          & 0.727          & 0.268          & 0.821          & 0.316 & 1.00 & $+0.026$ \\
\midrule
\multicolumn{9}{l}{\emph{Agentic (tool-use; up to 5 tool calls per entry)}} \\
GPT-5.1 + CrossRef/OpenAlex/arXiv                    & 0.967 & \cellcolor{cellbad}0.478 & 0.816 & 0.558 & 0.892 & 0.175 & 1.00 & $+0.080$ \\
GPT-5.1 + bibtex-updater (agentic; tool optional)    & 0.980 & \cellcolor{cellbad}0.470 & 0.824 & 0.584 & 0.900 & 0.125 & 1.00 & $\phantom{+}-0.114$ \\
Sonnet~4.6 + bibtex-updater (agentic; tool optional)$^{\ddagger}$ & 0.990 & \cellcolor{cellbad}0.431 & 0.841 & 0.630 & 0.913 & 0.118 & 1.00 & $\phantom{+}-0.088$ \\
\midrule
\rowcolor{gray!10}
\multicolumn{9}{l}{\emph{Co-designed (reference upper bound; see \cref{app:codesign})}} \\
\rowcolor{gray!10}
bibtex-updater (v1.2.0)                                     & 0.820 & 0.051 & 0.880 & 0.768 & 0.892 & 0.395 & 0.86 & $+0.029$ \\
\rowcolor{gray!10}
Cascade: bibtex-updater $\to$ Sonnet~4.6 (\cref{sec:cascade})$^{\sharp}$ & 0.990 & 0.037 & 0.980 & 0.955 & 0.986 & 0.117 & 1.00 & $+0.021$ \\
\rowcolor{gray!10}
Cascade: bibtex-updater $\to$ GPT-5.1 (\cref{sec:cascade}) & 0.974 & 0.039 & 0.970 & 0.935 & 0.976 & 0.091 & 1.00 & $+0.019$ \\
\rowcolor{gray!10}
Cascade: bibtex-updater $\to$ GPT-5.4 (\cref{sec:cascade})$^{\sharp}$ & 0.990 & 0.039 & 0.979 & 0.953 & 0.985 & 0.097 & 1.00 & $+0.019$ \\
\rowcolor{gray!10}
GPT-5.1 + bibtex-updater (always-call; output in prompt)    & 0.843 & 0.144 & 0.856 & 0.698 & 0.872 & 0.078 & 0.93 & $+0.112$ \\
\bottomrule
\end{tabular}%
}
\par\smallskip
{\footnotesize\noindent $^{\P}$DeepSeek-R1's $\Delta$FPR of $-0.303$ is measured over the 651 \texttt{test\_public} entries it answered, the other 180 being API error fallbacks (\cref{tab:test_public_full}). $^{\S}$Collected 2026-09-02. $^{\ddagger}$Contains Sonnet~4.6 from the earlier 2026-05-05 collection. $^{\sharp}$Collected 2026-09-21 on the Stage~1 shared by all three cascades. Comparing a $^{\S}$ row against a $^{\ddagger}$ row crosses endpoint snapshots, and the drift (\cref{tab:endpoint_drift}) can change the reading. Unmarked rows: 2026-05-05.}
\end{table}

\paragraph{Scoring conventions.}
 Rows in \cref{tab:results}---which reports \texttt{dev\_public}---are ranked by point estimate, with a paired test only where per-entry predictions allow (\cref{app:bootstrap}); an \gls{llm}'s \textsc{uncertain} verdicts are excluded from the \gls{dr}/\gls{fpr}/\gls{f1} triple, while \texttt{bibtex-updater} resolves its abstentions to \textsc{valid} before scoring (\cref{app:bootstrap,app:coverage}); predictions were collected 2026-05-05 apart from the two rows marked $^{\S}$ in \cref{tab:results}, from a later collection (2026-09-02), and hosted-\gls{llm} rows are dated snapshots subject to endpoint drift (\cref{sec:limitations}).

\begin{figure}[t]
\centering
\includegraphics[width=0.98\linewidth]{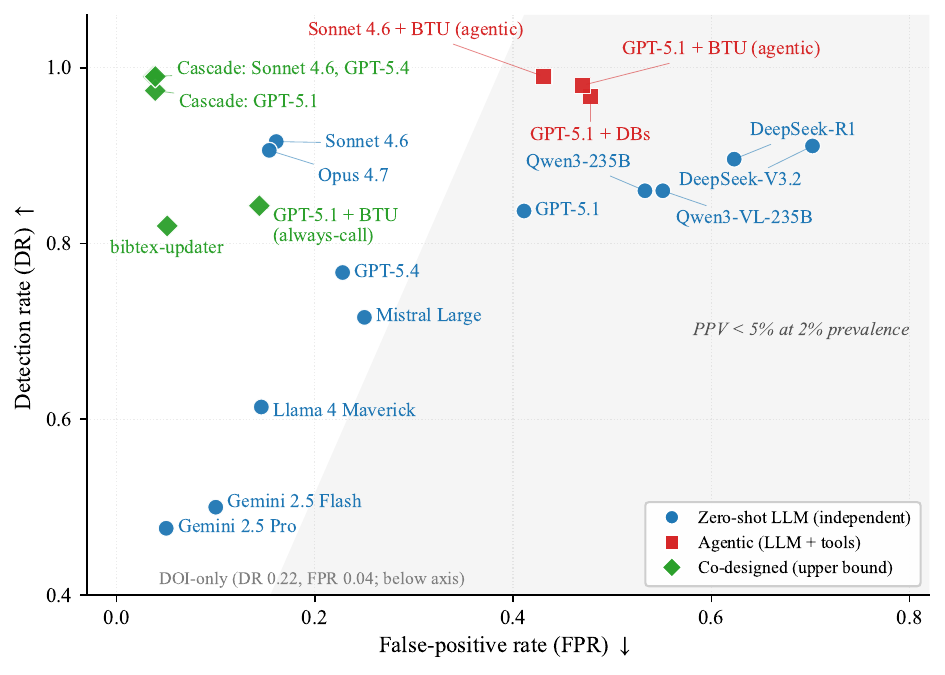}
\caption{\textbf{\Gls{dr}--\gls{fpr} operating points on \texttt{dev\_public}.} Each point is a tool's operating point, the (\gls{fpr}, \gls{dr}) pair it reaches, one per row of \cref{tab:results}. BTU: \texttt{bibtex-updater}; DBs: CrossRef/OpenAlex/arXiv. \emph{Agentic} (red): the \gls{llm} may call BTU or the DBs as a tool, up to five calls per entry; \emph{always-call} (green): BTU runs on every entry and its output is given to the \gls{llm}. In the shaded region, more than 19 out of every 20 flags are false alarms at a venue-realistic rate of $2\%$ hallucinated citations (\cref{sec:ppv}), so high recall alone does not make a tool usable. The co-designed cascades reach high recall at low \gls{fpr} (an upper bound).}
\label{fig:dr_fpr}
\end{figure}

\paragraph{\Glspl{llm} span a wide recall--precision spectrum.}
The \gls{fpr} ordering of \cref{tab:results} (\gls{dr} vs.\ \gls{fpr} tradeoff in \cref{fig:dr_fpr}) runs from models that flag rarely (Gemini~2.5~Pro) to models that flag most entries (DeepSeek-V3.2).
The two later-cutoff Anthropic models give the best balance---Opus~4.7 and Sonnet~4.6 have the highest \gls{f1} and the best calibration of the independent tools, at a middling \gls{fpr}. Their \gls{f1} gap is only 0.3\,pp (paired bootstrap $p{=}0.74$, \cref{app:bootstrap}), and it reverses direction between the May and September runs: Opus~4.7 leads on the 2026-05-05 run, Sonnet~4.6 on the 2026-09-02 one (\cref{tab:endpoint_drift}).
Neither model beats the other on both axes---Opus~4.7 has the lower \gls{fpr} ($.154$ vs.\ $.161$), Sonnet~4.6 the higher recall ($.916$ vs.\ $.906$)---and on the earlier 2026-05-05 collection both sat at the low-\gls{fpr} end ($.072$, $.127$; \cref{tab:endpoint_drift}). Sonnet's extra recall is also partly over-flagging: of the 27 \texttt{dev\_public} papers the ground-truth audit reclassified as real (\cref{sec:dataset}), Sonnet flagged 19 as hallucinated and Opus~8.
Both clear GPT-5.1 and the open-weight models that flag most entries, and GPT-5.4 (cutoff Aug~2025) sits between them at lower \gls{fpr} and comparable recall, consistent with \cref{sec:temporal_robustness}.

\Cref{app:coverage} examines whether \texttt{bibtex-updater}'s low \gls{fpr} depends on abstention. Because \gls{dr} and \gls{fpr} are prevalence-independent, we compare on them, and on \gls{f1} only within a split.

\begin{figure}[!t]
\centering
\includegraphics[width=0.98\linewidth]{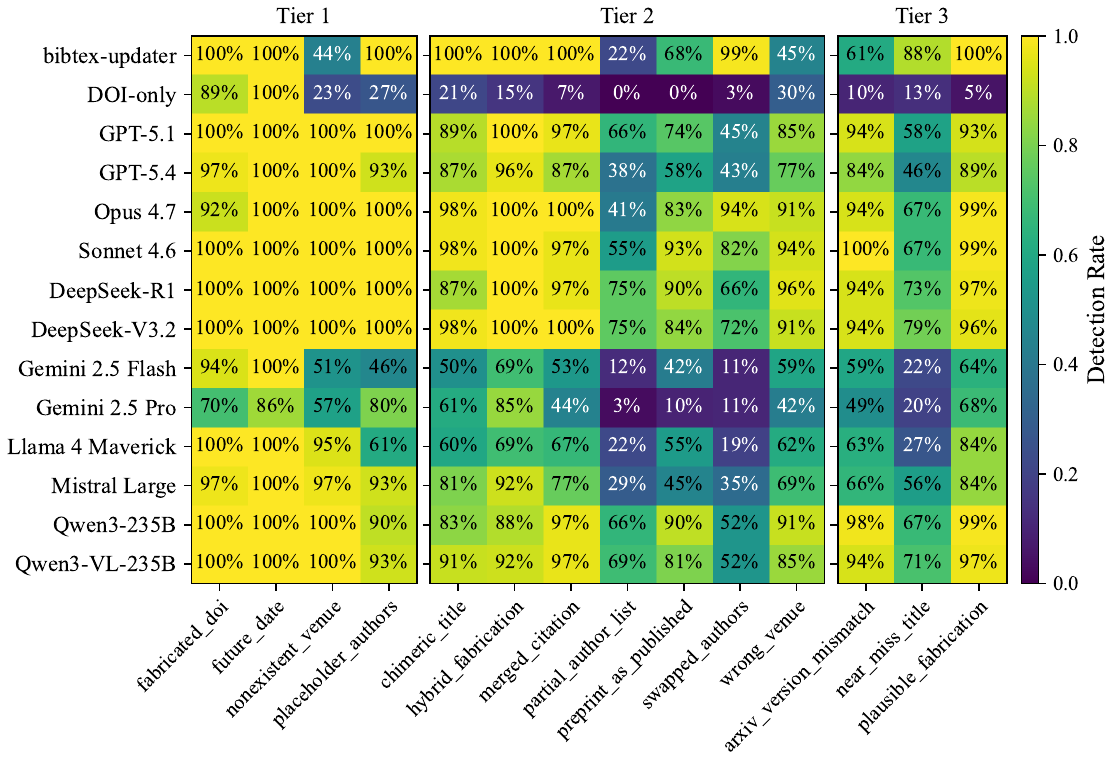}
\caption{\textbf{Per-type detection rates across all full-coverage tools.} The 14 types are grouped into three panels by difficulty tier (\cref{tab:taxonomy}). HaRC and verify-citations are excluded because neither produces its own verdicts at benchmark scale (\cref{app:harc_disclaimer}). \texttt{swapped\_authors} and \texttt{near\_miss\_title} remain the hardest types across the cohort, and the tier grouping makes the pattern visible: most \glspl{llm} catch nearly every Tier-1 hallucination, while Tiers~2--3 separate the cohort; tiers rank the effort a human verifier needs, not difficulty for an \gls{llm}.}
\label{fig:heatmap}
\end{figure}

\paragraph{Per-type analysis.}
\label{sec:per_type}
Per-type detection rates (heatmap in \cref{fig:heatmap}; full table in \cref{tab:pertype_full}) show the same pattern across models: most \glspl{llm} catch nearly every Tier-1 hallucination, while \texttt{swapped\_authors} and \texttt{near\_miss\_title} are the hardest types for almost every model.
GPT-5.1 reaches at least 74\% on 9 of 11 main types and is perfect on \texttt{nonexistent\_venue} and \texttt{placeholder\_authors}.
Its two weak types, \texttt{swapped\_authors} and \texttt{near\_miss\_title}, both turn on exact bibliographic knowledge and map to the \emph{Authors match} and \emph{Title exists} sub-tests respectively (\cref{sec:subtests}; \cref{tab:taxonomy}).

\paragraph{Small models.}
As not every researcher can access the largest (and, often, costliest) models, we investigate whether a small local model can be a good substitute. We test on Claude Haiku~4.5 and four Qwen3 open-weight models at 4, 8, 14 and 32B, under the same protocol as \cref{tab:results}, and report the details in \cref{app:small_models}.
Haiku performs substantially worse than the tested Anthropic Sonnet and Opus models, and the smaller open-weight models flag more than $95\%$ of the split as hallucinations, yielding a high \gls{dr} carrying no information: at venue-realistic prevalence their \gls{ppv} is indistinguishable from the base rate (\cref{app:ppv}), so nearly every entry requires manual checks. Thus, cheap inference does not make deployment cheap.

\paragraph{Cross-split robustness.}
\label{sec:crosssplit_robustness}
The held-out \texttt{test\_public} (831 entries) tells a complementary story to the \texttt{dev\_public} results ($\Delta$\gls{fpr} in \cref{tab:results}; full results in \cref{app:test_public_full}).
 Among standalone tools with \gls{dr}~${\geq}0.80$, \texttt{bibtex-updater} keeps the lowest false-positive rate on both splits ($0.051$ on \texttt{dev\_public}, $0.080$ on \texttt{test\_public}), which we read as an upper bound given the co-design; out of domain its \gls{fpr} stays low but its coverage drops (\cref{app:codesign}).

\paragraph{Prompt sensitivity.}
A prompt-sensitivity ablation on 150 \texttt{dev\_public} entries (four prompt variants, four models; \cref{app:ablation_prompt}) moves the same model's absolute \gls{fpr} by $10$--$37$\,pp on wording alone, while the model ranking barely moves (mean pairwise Spearman rank correlation $\rho{=}0.90$): absolute \gls{fpr} values depend on the prompt, the model ordering does not.

\begin{takeaway}
\textbf{Takeaway.} 
Even the highest-recall independent model (Sonnet~4.6) misses ${\sim}8\%$ of hallucinations, and its misses concentrate on subtle types: \texttt{near\_miss\_title} and \texttt{swapped\_authors} are its weakest, both demanding exact bibliographic knowledge (\cref{tab:pertype_full}).
Among independent tools, Sonnet~4.6 and Opus~4.7 give the best \gls{f1} and calibration at a middling \gls{fpr}, a ranking that survives a prompt-paraphrase ablation (\cref{app:ablation_prompt}).
\end{takeaway}

\subsection{Failure modes}
\label{sec:analysis}
Three failure modes recur across the cohort (\cref{tab:failure-modes}); we diagnose each in turn.

\paragraph{Agentic lookups give diminishing returns (failure mode~(i)).}
Five tool calls push GPT-5.1's recall past \texttt{bibtex-updater}'s, but its false-positive rate rises to roughly nine times \texttt{bibtex-updater}'s, because the prompted model---cross-referencing against CrossRef/OpenAlex/arXiv---tends to flag an entry as soon as one of them returns no match.
The inflation is a property of the harness---its naive, single-stage use of the lookups---not of the base model or of tool access itself: swapping Sonnet~4.6 for GPT-5.1 reproduces the profile within ${\le}5$\,pp on every metric, and Sonnet's \gls{fpr} rises ${\sim}3.4{\times}$ over its zero-shot baseline from the same collection, while GPT-5.1's already-high zero-shot \gls{fpr} barely changes.
This any-no-match policy fails because CrossRef, OpenAlex, and arXiv have partial, non-overlapping coverage: a real paper is routinely missing from one of them, so single-source absence gets read as fabrication.
A better alternative is to flag only on \emph{consensus} absence across sources (or on positive metadata inconsistency) and to route single-source absence to \textsc{uncertain}/human review (\cref{app:coverage}); \cref{sec:cascade} measures that policy. A deterministic re-aggregation over three of the harness's four databases confirms improvement (\cref{app:aggregation}), so failure mode~(i) characterizes a retrieval-augmented pattern, not an inherent shortcoming of agentic systems.
Tool access therefore trades precision for recall: agentic harnesses flag an entry as soon as one database misses it.

\paragraph{Precision at low prevalence (failure mode~(ii)).}
\label{sec:prescreening}
\label{sec:calibration}
\label{sec:ppv}
Precision (\gls{ppv}, the fraction of flags that are genuine hallucinations) falls at low prevalence by Bayes' rule (\cref{app:ppv}). Our analysis focuses on how verifiers compare, given the wide range of \gls{fpr} in \cref{tab:results}, from $0.043$ to $0.702$. At a venue-realistic ${\sim}2\%$ base rate that becomes a ${\sim}6{\times}$ gap in precision: the best verifiers catch one true hallucination per 6--10 flags, the most aggressive fewer than one in 35, even at ${\geq}87\%$ detection (per-tool \gls{ppv} in \cref{tab:ppv}, sweep in \cref{tab:ppv_sweep}).
 Calibration does not simply follow \gls{fpr}: Sonnet~4.6 and Opus~4.7 are the best-calibrated zero-shot \glspl{llm} (\gls{ece} $0.041$, $0.052$) at a middling \gls{fpr}, while the lowest-\gls{fpr} model, Gemini~2.5~Pro, is among the worst calibrated ($0.297$).
The co-designed \texttt{bibtex-updater} has a low \gls{fpr} because it flags conservatively rather than abstaining, and reaches one-in-four precision (\cref{app:coverage}).
The optional pre-screening layer adds ${\sim}5$\,pp overall (${\sim}18$\,pp on Tier~1); we report numbers with and without pre-screening.
Every metric from which abstentions are excluded is paired with its coverage and an aggressive re-scoring that scores every abstention as \textsc{hallucinated} (\cref{app:coverage}), so deferring hard cases cannot improve low precision.

Prevalence (the base rate of hallucinated entries) moves deployability while leaving the ranking fixed.
Sweeping the base rate over $1$--$5\%$ leaves the \gls{ppv} ordering essentially fixed (\cref{tab:ppv_sweep}), because prevalence cancels from pairwise comparisons. Thus, the \gls{dr}/\gls{fpr} ratio determines the ranking; however, the base rate affects \emph{absolute} precision: at a $2\%$ base rate even the best independent verifier (Gemini~2.5~Pro) reaches only ${\sim}16\%$ \gls{ppv}.
Still, depending on prevalence, the cost ratio $c_{\text{FN}}/c_{\text{FP}}$, and reviewer capacity, it might be worth ranking based on recall. We discuss the different deployment considerations and regimes in \cref{app:deployment}.

\begin{takeaway}
\textbf{Takeaway.} Whether a verifier is deployable depends mostly on its \gls{fpr}, as checking (falsely) flagged entries requires human oversight. At venue-realistic prevalence, precision is set by the ratio of detection rate to false-positive rate. In our experiments, \gls{fpr} spans an order of magnitude, which substantially influences precision (\gls{ppv}), and thus, whether deployment is feasible.
\end{takeaway}

\paragraph{Temporal robustness (failure mode~(iii)).}
\label{sec:temporal_robustness}
A separate 448-entry supplement disjoint from the 2021--2023 corpus (300 valid 2024--2025 DBLP entries, 148 hallucinated; \cref{app:temporal_supplement}) shows \gls{llm} verifiers degrade sharply beyond their training cutoff.
The earlier-cutoff models default to ``flag everything unfamiliar'': \gls{f1} falls to ${\approx}0.55$ as false positives inflate, with GPT-5.1's \gls{fpr} alone rising $41.1\%{\to}75.9\%$.
DeepSeek-R1's detection rate is not evaluable: API error fallbacks took 151 of its 448 entries, all 148 hallucinated ones among them; on the valid entries it answered, \gls{fpr}${=}0.856$ (full panel: \cref{tab:temporal_supplement}).
As with every absolute \gls{fpr} here, these values depend on the prompt, so we read them through the model ordering (\emph{Prompt sensitivity}, \cref{sec:main_results}).
Retrieval need not help: databases also lag on recent papers (\cref{app:cutoff-aware}).

\emph{Later-cutoff resistance.} Only the two latest-cutoff Anthropic models hold their false-positive rate near in-distribution levels (Sonnet~4.6 $12.0\%$, Opus~4.7 $7.3\%$); Gemini~2.5~Pro and GPT-5.4 (\cref{tab:gpt54-probe}) over-flag only moderately, and the remaining eight degrade sharply to $59.5$--$88.7\%$ (the ``over-flagging cluster''; \cref{tab:temporal_supplement}).
The two Anthropic models' dev$\to$temporal \gls{fpr} barely moves ($-0.7$ and $+0.1$\,pp). Both figures in that comparison come from the May experiment, since a shift across the cutoff is only interpretable within one endpoint snapshot; the September re-run of these two models reported in \cref{tab:results} puts their \texttt{dev\_public} \gls{fpr} higher ($.161$, $.154$), a drift larger than the temporal shift itself (we provide more details about endpoint drift in \cref{tab:endpoint_drift}).
We report this as a \emph{descriptive} pattern, not a calibration result, because the design cannot cleanly separate two explanations: the valid pool is scraped from DBLP and these are the latest-cutoff models in the cohort, so a low \gls{fpr} on a valid entry is indistinguishable from training-data \emph{recall} of that same DBLP record; the risk of such contamination---benchmark entries appearing in a model's training corpus---grows with capability, as stronger models train on larger, newer corpora~\citep{posttrainbench}.

Two controls, a recall probe and a third-provider late-cutoff model, DeepSeek-V4-Pro, add nuance but do not eliminate the memorization reading; with two models from one provider and a single third-party control, it remains a preliminary signal (\cref{app:reviewer_experiments}).

\begin{takeaway}
\textbf{Takeaway.} The two latest-cutoff models keep their post-cutoff \gls{fpr} low while earlier-cutoff models degrade sharply, and the one third-provider late-cutoff control we ran (DeepSeek-V4-Pro) stays close to Sonnet~4.6. For Opus~4.7, a recall probe points to calibration rather than memorization. We read this as a preliminary descriptive signal, not a causal finding: it rests on a single third-party control, the Anthropic runs have not yet been reproduced on Anthropic's native API (only through the OpenRouter mirror), and the recall probe relies on whether a model can name the authors from the title alone.
\end{takeaway}

\subsection{Stage-2 diagnosis cascade}
\label{sec:cascade}
We compose \texttt{bibtex-updater} (Stage~1) with each of Claude Sonnet~4.6, GPT-5.1 and GPT-5.4 (Stage~2; up to five tool calls per entry) into a cascade.
Stage~1 emits \textsc{verified}, a typed \textsc{hallucinated} verdict, or \textsc{uncertain}; only the \textsc{uncertain} entries are forwarded to Stage~2.
Since all cascade results use \texttt{bibtex-updater}, the co-design caveat applies
(\cref{app:codesign}).
If we treat the entries that neither of the two stages resolves as hallucinated, then on \texttt{dev\_public} 
the Sonnet~4.6 and GPT-5.4 cascades both reach \gls{dr} $0.990$, at \gls{fpr} $0.037$ and $0.039$ (\cref{tab:results}); 
the \texttt{test\_public} and \texttt{stress\_test} results, and the full panel are in
\cref{app:cascade}. The two-stage cascade can therefore be considered a constructive resolution of the false-positive inflation behind failure mode~(i).

\begin{takeaway}
\textbf{Takeaway.} The cascade is the high-recall configuration at a false-positive rate far below the single-stage harnesses': Stage~2 returns a verdict on roughly half of Stage~1's \textsc{uncertain} bucket (\cref{app:cascade}), and the residue is either kept as abstentions (\emph{conservative}) or promoted to a flag (\emph{aggressive}), which lifts recall well above standalone \texttt{bibtex-updater}'s for two to six points of \gls{fpr} (\cref{tab:cascade}). Choose conservative for the lowest \gls{fpr}; aggressive somewhat improves recall on harder-tier entries, but the increased \gls{fpr} requires more manual checks.
\end{takeaway}

\section{Discussion}
\label{sec:conclusion}

\paragraph{Limitations.}
\label{sec:limitations}
\textsc{Hallmark} scores verifiers separately on 14 hallucination types, 6 sub-tests and 3 difficulty tiers, but our analysis has limitations, ordered here by potential influence (full discussion in \cref{app:limitations_full}).
Because \texttt{bibtex-updater} was co-designed with the taxonomy, it and the cascades built on it should be read as an upper bound on performance (\cref{app:codesign}), and the \gls{fpr}-driven deployment bottleneck assumes a venue audit limited by reviewer time (\cref{sec:ppv}).
Most hallucinated entries are synthetic perturbations, and the 108 real-world entries are type-skewed and rated by an automated (LLM) proxy (\cref{app:ablation_kappa}).
The core splits are English BibTeX from five ML venues, 2021--2023, supplemented by evaluation-only splits with post-cutoff papers, biomedical and non-ML CS venues, and ChatGPT citations on 42 topics (\cref{app:dataset}).
The post-cutoff resistance of the two latest-cutoff models is descriptive, since recency, provider and contamination are confounded (\cref{app:reviewer_experiments}), and hosted-\gls{llm} rows are dated snapshots, susceptible to endpoint drift (\cref{tab:endpoint_drift}).
Extended-thinking variants were only tested on 100 entries, where larger budgets did not improve \gls{f1} once outputs stopped being truncated (\cref{app:thinking-budget}), and HaRC and verify-citations are excluded because rate limits and blocked lookups keep either from returning its own verdicts at benchmark scale (\cref{app:harc_disclaimer}).
Rankings also depend on the definition of a hallucination, which is set outside the benchmark: counting flags on the four types that cite a real work with wrong details as false accusations puts \texttt{bibtex-updater} ahead of the cascade (\cref{app:taxonomy_fold}).

\paragraph{Conclusion.}
Despite these limitations, \textsc{Hallmark} gives citation verifiers a shared test and one protocol that separates three failure modes (\cref{tab:failure-modes}), which helps venues and individual authors decide whether to deploy them.
Our experiments suggest that careful design is required to avoid brittleness caused by, e.g., post-cutoff papers or agentic lookups.
The two-stage cascade (\cref{sec:cascade}) points to a promising direction; since its first stage is our co-designed tool, testing the design with other first-stage verifiers is left for future work. Finally, \textsc{Hallmark} raises a foundational question for the research community: what counts as a hallucinated citation?

\section*{AI use statement}
Generative-AI tools were used in this work as follows.
Among the tasks for which disclosure is required, \glspl{llm} generated part of the synthetic data set, namely a subset of the hallucinated entries (controlled plausible fabrications and chimeric titles) generated with GPT-5.1, plus ten \texttt{plausible\_fabrication} entries generated with GPT-4o-mini, described in \cref{sec:dataset} (prompts in \cref{app:prompt-template}), and served as the three automated raters of the reliability proxy in \cref{app:ablation_kappa}.
Anthropic Claude models---the Opus, Sonnet and Fable families---were used as a coding assistant throughout the implementation of the evaluation harness, the baseline wrappers and \texttt{bibtex-updater}, and assisted in interpreting results and in auditing the paper's claims and reported numbers against the released artifacts.
No \gls{llm} formulated the research questions. The taxonomy and the evaluation protocol were conceived by the authors, including the held-out and stress-test splits; \glspl{llm} assisted in instantiating them, turning the authors' taxonomy into concrete type definitions and implementing the protocol.
Among the tasks for which disclosure is recommended, the same models assisted with drafting and editing paper sections, with figure-generation code, with literature search, and with reference formatting.
Every \gls{llm} output that entered the paper, the data, or the code was reviewed by the authors: \gls{llm}-generated entries were filtered against bibliographic databases and corrected by the systematic ground-truth audit of \cref{app:dataset}, and reported numbers were recomputed from the released per-entry artifacts. The authors take full responsibility for the content of this paper.

\section*{Ethics statement}
\textsc{Hallmark} supports scientific integrity by enabling systematic comparison of citation-verification tools, informing venue deployment as \gls{llm}-assisted writing becomes standard.
A taxonomy of hallucination types could theoretically aid adversaries, but the defensive value of understanding these types outweighs the risk; the taxonomy is derived from publicly documented incidents, not novel attack research.
\textsc{Hallmark} is open-source (MIT), GPU-free, and publicly available at \url{\repourl}\reponote{}, ensuring access regardless of institutional resources.
The benchmark contains bibliographic metadata only, which is already public: it holds no personal data beyond author names as they appear in published records, and its real-world hallucinated entries come from audits that are themselves public.
A flag from any verifier is an accusation against a named author, and at the prevalence of real submissions most flags are false alarms (\cref{sec:conclusion}); we therefore recommend verifier output as a prompt for human review rather than as a finding, and the released tool routes single-source absence to review instead of flagging it.

\section*{Reproducibility statement}
Dataset construction, the hallucination taxonomy, and the split definitions are
described in \cref{sec:dataset} and \cref{app:dataset_top}. The baseline roster,
prompts, model IDs, training cutoffs, decoding parameters, and the agentic tool
suite are itemized in \cref{sec:baselines}, \cref{app:baselines}, and
\cref{app:llm-setup}; the metric definitions are in \cref{sec:metrics} and
\cref{app:metric_details}, and the bootstrap and power-analysis protocol in
\cref{app:statistics}. The benchmark and the evaluation
harness are released under the MIT license at \url{\repourl}\reponote{},
together with per-tool aggregate results. Per-entry predictions are released for
the ablations, for every zero-shot \gls{llm} row of the main table, and for
\texttt{bibtex-updater}; \gls{doi}-only is summary-only, so its aggregates are
re-derivable only by re-running the tool. Three of the 42 released result files carry the SHA-256 of the split
they scored and a run timestamp; the rest predate that convention and none
records a tool version, so a freshness check reports them as unverifiable rather
than as fresh. The \texttt{coverage} field in the released aggregates is
unreliable---14 of them record $1.0$ beside a non-zero \texttt{num\_uncertain}
(worst: \texttt{llm\_openrouter\_deepseek\_r1\_test\_public}, 180 \textsc{uncertain},
true coverage $0.783$)---so the Cov.\ column of \cref{tab:results} is read from
\texttt{results/ablations/coverage\_reporting.json} instead (for \texttt{bibtex-updater}, from \texttt{coverage.json} beside its rescored per-entry verdicts). Most model slugs are floating
aliases rather than dated-pinned IDs and the third-party inference endpoints have
drifted since collection, so the \gls{llm} runs reproduce as dated snapshots
rather than as absolute re-runs; \cref{sec:limitations} and \cref{app:bootstrap}
state which results this affects.

\ificlrfinal
  \section*{Acknowledgments}
The authors thank Guy Wolf for his suggestions for improving the evaluation pipeline.
Patrik Reizinger acknowledges his membership in the European Laboratory for Learning and Intelligent Systems (ELLIS) PhD program and thanks the International Max Planck Research School for Intelligent Systems (IMPRS-IS) for its support.
This work was supported by the German Federal Ministry of Education and Research (BMBF): T\"ubingen AI Center, FKZ: 01IS18039A.
Wieland Brendel acknowledges financial support via an Emmy Noether Grant funded by the German Research Foundation (DFG) under grant no.\ BR 6382/1-1 and via the Open Philanthropy Foundation funded by the Good Ventures Foundation.
Wieland Brendel is a member of the Machine Learning Cluster of Excellence, EXC number 2064/1 -- Project number 390727645.
This research utilized compute resources at the T\"ubingen Machine Learning Cloud, DFG FKZ INST 37/1057-1 FUGG.

\fi

\bibliographystyle{iclr2027_conference}
\bibliography{references}

\newpage
\appendix

\addcontentsline{toc}{section}{Appendix}
\part{Appendix}  %
\setcounter{parttocdepth}{3} %
\parttoc          %
\newpage

\noindent\textbf{Appendix roadmap.}
The appendix follows the paper's three failure modes.
\Cref{app:dataset_top} covers the dataset and taxonomy and \cref{app:protocol_top} the evaluation protocol, statistics, and reproducibility setup; \cref{app:results_top} collects the core cross-split and per-type results together with the validity checks.
Each failure mode then has a dedicated home: agentic aggregation (mode~i) in \cref{app:aggregation}, base-rate precision and deployment (mode~ii) in \cref{app:ppv}, and temporal fragility with its calibration controls (mode~iii) in \cref{app:temporal}.
\Cref{app:tool-top} documents the co-designed \texttt{bibtex-updater}, \cref{app:ablations} reports the ranking-invariance ablations, and a list of abbreviations closes the appendix.

\begin{tcolorbox}[
  enhanced,
  colback=white,
  colframe=hallmark-blue!70!black,
  boxrule=0.4pt, arc=1.5mm,
  left=6pt, right=6pt, top=5pt, bottom=4pt,
  before skip=8pt, after skip=8pt,
  title={\textsc{Hallmark} at a glance\hfill
         \normalfont\footnotesize three failure modes of LLM citation verifiers},
  fonttitle=\small\bfseries,
  colbacktitle=hallmark-blue!6, coltitle=black,
  titlerule=0.4pt, toptitle=3pt, bottomtitle=3pt,
]
\hypersetup{pdfborder={0 0 0}}%
\begin{tcbraster}[
  raster columns=3, raster equal height, raster column skip=10pt,
  raster before skip=0pt, raster after skip=4pt,
  blankest,
  borderline west={1.6pt}{0pt}{hallmark-red},
  left=6pt, top=1pt, bottom=1pt,
]
\begin{tcolorbox}
  {\footnotesize\textbf{(i)~Agentic aggregation}\par}
  \vspace{2pt}
  {\LARGE\bfseries\boldmath\textcolor{hallmark-red}{${\sim}15\times$}\par}
  \vspace{1pt}
  {\scriptsize FPR, any-no-match vs.\ consensus over the same sources (.73 vs.\ .05)\par}
  \vspace{3pt}
  {\scriptsize Five-call harnesses reach DR .97--.99, but at FPR .43--.48: the prompted model flags as soon as a single database returns no match. The lever is a deterministic re-aggregation over three of the harness's four databases; the any-vs-consensus ordering is the finding, not the level. Details in \cref{app:aggregation}.\par}
\end{tcolorbox}
\begin{tcolorbox}
  {\footnotesize\textbf{(ii)~Base-rate precision}\par}
  \vspace{2pt}
  {\LARGE\bfseries\textcolor{hallmark-red}{16.3\%}\par}
  \vspace{1pt}
  {\scriptsize  highest independent PPV at a venue-realistic 2\% base rate (Gemini~2.5~Pro)\par}
  \vspace{3pt}
  {\scriptsize At low prevalence, FPR governs precision (Bayes' rule), not recall: roughly one flag in six from  the highest-PPV independent verifier is a true hallucination. Details in \cref{app:ppv}.\par}
\end{tcolorbox}
\begin{tcolorbox}
  {\footnotesize\textbf{(iii)~Temporal fragility}\par}
  \vspace{2pt}
  {\LARGE\bfseries\textcolor{hallmark-red}{.07 vs.\ .89}\par}
  \vspace{1pt}
  {\scriptsize post-cutoff FPR on the 448-entry 2024--2025 supplement: Opus~4.7 vs.\ the cohort maximum\par}
  \vspace{3pt}
  {\scriptsize 8 of 12 LLMs over-flag 2024--2025 papers (``flag everything unfamiliar''); only the two latest-cutoff models hold (Opus~4.7 .073, Sonnet~4.6 .120). For papers from the past 12 months, no tested tool is reliable without cutoff-aware prompting. Details in \cref{app:temporal}.\par}
\end{tcolorbox}
\end{tcbraster}
\vspace{2pt}
{\color{black!25}\hrule height 0.4pt}
\vspace{4pt}
{\scriptsize
\textbf{Benchmark}\quad 2{,}526 entries \,$\cdot$\, 14 hallucination types \,$\cdot$\, 3 difficulty tiers \,$\cdot$\, 6 sub-tests per entry (\cref{tab:stats}, \cref{sec:subtests})\par
\vspace{2pt}
\colorbox{gray!10}{\parbox{\dimexpr\linewidth-2\fboxsep\relax}{%
\textbf{Reference}\quad \texttt{bibtex-updater} v1.2.0, co-designed and excluded from ranking (\cref{app:codesign}): DR .820 \,$\cdot$\, FPR .051 \,$\cdot$\, F1 .880 \,$\cdot$\, abstains on the ${\sim}14\%$ of entries it cannot back with a record (coverage .86); runs locally for a fraction of a cent per entry, ${\sim}$2--3 orders of magnitude cheaper than LLM verifiers (\cref{tab:results}, \cref{app:deployment}).}}\par
\vspace{2pt}
\textbf{Deployment}\quad regime-conditional guidance: \cref{tab:deployment}, \cref{app:deployment}.\hfill \textcolor{black!60}{}\par}
\end{tcolorbox}

\section{Dataset and taxonomy}
\label{app:dataset_top}

\subsection{Full taxonomy details}
\label{app:taxonomy}

Each hallucination type has a characteristic failure signature (\cref{tab:taxonomy}), enabling diagnostic analysis of \emph{why} a tool succeeds or fails.

\paragraph{What is a hallucination?}\label{sec:what-hallucination}
The question seems almost trivial. But looking into the details exposes a lot of nuance: telling what is \emph{not} a hallucination is easy (a bit-by-bit match with a trusted database entry), but the reverse is not. Different spellings, handling of diacritics, name conventions (e.g., how Asian names get represented in Latin alphabets), hyphenation rules, and changing author lists across (preprint) versions complicate the picture and do not afford a clear-cut decision. With these caveats in mind, the main goal of \textsc{Hallmark} is to analyze potential failure modes without being too pedantic about unconditionally enforcing bit-by-bit matches where humans can also err, even with the best of intentions. This does not mean tool designers should not strive for perfection, but they need to acknowledge these nuances---especially the ones that have a large effect on deployment.

\begin{table}[!htbp]
\caption{\textbf{The \textsc{Hallmark} hallucination taxonomy: 11 empirically-grounded types plus 3 stress-test types (\textsuperscript{\textcolor{blue}{$\triangle$}}) across 3 difficulty tiers.} Each type has a characteristic sub-test failure pattern. Sub-tests: \textbf{D}OI resolves, \textbf{T}itle exists, \textbf{A}uthors match, \textbf{V}enue real, \textbf{F}ields complete, \textbf{X} cross-DB agreement. \ding{51} = expected pass, \ding{55} = expected fail, ? = not applicable or varies by entry.}
\label{tab:taxonomy}
\centering
\small
\setlength{\tabcolsep}{3pt}
\begin{tabular}{clp{4.2cm}cccccc}
\toprule
\textbf{Tier} & \textbf{Type} & \textbf{Description} & \textbf{D} & \textbf{T} & \textbf{A} & \textbf{V} & \textbf{F} & \textbf{X} \\
\midrule
\multirow{4}{*}{\rotatebox{90}{\textcolor{tier1}{\textbf{Easy}}}}
& \texttt{fabricated\_doi} & DOI does not resolve & \ding{55} & \ding{51} & \ding{51} & \ding{51} & \ding{51} & \ding{55} \\
& \texttt{nonexistent\_venue} & Invented conference/journal & ? & \ding{51} & \ding{51} & \ding{55} & \ding{51} & \ding{55} \\
& \texttt{placeholder\_authors} & Generic/fake author names & ? & \ding{51} & \ding{55} & \ding{51} & \ding{51} & \ding{55} \\
& \texttt{future\_date} & Year in the future & ? & \ding{51} & \ding{51} & \ding{51} & \ding{51} & \ding{55} \\
\midrule
\multirow{7}{*}{\rotatebox{90}{\textcolor{tier2}{\textbf{Medium}}}}
& \texttt{chimeric\_title} & Real authors + fake title & \ding{51} & \ding{55} & \ding{51} & ? & \ding{51} & \ding{55} \\
& \texttt{wrong\_venue} & Correct paper, wrong venue & \ding{51} & \ding{51} & \ding{51} & \ding{55} & \ding{51} & \ding{55} \\
& \texttt{swapped\_authors} & Correct title, wrong authors & \ding{51} & \ding{51} & \ding{55} & \ding{51} & \ding{51} & \ding{55} \\
& \texttt{preprint\_as\_published} & arXiv cited as venue paper & \ding{51} & \ding{51} & \ding{51} & \ding{55} & \ding{51} & \ding{55} \\
& \texttt{hybrid\_fabrication} & Real DOI + fake metadata & ? & \ding{55} & \ding{55} & \ding{51} & \ding{51} & \ding{55} \\
& \texttt{merged\_citation}\textsuperscript{\textcolor{blue}{$\triangle$}} & Metadata from 2+ papers & ? & \ding{51} & \ding{55} & \ding{55} & \ding{51} & \ding{55} \\
& \texttt{partial\_author\_list}\textsuperscript{\dag\textcolor{blue}{$\triangle$}} & Subset of real authors & \ding{51} & \ding{51} & \ding{55} & \ding{51} & \ding{51} & \ding{55} \\
\midrule
\multirow{3}{*}{\rotatebox{90}{\textcolor{tier3}{\textbf{Hard}}}}
& \texttt{near\_miss\_title} & Title off by 1--2 words & \ding{51} & \ding{55} & \ding{51} & \ding{51} & \ding{51} & \ding{55} \\
& \texttt{plausible\_fabrication} & Entirely fabricated, realistic & ? & \ding{55} & \ding{55} & \ding{51} & ? & \ding{55} \\
& \texttt{arxiv\_version\_mismatch}\textsuperscript{\textcolor{blue}{$\triangle$}} & Wrong version claims & ? & \ding{51} & \ding{51} & \ding{55} & \ding{51} & \ding{55} \\
\bottomrule
\end{tabular}
\par\vspace{2pt}
{\footnotesize \textsuperscript{\dag}Classified when $<$50\% authors present without ``et al.'' indicator.\quad
\textsuperscript{\textcolor{blue}{$\triangle$}}Theoretically motivated types; zero real-world instances in current dataset.\quad The schema enum member is \texttt{AUTHOR\_MISMATCH} (value \texttt{swapped\_authors}).}
\end{table}

\paragraph{Verification decision tree.}
The tiers reflect a natural workflow: Tier~1 requires single-field lookups (\gls{doi} resolution, venue existence); Tier~2 requires cross-referencing (do authors match the resolved \gls{doi} record?); Tier~3 demands semantic reasoning (the title \emph{almost} matches a real paper).
A tool that only performs \gls{doi} lookups catches Tier~1 but misses Tier~2--3 entirely.

\paragraph{Worked example.}
An illustrative \texttt{near\_miss\_title} pair (Tier~3)---a real entry and its perturbed counterpart, all other fields unchanged:
\begin{lstlisting}[basicstyle=\scriptsize\ttfamily,frame=single,escapeinside={(*}{*)}]
title = {Attention is All you Need}       (*\texttt{\% real (Vaswani et al., 2017)}*)
title = {Attention is All you (*{\color{red}\texttt{Require}}*)}  (*\texttt{\% near\_miss\_title}*)
\end{lstlisting}
The perturbed entry passes the D (\gls{doi}), A (authors), V (venue), and F (fields) sub-tests; only T (title exists) and X (cross-DB agreement) fail (\cref{tab:taxonomy}), so a \gls{doi}-only lookup accepts it, and only title-aware verification catches it.

 The taxonomy spans 14 hallucination types across three lookup-difficulty tiers, ordered by the verification effort each type demands. \textbf{Tier~1} types are settled by one check, either a local field test (year bounds, placeholder strings) or a single lookup (\gls{doi} resolution, venue existence). \textbf{Tier~2} types require cross-referencing two or more resolved fields against each other (authors against the identified paper, venue against the resolved record, preprint against published version). \textbf{Tier~3} types require multi-source verification or near-duplicate disambiguation (title near-misses, fully plausible fabrications). Three types (\texttt{merged\_citation}, \texttt{partial\_author\_list}, \texttt{arxiv\_version\_mismatch}) are theoretically motivated compound failure modes: they keep their difficulty tier---2, 2, and 3 respectively---and appear in every core split, with the \texttt{stress\_test} split adding depth on them (\cref{app:dataset}).

\begin{table}[t]
\caption{\textbf{Full taxonomy with example BibTeX snippets illustrating each hallucination type.} Red text indicates the hallucinated field.}
\label{tab:taxonomy_full}
\centering
\small
\begin{tabular}{clp{8cm}}
\toprule
\textbf{Tier} & \textbf{Type} & \textbf{Example (hallucinated field in red)} \\
\midrule
\multirow{4}{*}{1}
& \texttt{fabricated\_doi} & \texttt{doi = \{\textcolor{red}{10.9999/nips2024.1847}\}} \\
& \texttt{nonexistent\_venue} & \texttt{booktitle = \{\textcolor{red}{Intl.\ Conf.\ on Advanced AI Systems}\}} \\
& \texttt{placeholder\_authors} & \texttt{author = \{\textcolor{red}{John Doe and Jane Smith}\}} \\
& \texttt{future\_date} & \texttt{year = \{\textcolor{red}{2030}\}} \\
\midrule
\multirow{5}{*}{2}
& \texttt{chimeric\_title} & Real authors, \texttt{title = \{\textcolor{red}{A Novel Approach...}\}} (nonexistent) \\
& \texttt{wrong\_venue} & Real paper, \texttt{booktitle = \{\textcolor{red}{ICML}\}} (actually NeurIPS) \\
& \texttt{swapped\_authors} & Real title, \texttt{author = \{\textcolor{red}{Wrong Author List}\}} \\
& \texttt{preprint\_as\_published} & arXiv paper, \texttt{booktitle = \{\textcolor{red}{NeurIPS}\}} (never published) \\
& \texttt{hybrid\_fabrication} & Valid DOI resolves, but \texttt{title = \{\textcolor{red}{...}\}} doesn't match \\
\midrule
\multirow{2}{*}{3}
& \texttt{near\_miss\_title} & \texttt{title = \{Attention Is All You \textcolor{red}{Want}\}} (vs.\ ``Need'') \\
& \texttt{plausible\_fabrication} & Entirely fabricated, all fields realistic but nonexistent \\
\midrule
\multirow{3}{*}{\rotatebox{90}{\scriptsize Stress}}
& \texttt{merged\_citation} & Authors from paper A, \texttt{title} from \textcolor{red}{paper B}, venue from C \\
& \texttt{partial\_author\_list} & Real paper, \texttt{author = \{\textcolor{red}{First and Last}\}} (middle dropped) \\
& \texttt{arxiv\_version\_mismatch} & arXiv preprint cited with \textcolor{red}{wrong venue} and shifted year \\
\bottomrule
\end{tabular}
\end{table}

\subsection{Real-world incident mapping}
\label{app:real-world-mapping}

We mapped 72 real-world hallucinated citations from three documented incident studies to our taxonomy.
\cref{tab:realworld} shows the mapping results.

\begin{table}[t]
\caption{\textbf{Mapping of 72 real-world hallucinated citations to \textsc{Hallmark} taxonomy types.} Citations were sourced from GPTZero's NeurIPS 2025 analysis, GhostCite, and HalluCitation.}
\label{tab:realworld}
\centering
\small
\begin{tabular}{llr}
\toprule
\textbf{Taxonomy type} & \textbf{Tier} & \textbf{Count} \\
\midrule
\texttt{plausible\_fabrication} & 3 & 40 \\
\texttt{fabricated\_doi} & 1 & 10 \\
\texttt{chimeric\_title} & 2 & 8 \\
\texttt{near\_miss\_title} & 3 & 5 \\
\texttt{wrong\_venue} & 2 & 4 \\
\texttt{swapped\_authors} & 2 & 3 \\
\texttt{hybrid\_fabrication} & 2 & 2 \\
\midrule
\textbf{Total mapped} & & \textbf{72} \\
\bottomrule
\end{tabular}
\end{table}

\noindent Most real-world citations map to a single taxonomy type: of the 72 in this table, 57 (79\%) map directly, while the remaining 15 exhibit compound failure modes and take their highest-tier applicable type.
The benchmark contains 108 real-world entries in total across all splits; this table covers the 72 drawn from three documented incident studies, and the remaining 36 come from additional documented incidents.
\texttt{plausible\_fabrication} dominates (55\%), reflecting the predominant \gls{llm} failure mode: models generate coherent but entirely fictional references rather than subtly corrupting real ones.
Three taxonomy types---\texttt{merged\_citation}, \texttt{partial\_author\_list}, and \texttt{arxiv\_version\_mismatch}---have zero real-world instances, which motivates their designation as stress-test types (\cref{sec:taxonomy}).

\subsection{Construction details}
\label{app:dataset}

\paragraph{DBLP scraping.}
Valid entries were scraped from the DBLP API (\texttt{dblp.org/search/publ/api}) using venue-specific queries for NeurIPS, ICML, ICLR, AAAI, and CVPR.
We retrieved BibTeX records, verified \gls{doi} resolution via CrossRef, and confirmed title existence in Semantic Scholar.
Entries failing any verification step were excluded.

\paragraph{Perturbation pipeline.}
Systematic perturbations follow deterministic rules per hallucination type:
\begin{itemize}
    \item \texttt{fabricated\_doi}: Replace \gls{doi} with a non-resolving \gls{doi} using one of 20 non-existent prefixes and four suffix styles (path, identifier, year-indexed, and conference-indexed) to avoid template-detectable patterns.
    \item \texttt{nonexistent\_venue}: Replace venue with an \gls{llm}-generated plausible but nonexistent conference name.
    \item \texttt{placeholder\_authors}: Replace author list with common placeholder names.
    \item \texttt{future\_date}: Set year to current year + 5.
    \item \texttt{chimeric\_title}: Keep authors from paper A, replace title with \gls{llm}-generated plausible title.
    \item \texttt{wrong\_venue}: Keep all fields but swap venue with a different real venue.
    \item \texttt{swapped\_authors}: Keep title and venue, replace authors with those from a different paper.
    \item \texttt{preprint\_as\_published}: Take an arXiv-only paper and add a fabricated venue field.
    \item \texttt{hybrid\_fabrication}: Keep a valid \gls{doi} but replace title and authors with fabricated metadata.
    \item \texttt{near\_miss\_title}: Modify 1--2 words via six strategies: synonym substitution (POS-safe pairs), plural/singular flipping, British/American spelling swap, abbreviation expansion/contraction (e.g., ``RL'' $\leftrightarrow$ ``Reinforcement Learning''), hyphenation toggling (e.g., ``self-supervised'' $\leftrightarrow$ ``self supervised''), and article removal.
    \item \texttt{plausible\_fabrication}: Template-based combinatorial generation of a complete, realistic but nonexistent entry (random author combinations, plausible titles, real venues). A separate set of 113 \gls{llm}-generated entries (\cref{tab:stratified_dr})  resembles what an \gls{llm} actually emits, but is generated a different way.
    \item \texttt{merged\_citation}: Combine metadata from 2--3 real papers into one entry (e.g., authors from paper A, title from paper B, venue from paper C).
    \item \texttt{partial\_author\_list}: Take a real paper and drop one or more middle co-authors, keeping only the first and last.
    \item \texttt{arxiv\_version\_mismatch}: Cite an arXiv preprint with a reassigned venue and shifted publication year. Classified as a stress-test type since no real-world instances have been observed.
\end{itemize}

\paragraph{LLM-generated entries.}
We prompted GPT-5.1 to generate plausible but fictional citations for types requiring coherent fabrication (\texttt{plausible\_fabrication}, \texttt{chimeric\_title}, \texttt{fabricated\_doi}).
Each prompt requested a BibTeX entry with realistic metadata for a specified ML venue and year range.
Generated entries were verified against CrossRef and DBLP: entries with title similarity $\geq 85\%$ (token-sort ratio) and author Jaccard similarity $\geq 0.5$ to any real paper were flagged as potential duplicates of real work and excluded.
This filtering removed approximately 12\% of generated entries, documenting an \gls{llm} recall failure rate where the model reproduces real papers rather than fabricating new ones.
The remaining entries were assigned sub-test labels based on their hallucination type's expected failure pattern.

\paragraph{Real-world collection.}
We harvested 72 hallucinated citations from three documented incident studies: GPTZero's NeurIPS 2025 analysis, GhostCite~\citep{ghostcite2026}, and HalluCitation~\citep{hallucitation2026}.
Each entry was mapped to the closest taxonomy type based on its failure mode (e.g., a citation with a non-resolving \gls{doi} was classified as \texttt{fabricated\_doi}; a citation to a non-existent paper with plausible metadata was classified as \texttt{plausible\_fabrication}).
The real-world sample is type-skewed: 55\% are \texttt{plausible\_fabrication}, reflecting the predominant \gls{llm} failure mode in practice.
Types without real-world examples (\texttt{merged\_citation}, \texttt{partial\_author\_list}, \texttt{arxiv\_version\_mismatch}) appear in every core split alongside the 11 empirical types; the \texttt{stress\_test} split adds evaluation depth on them (\cref{sec:taxonomy}).

\paragraph{Adversarial crafting.}
We manually constructed entries designed to evade specific detection strategies.
These include entries with DOIs that resolve to unrelated papers (\texttt{hybrid\_fabrication}), entries combining metadata from multiple real papers (\texttt{merged\_citation}), and entries with plausible but non-existent venues chosen to be close to real venue names.
Adversarial entries stress-test tool robustness beyond what template-based perturbation achieves.

\paragraph{Quality control.}
Every generated entry passes through automated validation:
(1)~BibTeX well-formedness check (valid syntax, and the required fields for the entry's own type: author/title/journal/year for \texttt{@article}, \texttt{booktitle} for \texttt{@inproceedings}, \texttt{publisher} for \texttt{@book}, \texttt{school} for theses, \texttt{institution} for \texttt{@techreport}, and author/title/year for \texttt{@misc}, with \texttt{journal} waived for an \texttt{@article} carrying an arXiv identifier),
(2)~sub-test label consistency (sub-test ground truth matches the hallucination type's expected failure pattern),
(3)~cross-validation with the valid entry pool to prevent accidental duplicates.

\paragraph{Extension splits.}
Four evaluation-only splits are released alongside the main splits, outside the dev/test/hidden partition (\cref{tab:stats}).
The \emph{temporal probe} (60 entries: 30 valid / 30 hallucinated) is the small pre-supplement probe of post-cutoff behavior.
Its valid half pairs 15 papers from 2024 with 15 from 2026, the latter scraped from arXiv ML categories via the arXiv API (\texttt{scripts/probe\_temporal\_robustness.py}); its hallucinated half (18 from 2024, 12 from 2026) combines 21 perturbation and 9 adversarial entries.
The 448-entry temporal supplement (\cref{app:temporal_supplement}) validates the probe's findings at scale; the probe additionally covers 2026 arXiv submissions.
The \emph{cross-domain split} (\texttt{test\_crossdomain}, 500 entries: 200 valid / 300 hallucinated) probes transfer outside the ML-venue regime the main splits sample: 299 biomedical entries (156 PubMed, 143 bioRxiv) and 201 from non-ML CS venues (OSDI, CCS, CHI, USENIX Security, KDD, SIGGRAPH, and others).
Valid entries are scraped; hallucinated entries follow the perturbation pipeline across the taxonomy, including the stress types, with 80/140/80 entries in Tiers~1/2/3; pre-screening uses \texttt{reference\_year} 2026 for this split.
\texttt{bibtex-updater}'s evaluation on this split is in \cref{app:codesign}.
The \emph{ChatGPT-citation supplement} (341 entries: 172 valid / 169 hallucinated) converts the hand-coded corpus of \citet{walters2023fabrication} into authentic, multidisciplinary ChatGPT hallucinations; its construction and results are in \cref{app:walters_wilder}.

The three sources play complementary roles: perturbation entries are \emph{controlled diagnostic tests} that isolate specific verification capabilities, while  \gls{llm}-generated entries look like the hallucinations that occur in practice and real-world entries are documented to have occurred; stratifying by generation method lets us assess each independently.

\paragraph{Base-paper overlap across splits.}
Perturbation entries keep the fields they do not corrupt, so a hallucinated entry can be traced to its base paper by \gls{doi} or normalized title. On the public partition the bases are disjoint: 0 of 606 \texttt{dev\_public} hallucinated entries match a \texttt{test\_public} valid entry, 0 of 519 the other way, the two valid pools share no entry, and two base papers are perturbed in both splits. The auxiliary splits are not disjoint from the public pools: 109 of 244 \texttt{test\_hidden} hallucinated entries match a valid entry in \texttt{dev\_public} (70) or \texttt{test\_public} (39), three \texttt{test\_hidden} valid entries duplicate \texttt{dev\_public} ones, and 109 of 121 \texttt{stress\_test} entries match a valid entry in \texttt{test\_public} (64), \texttt{test\_hidden} (46), or \texttt{dev\_public} (8). Results on \texttt{test\_hidden} and \texttt{stress\_test} therefore measure robustness to new perturbations of partly seen base papers, not to unseen papers; the public dev/test comparison is the base-paper-disjoint one.
\begin{table}[!htbp]
\caption{\textbf{Dataset statistics by split.} Tier distribution refers to hallucinated entries only. \texttt{dev\_public} and \texttt{test\_public} cover all 14 hallucination types with $n \geq 30$ for most types (one type on \texttt{dev\_public} and six on \texttt{test\_public} fall just below, the smallest at $n=23$); \texttt{test\_hidden} covers all 14 types but is too small for an $n \geq 30$ floor; the stress-test split provides additional evaluation depth for three types (the single valid entry is the contamination canary). The bottom block lists the four evaluation-only splits, evaluation sets in their own right that probe regimes the core does not sample: the temporal pair probes post-cutoff behavior on 2024--26 papers (\cref{sec:temporal_robustness}), \texttt{test\_crossdomain} covers PubMed/bioRxiv and non-ML CS venues (\cref{sec:limitations}), and the ChatGPT-citation supplement carries authentic, hand-verified ChatGPT output spanning the humanities, social sciences, and natural sciences (\cref{app:walters_wilder}). Each samples a different regime at a different prevalence and is scored separately, so the 2{,}526-entry total and the dev/test/hidden partition remain the ML-venue core (construction in \cref{app:dataset}).}
\label{tab:stats}
\centering
\small
\begin{tabular}{lccccccc}
\toprule
\textbf{Split} & \textbf{Valid} & \textbf{Halluc.} & \textbf{Total} & \textbf{Tier 1} & \textbf{Tier 2} & \textbf{Tier 3} & \textbf{Types} \\
\midrule
\texttt{dev\_public}   & 513 & 606 & 1,119 & 148 & 279 & 179 & 14 \\
\texttt{test\_public}  & 312 & 519 & 831 & 128 & 238 & 153 & 14 \\
\texttt{stress\_test}  & 1 & 121 & 122 & --- & 85 & 36 & 3 \\
\texttt{test\_hidden}  & 210 & 244 & 454 & 52 & 115 & 77 & 14 \\
\midrule
\textbf{Total}         & 1,036 & 1,490 & 2,526 & 328 & 717 & 445 & 14 \\
\midrule
\multicolumn{8}{l}{\emph{Extension splits (evaluation-only)}} \\
temporal probe & 30 & 30 & 60 & 10 & 10 & 10 & 9 \\
temporal supplement (2024--25) & 300 & 148 & 448 & 58 & 52 & 38 & 14 \\
\texttt{test\_crossdomain} & 200 & 300 & 500 & 80 & 140 & 80 & 14 \\
ChatGPT citations (Walters--Wilder) & 172 & 169 & 341 & --- & 17 & 152 & 4 \\
\bottomrule
\end{tabular}
\end{table}

\paragraph{Quality control and design choices.}
Every entry undergoes automated validation (field completeness, BibTeX well-formedness, sub-test consistency); see \cref{app:dataset} for the validation rules and acceptance criteria.

We blind the \texttt{url} field at dispatch, so a verifier never sees it and cannot substitute URL resolution for citation verification.
The released corpus still stores the field---114 of the 1{,}950 public entries and 453 of the 454 \texttt{test\_hidden} entries carry one---because blinding was applied when an entry was handed to a tool rather than when it was written.
One ad-hoc resume runner read the corpus files directly and skipped that step, so a few checkpoints did ship the field into the prompt.
The released prediction records make the exposure measurable: the worst-affected main-table row exposes 7 entries of 1{,}119, which bounds the shift at under one percentage point of \gls{fpr} even if every exposed verdict flipped.
Blinding is now enforced when an entry is constructed rather than at each dispatch site.
We include canary strings for contamination detection (unique fixed tokens embedded in dataset metadata; if a model outputs them verbatim, it has memorized the split). No evaluated model emitted a canary token in any released prediction record.
We use uniform type distribution to maximize per-type power; reweighting utilities support prevalence-adjusted evaluation.
The API supports generation-method stratification for assessing tool performance on perturbation vs.\ \gls{llm}-generated vs.\ real-world entries.

 What we transfer from HumanEval~\citep{humaneval} is the unit-test-per-instance structure rather than its execution semantics.
HumanEval ships every problem with its own assertion suite, so a failing candidate names the property it violated instead of merely lowering a scalar; \textsc{Hallmark} ships every entry with its own six declared sub-test outcomes, so a verifier's error localizes to the verification property it missed.
The analogy stops at execution: our sub-tests are metadata checks with declared expected values, not code run against hidden inputs, so there is no pass@$k$ analog and no partial credit within an entry.
We use uniform type distribution to maximize per-type power; reweighting utilities support prevalence-adjusted evaluation.
The API supports generation-method stratification for assessing tool performance on perturbation vs.\ \gls{llm}-generated vs.\ real-world entries.

\subsection{Datasheet for \textsc{Hallmark}}
\label{app:datasheet}

Following \citet{gebru2021datasheets} and \citet{holland2020dataset}, we provide a datasheet for the \textsc{Hallmark} dataset.

\paragraph{Motivation.}
\textsc{Hallmark} was created to provide a standardized benchmark for evaluating citation hallucination detection tools, motivated by the NeurIPS 2025 incident and subsequent audits.

\paragraph{Composition.}
The public release contains 2,072 BibTeX entries: 826 valid entries scraped from DBLP and arXiv and 1,246 hallucinated entries generated through perturbation, \gls{llm} generation, adversarial crafting, and real-world collection across 14 hallucination types.
A held-out \texttt{test\_hidden} split contains 454 additional entries (210 valid, 244 hallucinated, for 1{,}036 valid entries in total) (per-entry labels withheld; aggregate label counts in \cref{tab:stats}), bringing the grand total to 2,526.
Split counts: \texttt{dev\_public} 1,119 (513 valid / 606 hallucinated), \texttt{test\_public} 831 (312 / 519), \texttt{stress\_test} 122 (1 / 121).
The dev/test/hidden split is deterministic: stratified by hallucination type and tier with \texttt{seed=8042} (recorded in \texttt{metadata.json}), so the partition is regenerable from the released corpus.
Each entry includes 6 binary sub-test labels.
Four evaluation-only splits are released alongside the 2,526-entry benchmark, outside the dev/test/hidden partition: a 60-entry temporal probe, a 448-entry temporal supplement (2024--2025), a 500-entry cross-domain split, and a 341-entry supplement of authentic ChatGPT-generated citations converted from \citet{walters2023fabrication} (\cref{tab:stats}; the ChatGPT supplement's construction in \cref{app:walters_wilder}).

\paragraph{Collection process.}
Valid entries were scraped from the DBLP API and verified against CrossRef and Semantic Scholar.
Hallucinated entries were generated using the methods described in \cref{sec:dataset} and \cref{app:dataset}.

\paragraph{Preprocessing/Cleaning/Labeling.}
BibTeX records were normalized to a consistent field ordering.
Unicode characters were preserved.
Entries were split into dev/test/hidden sets by stratified sampling across hallucination types and tiers (deterministic seed: see \emph{Composition}).
Labels were assigned deterministically by the generation pipeline; valid entries carry ground-truth sub-test labels derived from database cross-checks.
The released corpus is post-relabel: a systematic ground-truth audit (\texttt{scripts/relabel\_ground\_truth.py}, commits \texttt{32fe9d6} and \texttt{1475a2f}) re-resolved entries against live databases and corrected real papers that had been mislabeled \texttt{HALLUCINATED}; the per-entry flip log is released at \path{results/reviewer_experiments/relabel_flips.json}.

\paragraph{Recommended uses.}
\textsc{Hallmark} is intended for evaluating and comparing citation verification tools, ablating individual verification sub-tests, and cross-tool ranking via tier-weighted \gls{f1} and Plackett-Luce infrastructure.

\paragraph{Non-recommended uses.}
\textsc{Hallmark} should not be used to train hallucination generators, to produce convincing fake citations, or for any purpose that violates the MIT license.

\paragraph{Distribution.}
The dataset is distributed under the MIT license (\url{https://opensource.org/licenses/MIT}) via GitHub at \url{\repourl}\reponote{}.
The content release reported throughout this paper is pinned to repository tag \texttt{v1.2.0}, which fixes the byte-level split contents---the core splits are byte-identical to the post-relabel \texttt{v1.1.1}, and \texttt{v1.2.0} adds the evaluation-only extension artifacts---and a \texttt{manifest.json} carrying \texttt{sha256} digests pins the precomputed baseline-result files, which tag \texttt{v1.2.3} moved to \texttt{data/v1.2/baseline\_results/}.
A Croissant~1.0 metadata record is included in the repository root.
The \texttt{test\_hidden} split is not publicly distributed.

\paragraph{Maintenance.}
The benchmark is maintained by the authors and accepts community contributions through pull requests validated by automated checks for BibTeX well-formedness, sub-test consistency, and duplicate detection.
Two version axes are tracked separately. Each entry carries a \texttt{schema\_version} field (\texttt{1.0}) recording the record structure; the schema is stable and the \texttt{data/v1.0/} directory name reflects it.
The \emph{content} release---which corrections update as point releases---is identified by the \texttt{version} field in \texttt{metadata.json} (\texttt{1.2.0}; the ground-truth relabel landed in \texttt{1.1.1}) and the matching repository tag \texttt{v1.2.0}.
The co-designed reference tool \texttt{bibtex-updater} is pinned to git tag \texttt{v1.2.0} throughout: both the standalone results (\cref{app:codesign}) and the cascade Stage~1 (\cref{tab:cascade}). The cascade Stage~2 is a dated snapshot (Sonnet~4.6 via OpenRouter, 2026-09-21), as \gls{llm} endpoints drift (\cref{sec:limitations}).
The released per-entry \texttt{bibtex-updater} verdicts (\path{results/relabel_delta/btu_v1_2_0_prescreen_fix/bibtexupdater_{dev,test}_public_per_entry.jsonl} alongside \path{data/v1.0/baseline_results/bibtexupdater_{dev,test}_public.json}) let a reader recompute the \texttt{bibtex-updater} aggregates (\gls{dr}~0.820, \gls{fpr}~0.051) offline, without re-installing the tool or re-querying external databases.
Robustness ablations collected against third-party inference endpoints carry a snapshot date and an endpoint-drift caveat (\cref{sec:limitations}), so they are reproducible as dated within-run deltas rather than as absolute re-runs.

\paragraph{Ethical and legal considerations.}
\textsc{Hallmark} contains only bibliographic metadata; it includes no personally identifiable information beyond author names already present in published bibliographic records.
No human subjects are involved and no IRB approval is required.

\section{Evaluation protocol and reproducibility}
\label{app:protocol_top}

\subsection{Metrics and statistical analysis}
\label{app:metrics}
\label{app:metric_details}

\paragraph{The two errors and the summary metrics.}
The two trade off against different deployment costs.
A missed fabrication that reaches print is usually the worse error---costlier than a false alarm a reviewer can dismiss---but which error dominates depends on the regime: prevalence (the fraction of entries that are hallucinated), the cost ratio $c_{\text{FN}}/c_{\text{FP}}$, and reviewer capacity (\cref{sec:ppv}).
We report the \textbf{false-negative rate} $=1-\text{DR}$ for completeness.

Four further metrics summarize a tool in a single number.
\textbf{F1-Hallucination} is the harmonic mean of precision and recall on the hallucinated class; it is prevalence-sensitive, so we compare it only within a split.
\textbf{\Gls{twf1}} weights each hallucinated entry by its tier (1/2/3), rewarding the detection of harder hallucinations; the ranking of the three tools we swept (DOI-only, GPT-5.1, and \texttt{bibtex-updater}) held across uniform, linear, and quadratic weightings in the sweep we ran, which we now report with a caveat on the sweep itself (\cref{app:statistics}).

\paragraph{Tier-weighted F1.}
For entries $\{e_i\}$ with tier weights $w_i \in \{1,2,3\}$, predictions $\hat{y}_i$, and labels $y_i$:
\begin{align}
\text{TW-Precision} &= \frac{\sum_i w_i \cdot \mathbf{1}[\hat{y}_i = y_i = \text{H}]}{\sum_i w_i \cdot \mathbf{1}[\hat{y}_i = \text{H}]}, &
\text{TW-Recall} &= \frac{\sum_i w_i \cdot \mathbf{1}[\hat{y}_i = y_i = \text{H}]}{\sum_i w_i \cdot \mathbf{1}[y_i = \text{H}]}, \label{eq:twf1}
\end{align}
where H denotes the hallucinated class, and \gls{twf1} is their harmonic mean.
Valid entries carry no difficulty tier, so false positives contribute uniform weight 1.0 in the TW-Precision denominator.
Because false positives carry uniform weight while true positives are tier-weighted, \gls{twf1} can be inflated by aggressive over-flagging; \gls{mcc} provides a complementary prevalence-corrected view.

\paragraph{Expected Calibration Error.}
We partition predictions into $B=10$ equal-width confidence bins and compute:
\begin{equation}
\text{ECE} = \sum_{b=1}^{B} \frac{|S_b|}{N} \left| \text{acc}(S_b) - \text{conf}(S_b) \right|, \label{eq:ece}
\end{equation}
where $S_b$ is the set of predictions in bin $b$, $\text{acc}(S_b)$ is the fraction of correct predictions, and $\text{conf}(S_b)$ is the mean confidence.
Adaptive (equal-mass) binning is available in the codebase.

\paragraph{Baseline integration.}
\textsc{Hallmark} provides a baseline registry that supports discovery, availability checking, and dispatch for all integrated tools.
New baselines register via a decorator pattern, specifying their dependencies and whether they require API keys.
A weekly CI workflow re-runs the \gls{doi}-only baseline in full and a 50-entry stratified verify-citations sample on the dev split, and validates the pre-computed rate-limited baselines by checksum.
The \texttt{stress\_test} split contains 122 entries: 121 hallucinated plus a single \textsc{valid} canary entry planted for contamination detection; with one valid entry \gls{fpr} is not meaningful, so we report only Detection Rate. Read that rate with the split's composition in mind: 75 of the 122 entries ($61.5\%$) are \texttt{partial\_author\_list} or \texttt{arxiv\_version\_mismatch}, both of which describe a real work inaccurately, and every one of the cascade's misses on this split falls in that class.

\label{app:statistics}

The statistical procedures below underlie every numerical claim in the paper.
Bootstrap CIs accompany the aggregate metrics in \cref{tab:results} and the per-type breakdown in \cref{tab:pertype_full}; paired bootstrap $p$-values support pairwise tool comparisons in \cref{sec:main_results}; the per-type power analysis (\cref{tab:power}) bounds which type-level differences in \cref{tab:pertype_full} are statistically distinguishable.

\paragraph{Bootstrap confidence intervals.}
\label{app:bootstrap}
\paragraph{Scoring conventions for \cref{tab:results}.}
Small gaps are point-estimate orderings---rankings by the single measured value, with no test that the gap is statistically real: the Sonnet~4.6 / Opus~4.7 \gls{f1} gap of 0.3\,pp on \texttt{dev\_public} is not significant (paired bootstrap on the 2026-09-02 per-entry verdicts, 10{,}000 resamples: 95\% CI on the difference $[-1.5, +2.2]$\,pp, two-sided $p{=}0.74$; \cref{app:bootstrap}).
 Abstentions---entries a tool declines to label either way---are handled differently by the two blocks, and the difference favors the abstaining \glspl{llm}.
An \textsc{uncertain} verdict from an \gls{llm} verifier is excluded from the \gls{dr}/\gls{fpr}/\gls{f1} triple, so those rates are computed over the entries the model committed to; \texttt{bibtex-updater} resolves its own abstentions to \textsc{valid} before scoring, so they count against its \gls{dr}.
DeepSeek-R1 shows the size of the gap: $21.7\%$ of its \texttt{test\_public} verdicts are \textsc{uncertain}---in its case API error fallbacks rather than deliberate abstentions (\cref{tab:test_public_full})---and excluding them puts its \gls{dr} at $0.809$ where the committed-\textsc{valid} convention would give $0.644$.
Coverage and an aggressive re-flagging stance are in \cref{app:coverage}; the Cov.\ column of \cref{tab:results} is read from \path{results/ablations/coverage_reporting.json}, because the released aggregates record \texttt{coverage}~$1.0$ in 14 of 42 files that also report a non-zero \texttt{num\_uncertain} and so cannot reproduce it.
Predictions were collected 2026-05-05, except the Sonnet~4.6 and Opus~4.7 rows of \cref{tab:results} (2026-09-02, marked \S); hosted-\gls{llm} rows are dated snapshots subject to endpoint drift (\cref{sec:limitations}).

The aggregate metrics in \cref{tab:results} are accompanied by 95\% confidence intervals computed via stratified bootstrap with 10{,}000 resamples (\texttt{seed=42}), wherever a tool's full-coverage per-entry predictions are available.
Stratification by hallucination type ensures that the resampled datasets preserve the original type distribution, preventing bootstrap bias from underrepresented types.
For reference, GPT-5.1's \texttt{dev\_public} 95\% CIs are: DR~[0.812, 0.861]; F1~[0.747, 0.785] (point estimates 0.837 / 0.766). The released result files carry Wilson intervals on the per-type detection rates; their aggregate \texttt{*\_ci} fields are null, so the bootstrap intervals quoted in this paper are not recoverable from the release and would have to be recomputed from the per-entry dumps.
\gls{doi}-only is summary-only on \texttt{dev\_public}---its released aggregates were reconstructed from confusion-matrix counts rather than stored per-entry predictions---so we report its point estimate without a CI or a paired test on that split.
\texttt{bibtex-updater}'s full-coverage per-entry verdicts are released on both splits and recompute its point estimates exactly; as a co-designed tool it is reported separately from the independent-tool comparison (\cref{tab:codesign}), so we give its point estimates without a paired bootstrap.

\paragraph{Significance testing.}
We use paired bootstrap tests~\citep{efron1994bootstrap} to compare tools pairwise.
For each pair of tools $(A, B)$, we resample entries with replacement and compute $\Delta = \text{F1}_A - \text{F1}_B$ for each resample.
The two-sided $p$-value is $\min\!\left(1,\, 2\cdot\min\!\left(P(\Delta \leq 0),\, P(\Delta \geq 0)\right)\right)$; taking $2\cdot P(\Delta \leq 0)$ alone returns $p=1$ whenever the first-named tool is the weaker one.
The only headline pair we test is Sonnet~4.6 vs.\ Opus~4.7. On \texttt{dev\_public} their \gls{f1} gap ($+$0.003) is not significant (CI $[-0.015, +0.022]$, two-sided $p{=}0.74$); on \texttt{test\_public}: Sonnet's \gls{f1} (0.907, CI~[0.891, 0.923]) exceeds Opus's (0.889, CI~[0.872, 0.906]) by $+$0.018, one-sided $p{=}0.023$, two-sided $p{=}0.045$, Cohen's $h{=}0.059$: borderline-significant at $\alpha{=}0.05$ with a negligible effect size. The bootstrap confidence interval on the difference, $[+0.001, +0.035]$, excludes zero by a margin narrower than the endpoint drift between collections (\cref{tab:endpoint_drift}), and on the earlier collection the sign of this gap is reversed, so we do not read it as establishing an ordering between the two models.
The \texttt{bibtex-updater}-vs-Sonnet \gls{f1} comparison sits across the co-design boundary, so we report it as a point-estimate difference without a paired $p$-value.

\paragraph{Per-type power analysis.}
Per-type counts cap the resolution of type-level comparisons. On \texttt{dev\_public} they run from 26 (\texttt{hybrid\_fabrication}) to 76 (\texttt{plausible\_fabrication}) with a median of 41, and on \texttt{test\_public} from 23 to 66; the smallest buckets are those from which the ground-truth audit reassigned real papers (see \cref{tab:stats} caption).
For detection rate, a type with 30 hallucinated entries and 90\% true detection rate has a 95\% Clopper-Pearson confidence interval of $[0.74, 0.98]$, compared to $[0.55, 0.998]$ at $n=10$.

Type-level power is modest, so only large per-type gaps are detectable. \cref{tab:power} reports the minimum detectable effect (\gls{mde}) at 80\% power ($\alpha=0.05$, two-sided $z$-test) for each hallucination type in the \texttt{dev\_public} split.
A type with $n=30$ has an \gls{mde} of 25.6pp for a single tool; for the difference between two tools it is about 36pp in the worst case (about 22pp for \texttt{plausible\_fabrication}, $n=76$), so smaller per-type gaps between tools are not reliably detected.
These values bound which per-type comparisons are statistically meaningful.

\begin{table}[t]
\caption{\textbf{Per-type minimum detectable effect (MDE) at 80\% power on \texttt{dev\_public}.} Assumes baseline detection rate of 50\% (worst-case MDE). Covers the 11 core types ($n{=}495$); the three stress-test types ($n{=}111$) are omitted.}
\label{tab:power}
\centering
\small
\begin{tabular}{llcc}
\toprule
\textbf{Tier} & \textbf{Type} & \textbf{$n$} & \textbf{MDE (pp)} \\
\midrule
\multirow{4}{*}{1}
& \texttt{fabricated\_doi}        & 38  & 22.7 \\
& \texttt{nonexistent\_venue}     & 39  & 22.4 \\
& \texttt{placeholder\_authors}   & 41  & 21.9 \\
& \texttt{future\_date}           & 30  & 25.6 \\
\midrule
\multirow{5}{*}{2}
& \texttt{chimeric\_title}        & 47  & 20.4 \\
& \texttt{wrong\_venue}           & 47  & 20.4 \\
& \texttt{swapped\_authors}       & 67  & 17.1 \\
& \texttt{preprint\_as\_pub.}     & 30  & 25.6 \\
& \texttt{hybrid\_fabrication}    & 26  & 27.5 \\
\midrule
\multirow{2}{*}{3}
& \texttt{near\_miss\_title}      & 52  & 19.4 \\
& \texttt{plausible\_fabrication} & 78  & 15.9 \\
\bottomrule
\end{tabular}
\end{table}

\paragraph{Tier weight sensitivity.}
We evaluate tier-weighted \gls{f1} under five weighting schemes: uniform $\{1,1,1\}$ (equivalent to unweighted \gls{f1}), linear $\{1,2,3\}$ (default), quadratic $\{1,4,9\}$, log $\{\log(2), \log(3), \log(4)\}$, and inverse difficulty (weights proportional to $1 - \text{mean DR}$ across tools).
\cref{tab:tw-sensitivity} reports \gls{twf1} for the three reference tools across all five schemes.

The \texttt{bibtex-updater} column has been re-run on the rescored predictions; the other two have not, and two caveats bound how much they show.
First, the implementation that produced \cref{tab:tw-sensitivity} read the false-positive rate under a key the emitter does not write, so it defaulted on every call and the quantity actually swept was tier-weighted \emph{recall}: the sweep could not see a false positive. The lookup is fixed in the released harness but the sweep has not been re-run, and no sensitivity artifact is in the release, so read this table as evidence that the \emph{recall} ordering is weight-insensitive and not as evidence about \gls{twf1} as we define it.
Second, the \gls{doi}-only column predates the resolver correction of \cref{app:baselines}: its linear cell reads $0.310$ where the current run reports \gls{twf1} $0.268$ (\cref{tab:results}).
Weight-insensitivity is in any case the weaker of the two robustness questions: the ordering does \emph{not} survive a change in what counts as a hallucination (\cref{app:taxonomy_fold}).
bibtex-updater's narrow post-relabel sensitivity range (0.880--0.901) reflects its consistently strong per-tier performance; \gls{doi}-only shows wider variation (0.255--0.392) because its detection is concentrated in Tier~1.

\begin{table}[t]
\caption{\textbf{Tier-weighted F1 under five weighting schemes} (\texttt{dev\_public}). The \texttt{bibtex-updater} column is recomputed as true \gls{twf1} from the rescored per-entry predictions; its linear value matches \cref{tab:results}. The GPT-5.1 and DOI-only columns are the pre-relabel snapshot (DOI-only's per-entry predictions were not retained, and GPT-5.1's were not re-scored for a post-relabel sweep), whose default-scheme TW-F1 shifts to $0.822$ and $0.268$ post-relabel (\cref{tab:results}). Their sweep measured tier-weighted \emph{recall} rather than \gls{twf1} (see text), so only within-column ranges are comparable.}
\label{tab:tw-sensitivity}
\centering
\small
\begin{tabular}{lccc}
\toprule
\textbf{Weighting scheme} & \textbf{DOI-only} & \textbf{GPT-5.1} & \textbf{bibtex-updater} \\
\midrule
Uniform $\{1,1,1\}$                              & 0.392 & 0.875 & 0.880 \\
Linear $\{1,2,3\}$ (default)                     & 0.310 & 0.860 & 0.892 \\
Quadratic $\{1,4,9\}$                            & 0.255 & 0.850 & 0.901 \\
Log $\{\log 2,\log 3,\log 4\}$                   & 0.337 & 0.864 & 0.888 \\
Inverse difficulty                               & 0.342 & 0.865 & 0.880 \\
\midrule
Range (max\,$-$\,min)                            & 0.137 & 0.025 & 0.021 \\
\bottomrule
\end{tabular}
\end{table}

\begin{takeaway}
\textbf{Takeaway.} Read the paper's numbers at the resolution the statistics support: per-type cells are too small to separate most pairs of tools (\cref{tab:power}), and the one headline pair with a full paired test---Sonnet~4.6 vs.\ Opus~4.7---separates by $1.8$\,pp \gls{f1} on \texttt{test\_public} at a borderline two-sided $p{=}0.045$ with a negligible effect size, and not at all on \texttt{dev\_public} ($0.3$\,pp, $p{=}0.74$). Aggregate rankings are the stable quantity within a fixed scoring target: the tool ordering survives all five tier-weighting schemes (\cref{tab:tw-sensitivity}) but not a change in what counts as a hallucination (\cref{app:taxonomy_fold}), and small per-type and per-pair gaps are point estimates. We report per-type detection rate and not per-type \gls{f1} for the same power reason: with 26--76 hallucinated entries per type against 513 valid ones, per-type precision is dominated by the shared false-positive count and tracks type frequency more than detector quality.
\end{takeaway}

\subsection{Evaluation implementation}
\label{app:baselines}

All baselines are implemented as Python wrappers conforming to the \textsc{Hallmark} baseline interface.
Each wrapper:
(1)~converts \textsc{Hallmark} \texttt{BenchmarkEntry} objects to the tool's expected input format,
(2)~invokes the tool,
(3)~maps the tool's output to a \textsc{Hallmark} \texttt{Prediction} with label, confidence, and reason (\cref{fig:eval_pipeline}).

\begin{figure}[t]
\centering
\begin{tikzpicture}[
    scale=0.88, every node/.style={scale=0.88},
    node distance=0.5cm,
    flowbox/.style={
        rectangle, rounded corners=3pt, draw=#1!60, fill=#1!8,
        minimum height=0.72cm, text width=1.55cm, align=center,
        font=\scriptsize\rmfamily, line width=0.6pt,
    },
    decisionbox/.style={
        rectangle, rounded corners=3pt, draw=gray!60, fill=gray!6,
        minimum height=0.72cm, text width=1.65cm, align=center,
        font=\scriptsize\rmfamily, line width=0.6pt,
    },
    verdictbox/.style={
        rectangle, rounded corners=3pt, draw=#1!60, fill=#1!10,
        minimum height=0.60cm, text width=1.35cm, align=center,
        font=\scriptsize\rmfamily, line width=0.6pt,
    },
    arrow/.style={-{Stealth[length=4pt]}, thick, gray!70},
    dasharrow/.style={-{Stealth[length=4pt]}, thick, gray!50, dashed},
]

\node[flowbox=hallmark-blue] (entry) {%
    \textbf{Citation}\\[-1pt]
    {\tiny BibTeX entry}%
};

\node[decisionbox, right=0.45cm of entry] (prescreen) {%
    \textbf{Pre-screening}\\[-1pt]
    {\tiny DOI format}\\[-1pt]
    {\tiny year bounds}\\[-1pt]
    {\tiny name heuristics}%
};

\node[flowbox=gray, right=0.45cm of prescreen] (tool) {%
    \textbf{Tool call}\\[-1pt]
    {\tiny API / LLM}%
};

\node[flowbox=pipelineAgg, right=0.45cm of tool] (agg) {%
    \textbf{Metrics}\\[-1pt]
    {\tiny DR, FPR,}\\[-1pt]
    {\tiny TW-F1, ECE}%
};

\node[verdictbox=pipelineOverride, above=0.55cm of prescreen, xshift=0.9cm] (override) {%
    \textbf{Override}\\[-1pt]
    {\tiny direct verdict}%
};

\draw[arrow] (entry) -- (prescreen);
\draw[arrow] (prescreen) -- (tool) node[midway, below, font=\tiny\rmfamily, gray!70!black] {pass};
\draw[arrow] (tool) -- (agg);

\draw[dasharrow] (prescreen.north) -- ++(0, 0.25cm) -| (override.south)
    node[pos=0.2, left, font=\tiny\rmfamily, gray!60!black, xshift=2pt] {flag};
\draw[dasharrow] (override.east) -| ([xshift=0cm]agg.north)
    node[pos=0.0, right, font=\tiny\rmfamily, gray!60!black] {};

\node[font=\tiny\rmfamily, gray!60!black, anchor=north, yshift=-2pt]
    at (prescreen.south west) {\hspace{-0.9cm}\textit{shared layer}};

\end{tikzpicture}
\caption{%
\textbf{Evaluation pipeline.}
Each citation entry may pass through an optional shared pre-screening layer (DOI format, year bounds, name heuristics) that emits a final verdict directly; otherwise the entry is forwarded to the tool.
 The layer is applied to the database-lookup tools, whose reported rows therefore include its verdicts (\cref{sec:prescreening}).
The \gls{llm} verifiers are run without it, so an \gls{fpr} is comparable across the two blocks only up to that difference.%
}
\label{fig:eval_pipeline}
\end{figure}

\subsubsection{Pre-screening layer specification}
\label{app:prescreening}

The pre-screening layer implements four checks that run before external tool invocation:

\begin{enumerate}
    \item \textbf{\Gls{doi} resolution}: Normalizes the \gls{doi} (strips a URL prefix and, for arXiv \glspl{doi}, a version suffix such as \texttt{v1}, since DataCite registers only the unversioned form) and sends a live HTTP HEAD request to \texttt{doi.org}. Only a 404 or 410 served by \texttt{doi.org} itself flags the entry; a missing or malformed \gls{doi}, a network error, a rate limit, a block or a server error yields no verdict, so a transient failure never overrides the tool.
    \item \textbf{Year bounds checking}: Flags entries with publication years in the future or before 1900.\footnote{Genuinely valid pre-1900 references exist, of course (classical mathematics and physics citations); the bound is a corpus-tuned heuristic---no entry pool in the benchmark carries a publication year before 2021---so deployments verifying older literature should relax or drop this check.}
    \item \textbf{Author name heuristics}: Detects common placeholder patterns (``John Doe,'' ``A.\ Author,'' single-word author names, repeated names).
    \item \textbf{Capitalized-token placeholders}: Flags author fields built from placeholder tokens that the previous check misses: a capitalized word followed by digits (\texttt{Smith2}), a field made only of initials, or one where more than half the tokens are all-uppercase words.
\end{enumerate}

Pre-screening results are tagged with \texttt{[Pre-screening override]} in the reason string to maintain transparency about which detections come from the pre-screening layer vs.\ the external tool.

\subsubsection{HaRC and verify-citations: why neither is in the main table}
\label{app:harc_disclaimer}

We attempted HaRC~\citep{harc2024} and verify-citations~\citep{verifycitations2025} as full-coverage baselines and could not obtain a verdict from either at benchmark scale.
The binding constraint for HaRC is not Semantic Scholar throttling but \texttt{harcx}'s own lookup path, which queries Google Scholar through \texttt{scholarly}: Scholar blocks the client, the library retries rather than failing, and under the \texttt{-q --threshold 0.75} invocation our wrapper uses even a single-entry \texttt{.bib} does not complete within 150\,s. Dropping \texttt{-q} returns in seconds but emits no verdict line at all, exit~0. An authenticated S2 key (1 RPS, 5-hour budget) does not change either behavior, so 0 of 1{,}119 entries are actually checked by \texttt{harcx}.
verify-citations exhibits the same throughput pathology and we never ran it over a full split; the released artifact for it covers a 500-entry sample.
In both cases whatever verdicts a run does produce come from the shared pre-screening layer (\cref{app:prescreening}) rather than from the tool, so the numbers would not measure what the row claims to measure.

Both tools are therefore excluded from \cref{tab:results}.
The one keyed HaRC artifact we release (\texttt{harc\_with\_s2key\_dev\_public.json}) records DR\,=\,0.209, FPR\,=\,0.045, F1\,=\,0.335 over all 1{,}119 entries, but every one of those verdicts is a pre-screening verdict, and the file is scored against the pre-relabel ground truth (633 hallucinated / 486 valid, against the current 606/513), so it is comparable to no other number in this paper and we register it as known-stale rather than report it as a baseline.

\subsubsection{Baseline wrappers and prompt template}

\paragraph{LLM prompt template.}
\label{app:prompt-template}
The \gls{llm}-based baselines (GPT-5.1, Claude, and OpenRouter models) use a zero-shot prompt with no few-shot examples.
The exact template, reproduced verbatim from \texttt{hallmark/baselines/llm\_verifier.py} (\texttt{VERIFICATION\_PROMPT}; \texttt{\{bibtex\}} is the substituted entry), is:

\begin{lstlisting}[basicstyle=\footnotesize\ttfamily, breaklines=true, frame=single, captionpos=b]
You are a citation verification expert. Analyze the following BibTeX entry and determine if it is a VALID real publication or a HALLUCINATED (fabricated) citation.

BibTeX entry:
```bibtex
{bibtex}
```

Consider:
1. Is the title plausible and does it match known work by these authors?
2. Are the authors real researchers in this field?
3. Is the venue (journal/conference) real?
4. Does the year make sense?
5. If a DOI is present, does it look properly formatted?

When the entry is HALLUCINATED, classify the hallucination mode using exactly one of:
`fabricated_doi`, `nonexistent_venue`, `placeholder_authors`, `future_date`,
`chimeric_title`, `wrong_venue`, `swapped_authors`, `preprint_as_published`,
`hybrid_fabrication`, `near_miss_title`, `plausible_fabrication`,
`merged_citation`, `partial_author_list`, `arxiv_version_mismatch`.
Brief definitions:
- `fabricated_doi`: DOI does not resolve / is invented.
- `nonexistent_venue`: venue/journal does not exist.
- `placeholder_authors`: authors are placeholders ("Author1", "et al." alone, etc.).
- `future_date`: year is in the future relative to publication.
- `chimeric_title`: title combines fragments from multiple real works.
- `wrong_venue`: real paper but cited at wrong venue.
- `swapped_authors`: authors swapped or mismatched against the real paper.
- `preprint_as_published`: arXiv preprint cited as published in a venue.
- `hybrid_fabrication`: real DOI but other metadata (authors/title) doesn't match the DOI target.
- `near_miss_title`: title differs from a real paper by small but meaningful edits.
- `plausible_fabrication`: entirely fabricated yet plausible-sounding paper.
- `merged_citation`: metadata combined from two real papers.
- `partial_author_list`: real paper but author list is incomplete.
- `arxiv_version_mismatch`: arXiv version cited as a different version (or as published).

Respond with JSON only:
{
    "label": "VALID" or "HALLUCINATED" or "UNCERTAIN",
    "confidence": 0.0 to 1.0,
    "predicted_hallucination_type": "<one of the 14 types above, or null>",
    "reason": "brief explanation"
}
`predicted_hallucination_type` MUST be null when label is VALID or UNCERTAIN.
\end{lstlisting}

\subsubsection{Reproducibility: LLM experimental setup}
\label{app:llm-setup}

\paragraph{Model roster.}
The evaluation pipeline accepts any OpenRouter-, OpenAI-, or Anthropic-compatible model ID as a drop-in kwarg (\texttt{model=...}).
\cref{tab:model-roster} lists every zero-shot baseline reported in the paper, the API endpoint used for the predictions in \texttt{data/v1.0/baseline\_results/} and \texttt{results/temporal\_supplement/}, and the corresponding training cutoff (cross-referenced against \cref{tab:cutoffs}).
The two Anthropic models are reachable through both the native API (baselines \texttt{llm\_anthropic} and \texttt{llm\_anthropic\_opus\_4\_7}) and the OpenRouter mirror; we use the OpenRouter mirror for the released predictions.
Agentic baselines (\texttt{llm\_agentic\_openai}, \texttt{llm\_agentic\_btu\_openai}) use \texttt{gpt-5.1}; the reported Anthropic agentic-\texttt{bibtex-updater} run (\texttt{llm\_agentic\_btu\_sonnet\_4\_6}) uses the OpenRouter mirror \path{anthropic/claude-sonnet-4.6}, matching the zero-shot Sonnet endpoint. The registry also retains legacy native-API agentic variants (\texttt{llm\_agentic\_anthropic}, \texttt{llm\_agentic\_btu\_anthropic}) hard-coded to \texttt{claude-sonnet-4-5-20250929}; these are not used for any number reported in the paper.
OpenRouter is accessed via the OpenAI SDK with \texttt{base\_url=https://openrouter.ai/api/v1}.
All zero-shot, \gls{doi}-only, and \texttt{bibtex-updater} main-run predictions in \texttt{data/v1.0/baseline\_results/} were collected on 2026-05-05 via the endpoints above; hosted-\gls{llm} rows are dated snapshots subject to endpoint drift (\cref{sec:limitations}).

\begin{table}[t]
\caption{\textbf{Zero-shot LLM baseline roster.} Model references: GPT-5.1~\citep{openai2025gpt5}, DeepSeek-V3.2~\citep{deepseekv3}, Qwen3~\citep{qwen3}, Mistral Large~\citep{mistral2025large} and Gemini~2.5~\citep{google2025gemini}. Endpoint indicates the API used for the released predictions; \emph{native} = vendor SDK, \emph{OR} = OpenRouter mirror. Training cutoffs cross-reference \cref{tab:cutoffs}; ``${\leq}$'' marks inferred upper bounds. Run date is the collection date of the released predictions; the two Anthropic rows are a later re-run, collected after the endpoint had drifted (\cref{sec:limitations}). \texttt{gpt-5.1}, \texttt{claude-sonnet-4.6}, and \texttt{claude-opus-4.7} are floating aliases (only \texttt{gpt-5.4-2026-03-05} is a dated-pinned ID), so every row is reproducible only as a dated snapshot.}
\label{tab:model-roster}
\centering
\small
\setlength{\tabcolsep}{4pt}
\fitwidth{%
\begin{tabular}{lllll}
\toprule
\textbf{Model} & \textbf{Endpoint / model ID} & \textbf{Provider} & \textbf{Cutoff} & \textbf{Run date} \\
\midrule
GPT-5.1                       & native: \texttt{gpt-5.1}                                  & OpenAI    & Sep~2024          & 2026-05-05 \\
GPT-5.4                       & native: \texttt{gpt-5.4-2026-03-05}                       & OpenAI    & Aug~2025          & 2026-05-05 \\
Claude Sonnet~4.6             & native: \texttt{claude-sonnet-4-6}; OR: \path{anthropic/claude-sonnet-4.6} & Anthropic & ${\leq}$\,Aug~2025 & 2026-09-02$^{\dagger}$ \\
Claude Opus~4.7               & native: \texttt{claude-opus-4-7}; OR: \texttt{anthropic/claude-opus-4.7}     & Anthropic & ${\leq}$\,Oct~2025 & 2026-09-02$^{\dagger}$ \\
DeepSeek-R1                   & OR: \texttt{deepseek/deepseek-r1}                         & DeepSeek  & Jul~2024          & 2026-05-05 \\
DeepSeek-V3.2                 & OR: \texttt{deepseek/deepseek-v3.2}                       & DeepSeek  & ${\leq}$\,Oct~2024 & 2026-05-05 \\
Qwen3-235B-A22B-2507          & OR: \texttt{qwen/qwen3-235b-a22b-2507}                    & Alibaba   & Jun~2025          & 2026-05-05 \\
Qwen3-VL-235B-A22B-Instruct   & OR: \texttt{qwen/qwen3-vl-235b-a22b-instruct}             & Alibaba   & ${\leq}$\,Jun~2025 & 2026-05-05 \\
Mistral Large~2512            & OR: \texttt{mistralai/mistral-large-2512}                 & Mistral   & ${\leq}$\,mid-2025 & 2026-05-05 \\
Gemini~2.5~Flash              & OR: \texttt{google/gemini-2.5-flash}                      & Google    & Jan~2025          & 2026-05-05 \\
Gemini~2.5~Pro                & OR: \texttt{google/gemini-2.5-pro}                        & Google    & Jan~2025          & 2026-05-05 \\
Llama~4~Maverick              & OR: \texttt{meta-llama/llama-4-maverick}                  & Meta      & ${\leq}$\,Aug~2024 & 2026-05-05 \\
\bottomrule
\multicolumn{5}{l}{\footnotesize $^{\dagger}$ Re-run on 2026-09-02, later than the other rows, because the OpenRouter Anthropic endpoint drifted after the first run. The numbers can be recomputed from the released per-entry predictions; re-querying the endpoint may not reproduce them (\cref{sec:limitations}, \cref{app:bootstrap}).}
\end{tabular}%
}
\end{table}

\paragraph{Decoding configuration.}
Zero-shot calls request \texttt{temperature=0.0} on every endpoint that accepts it. The one forced exception is \texttt{gpt-5.5}, which the code sets to \texttt{temperature=1.0}; \texttt{gpt-5.1} and \texttt{gpt-5.4} are sent \texttt{temperature=0.0}, but the GPT-5 endpoint samples stochastically regardless of the requested value (E3 in \cref{app:reviewer_experiments} measures the resulting run-to-run variance).
The completion-token budget is \texttt{max\_completion\_tokens=1024} on OpenAI- and OpenRouter-compatible endpoints---sized to absorb hidden thinking tokens emitted by reasoning models---and \texttt{max\_tokens=256} on the Anthropic native endpoint. This single 1024-(non-Anthropic)/256-(Anthropic-native) split applies to every main-table, temporal-supplement, and cutoff-aware run; the thinking-budget smoke test (\cref{app:thinking-budget}) deliberately sweeps around this budget to locate the saturation boundary.
We pass \texttt{seed=42} on OpenAI-SDK-compatible endpoints; the Anthropic Messages API does not expose an equivalent.
OpenRouter is left on its default dynamic provider routing (no \texttt{provider} pin), so an open-weight slug may be served by different underlying providers (e.g.\ DeepInfra, Together, Fireworks) on different runs; the OpenRouter rows are therefore reproducible only as dated snapshots.
The zero-shot prompt (\cref{app:prompt-template}) is sent as a single user-role message with no system prompt.
We do not use JSON mode; responses are parsed with a tolerant extractor that handles bare JSON and fenced code blocks, with an \texttt{UNCERTAIN@0.5} fallback on parse failure (throughout, \texttt{label@}$c$ denotes the label assigned with confidence $c$).
Agentic calls use a 1024-token completion budget to accommodate tool-call turns.
Client-side retry is \texttt{max\_retries=5} for zero-shot and \texttt{3} for agentic baselines, with a 120\,s per-request timeout.

\paragraph{Agentic baselines: tools and iteration.}
We evaluate four agentic variants: OpenAI (\texttt{gpt-5.1}) and Anthropic (OpenRouter \path{anthropic/claude-sonnet-4.6}) backends, each with either a multi-source tool suite or \texttt{bibtex-updater}-as-tool. The OpenAI agentic and the Anthropic agentic-\texttt{bibtex-updater} runs are the variants reported in the paper; numbers come from \path{data/v1.0/baseline_results/llm_agentic_btu_sonnet_4_6_dev_public.json} and the corresponding OpenAI checkpoint.
Tools are exposed via the provider's native function-calling interface with \texttt{tool\_choice="auto"}: the model may answer without any tool call, and we track the parametric-vs-tool split per prediction.
Up to \texttt{MAX\_TOOL\_CALLS=5} invocations per entry, plus one final verdict turn.
If the cap is exceeded without a verdict, the entry is assigned \texttt{UNCERTAIN@0.5}.

The multi-source suite comprises 4 tools, each returning \texttt{\{authors, title, venue, year, doi\}} with fields truncated to 500 characters:
\begin{itemize}
    \item \texttt{resolve\_doi(doi)}: CrossRef \gls{doi} resolution (\texttt{api.crossref.org/works/\{doi\}}).
    \item \texttt{search\_crossref(query, limit=5)}: CrossRef bibliographic search (max 20 results).
    \item \texttt{search\_openalex(query, limit=5)}: OpenAlex polite pool (max 25).
    \item \texttt{search\_arxiv(query, limit=5)}: arXiv Atom API (max 20).
\end{itemize}
All HTTP requests use a \texttt{HALLMARK/1.0} User-Agent and 10--15\,s timeouts; tool exceptions are retried with exponential backoff (up to 3 retries, base delay 1\,s).

The \texttt{bibtex-updater}-as-tool variant exposes a single tool, \texttt{verify\_with\_bibtex\_updater(bibtex)}, that invokes the \texttt{bibtex-check} CLI with \texttt{--rate-limit 120 --academic-only}, returning \texttt{\{status, confidence, mismatched\_fields, api\_sources, errors\}}.
Subprocess timeout: 180\,s.

\paragraph{External-database access.}
The three database-backed endpoints---\gls{doi}-only, standalone \texttt{bibtex-updater}, and the agentic suite---query five external services: CrossRef (\texttt{api.crossref.org}, polite pool, no key), OpenAlex (polite pool, identified by a \texttt{mailto} contact), Semantic Scholar (the \texttt{graph/v1} paper-lookup API, accessed with an \texttt{S2\_API\_KEY}), the arXiv Atom API, and DBLP.
Semantic Scholar resolves a citation by matching its title, authors, venue, and identifiers against the S2 corpus and returning the canonical record---or none---which is the evidence \texttt{bibtex-updater} weighs to confirm or flag an entry.
Semantic Scholar requires an API key for reliable programmatic access: the free key moves requests off the throttled shared public pool onto a dedicated per-key rate limit. The key affects only throughput and reliability: an unauthenticated re-run hits the shared limit (frequent HTTP~429) and resolves the corpus more slowly, but queries the same records and returns the same matches, so it does not change the reported verdicts.
The main-run numbers were queried on 2026-05-04/05.
Agentic tool results are replayable offline from the released \texttt{sha256}-keyed tool-cache (mechanics under \emph{Tool-call caching} below), so the agentic baselines reproduce without live calls.
The \gls{doi}-only and standalone-\texttt{bibtex-updater} endpoints have no such cache, so their numbers are dated snapshots of the corresponding databases.

\paragraph{System prompts.}
The multi-source agentic system prompt:

\begin{lstlisting}[basicstyle=\small\ttfamily, breaklines=true, frame=single, captionpos=b, label={lst:system-multi}, caption={System prompt (multi-source)}]
You are a citation verification expert with access to bibliographic lookup tools.
Your task: determine whether a BibTeX entry is a VALID real publication or a HALLUCINATED (fabricated) citation.

Strategy:
1. Inspect the entry for obvious red flags (fake DOI prefix, future year, placeholder authors).
2. Use tools to cross-reference: resolve the DOI, or search by title/author.
3. After gathering evidence (or after finding sufficient signal), emit your verdict.

When you are ready to give your final answer, output ONLY valid JSON -- no prose, no markdown fences:
{
    "label": "VALID" or "HALLUCINATED",
    "confidence": 0.0 to 1.0,
    "reason": "concise explanation citing the evidence you found"
}

Do NOT output the JSON until you have used enough tools or determined that parametric knowledge is sufficient.
\end{lstlisting}

The \texttt{bibtex-updater}-as-tool agentic system prompt:

\begin{lstlisting}[basicstyle=\small\ttfamily, breaklines=true, frame=single, captionpos=b, label={lst:system-btu}, caption={System prompt (\texttt{bibtex-updater}-as-tool)}]
You are a citation verification expert with access to a specialized tool, `verify_with_bibtex_updater`, which cross-references a BibTeX entry against CrossRef, DBLP, and Semantic Scholar and returns a structured verdict.

Strategy:
1. For almost every entry, call `verify_with_bibtex_updater` once with the exact BibTeX string you were given.
2. Interpret the returned `status` field: statuses like `verified`, `url_verified`, or `published_version_exists` suggest VALID; statuses like `not_found`, `title_mismatch`, `swapped_authors`, `hallucinated`, `future_date`, `doi_not_found`, or `venue_mismatch` suggest HALLUCINATED.
3. If the tool returns `api_error` or you suspect the tool is wrong (e.g. it reports `verified` but the entry still looks suspicious on inspection), apply your own judgment -- the tool is evidence, not an oracle.
4. If the first call is unambiguous, do NOT call again. Extra calls waste budget.

When you are ready to give your final answer, output ONLY valid JSON -- no prose, no markdown fences:
{
    "label": "VALID" or "HALLUCINATED",
    "confidence": 0.0 to 1.0,
    "reason": "concise explanation referencing the tool status and any disagreement"
}
\end{lstlisting}

\paragraph{Tool-call caching.}
Agentic tool results are cached in SQLite keyed on \texttt{sha256(tool\_name::canonical\_args\_json)}.
TTL is disabled by default; cache hits bypass the network.
This is a reproducibility artifact, transparent to the \gls{llm}.

\paragraph{Failure handling.}
After three consecutive API errors, the baseline aborts and marks remaining entries as \texttt{UNCERTAIN@0.5} with reason tag \texttt{[Error fallback]}.
These entries count as abstentions in Cov.\ and contribute zero to detection rate.

\section{Core results and validity}
\label{app:results_top}

\subsection{Full \texttt{test\_public} results}
\label{app:test_public_full}
\paragraph{Cross-split observations (from the main text).}
\begin{itemize}
  \item \textbf{ Cross-split FPR shifts vary by model.}  Most \glspl{llm} gain or hold \gls{f1} on \texttt{test\_public}; \gls{fpr} shifts range from under $1$\,pp (GPT-5.4, both Geminis) to about $+8$\,pp (the Qwen variants), with Sonnet~4.6 at $+2.6$\,pp and Opus~4.7 at $+5.1$\,pp. DeepSeek-R1's $-30.3$\,pp is a denominator artifact---180 of its 831 \texttt{test\_public} verdicts are API error fallbacks recorded as \textsc{uncertain}---not a precision gain. All shifts sit within bootstrap-CI width at $n_{\text{valid}}{=}312$.
  \item \textbf{bibtex-updater is cross-split stable.} Its \gls{fpr} rises only $+2.9$\,pp ($0.051{\to}0.080$) and \gls{dr} and \gls{f1} hold: conservative matching flags so sparingly that \texttt{test\_public}'s harder valid pool does not inflate its false positives, and its flat risk--coverage curve---the error rate as a function of the fraction of entries it commits to, traced by varying the abstention threshold---confirms the low \gls{fpr} is structural, arising from conservative matching rather than from abstention (a reading the coverage re-scoring puts in question; \cref{app:coverage}).
  \item \textbf{Both agentic \texttt{bibtex-updater} harnesses lose part of their dev-side \gls{fpr} inflation on \texttt{test\_public}.} Sonnet+\texttt{bibtex-updater} \gls{fpr} drops $-8.8$\,pp and GPT-5.1+\texttt{bibtex-updater} $-11.4$\,pp cross-split, shrinking each harness's dev-side multiplier, but the harness-driven precision cost persists in both.
\end{itemize}

\cref{tab:test_public_full} reports all tools with both \texttt{dev\_public} and \texttt{test\_public} evaluations: the \gls{doi}-only baseline, twelve zero-shot \gls{llm} verifiers, three agentic harnesses, \texttt{bibtex-updater}, and the always-call evidence-injection co-designed variant.

\begin{table}[t]
\caption{\textbf{Full \texttt{test\_public} (831 entries) results vs.\ \texttt{dev\_public} (1{,}119 entries).} Zero-shot LLMs ordered by $|\Delta\text{FPR}|$. DeepSeek-R1 is the lone zero-shot outlier moving in the opposite direction ($-30.3$\,pp): it records 180 API error fallbacks on \texttt{test\_public} as \texttt{UNCERTAIN} (21.7\%, vs.\ 18/1119 on dev), so its test \gls{fpr} is measured over 651 answered entries against dev's 1{,}101. Both \texttt{bibtex-updater}-agentic harnesses move in the opposite direction (Sonnet+\texttt{bibtex-updater} $-8.8$\,pp, GPT-5.1+\texttt{bibtex-updater} $-11.4$\,pp), partly reversing the dev-side harness FPR penalty. In the $\Delta$FPR column, blue marks cross-split-stable tools ($|\Delta\text{FPR}|\le 1$\,pp) and red upward drift (${\ge}{+}6.9$\,pp); DeepSeek-R1's $-30.3$\,pp is an artifact of that truncated denominator and is left unshaded.}
\label{tab:test_public_full}
\centering
\small
\setlength{\tabcolsep}{2pt}
\begin{tabular*}{\textwidth}{@{\extracolsep{\fill}}lccccccc@{}}
\toprule
 & \multicolumn{3}{c}{\texttt{dev\_public}} & \multicolumn{3}{c}{\texttt{test\_public}} & \\
\cmidrule(lr){2-4}\cmidrule(lr){5-7}\cmidrule(lr){8-8}
\textbf{Tool} & \textbf{DR} & \textbf{FPR} & \textbf{F1} & \textbf{DR} & \textbf{FPR} & \textbf{F1} & \textbf{$\Delta$FPR} \\
\midrule
\multicolumn{8}{l}{\emph{Citation-database baseline}} \\
DOI-only          & 0.218 & 0.043 & 0.347 & 0.191 & 0.042 & 0.314 & \cellcolor{cellgood}$\phantom{+}-0.001$ \\
\midrule
\multicolumn{8}{l}{\emph{Zero-shot LLMs}} \\
GPT-5.4           & 0.767 & 0.228 & 0.783 & 0.780 & 0.224 & 0.815 & \cellcolor{cellgood}$\phantom{+}-0.004$ \\
Gemini~2.5~Flash  & 0.500 & 0.100 & 0.631 & 0.505 & 0.106 & 0.644 & \cellcolor{cellgood}$+0.006$ \\
Gemini~2.5~Pro    & 0.476 & 0.050 & 0.627 & 0.458 & 0.059 & 0.613 & \cellcolor{cellgood}$+0.009$ \\
Llama 4 Maverick  & 0.614 & 0.146 & 0.707 & 0.631 & 0.167 & 0.729 & $+0.020$ \\
DeepSeek-V3.2     & 0.911 & 0.702 & 0.727 & 0.911 & 0.728 & 0.776 & $+0.026$ \\
Claude Sonnet~4.6 & 0.916 & 0.161 & 0.893 & 0.922 & 0.188 & 0.907 & $+0.026$ \\
Mistral Large     & 0.716 & 0.250 & 0.742 & 0.688 & 0.282 & 0.741 & $+0.032$ \\
Claude Opus~4.7   & 0.906 & 0.154 & 0.890 & 0.900 & 0.205 & 0.889 & $+0.051$ \\
GPT-5.1           & 0.837 & 0.411 & 0.766 & 0.852 & 0.481 & 0.796 & \cellcolor{cellbad}$+0.069$ \\
Qwen3-VL-235B     & 0.860 & 0.551 & 0.740 & 0.927 & 0.628 & 0.804 & \cellcolor{cellbad}$+0.077$ \\
Qwen3-235B        & 0.860 & 0.533 & 0.744 & 0.909 & 0.615 & 0.798 & \cellcolor{cellbad}$+0.082$ \\
DeepSeek-R1       & 0.896 & 0.623 & 0.739 & 0.809 & 0.319 & 0.812 & $\phantom{+}-0.303$ \\
\midrule
\multicolumn{8}{l}{\emph{Agentic harnesses (tool-use; up to 5 tool calls per entry)}} \\
GPT-5.1 + CrossRef/OpenAlex/arXiv                    & 0.967 & 0.478 & 0.816 & 0.942 & 0.558 & 0.827 & \cellcolor{cellbad}$+0.080$ \\
GPT-5.1 + bibtex-updater (agentic; tool optional)    & 0.980 & 0.470 & 0.824 & 0.960 & 0.356 & 0.883 & $\phantom{+}-0.114$ \\
Sonnet~4.6 + bibtex-updater (agentic; tool optional) & 0.990 & 0.431 & 0.841 & 0.990 & 0.343 & 0.902 & $\phantom{+}-0.088$ \\
\midrule
\rowcolor{gray!10}
\multicolumn{8}{l}{\emph{Co-designed (reference upper bound; \cref{app:codesign})}} \\
\rowcolor{gray!10}
bibtex-updater (v1.2.0)                                     & 0.820 & 0.051 & 0.880 & 0.838 & 0.080 & 0.889 & $+0.029$ \\
\rowcolor{gray!10}
GPT-5.1 + bibtex-updater (always-call; output in prompt)    & 0.843 & 0.144 & 0.856 & 0.855 & 0.256 & 0.851 & $+0.112$ \\
\bottomrule
\end{tabular*}
\end{table}

\begin{takeaway}
\textbf{Takeaway.}  How much a model's \gls{fpr} shifts across splits does not depend on how aggressively it flags: GPT-5.4 and both Geminis move under $1$\,pp, Sonnet~4.6 and Opus~4.7 $+2.6$ and $+5.1$\,pp, the recall-aggressive open-weight models $+2.6$ to $+8.2$\,pp. Sonnet~4.6 now edges past \texttt{bibtex-updater}'s \gls{f1} on both splits ($0.893$ vs.\ $0.880$; $0.907$ vs.\ $0.889$), at about three times its \gls{fpr} ($0.161$ vs.\ $0.051$).
\end{takeaway}

\subsection{LLM baseline agreement and calibration}
\label{app:llm_comparison}

The aggregate and per-type numbers for every LLM baseline are in \cref{tab:results,tab:pertype_full}. Here we add three diagnostics that the released per-entry \texttt{dev\_public} predictions support for the open-weight pair Qwen3-235B and DeepSeek-V3.2, which characterize how these verifiers fail rather than how often.

\paragraph{Generation-method stratification.}
\cref{tab:llm_gen_method} stratifies detection rate by the entry's generation method, probing whether an LLM verifier has an edge on LLM-generated entries.

\begin{table}[h]
\caption{\textbf{Detection rate stratified by generation method on \texttt{dev\_public}.} LLM-generated entries are the hardest source for all three models, the opposite of what a generator-familiarity advantage would predict.}
\label{tab:llm_gen_method}
\centering
\small
\begin{tabular}{lcccc}
\toprule
\textbf{Generation method} & \textbf{$n$} & \textbf{Qwen3-235B} & \textbf{DeepSeek-V3.2} & \textbf{GPT-5.1} \\
\midrule
Adversarial       & 60  & 0.917 & 1.000 & 1.000 \\
Perturbation      & 410 & 0.866 & 0.937 & 0.846 \\
Real-world        & 46  & 0.978 & 0.935 & 0.891 \\
LLM-generated     & 90  & 0.730 & 0.722 & 0.656 \\
All valid (FPR)     & 513 & 0.533 & 0.702 & 0.411 \\
\bottomrule
\end{tabular}
\par\vspace{2pt}
{\footnotesize For hallucinated entries the metric is detection rate; for valid entries it is the false positive rate over all 513 (486 scraped $+$ 27 relabeled). Values are regenerated on the relabeled \texttt{dev\_public} split; the relabel reclassified 23 recovered real papers out of the LLM-generated hallucinated bucket ($n$ 113$\to$90). The GPT-5.1 column repeats \cref{tab:stratified_dr}; GPT-5.1 also generated the LLM-generated entries, so its column is the same-generator case.}
\end{table}

LLM-generated hallucinations remain the hardest source for both models (DR 0.722--0.730) relative to perturbation-based entries (DR 0.866--0.937), consistent with the GPT-5.1 finding in \cref{app:statistics}.
Were verifiers systematically attuned to LLM-generated text, those entries would be easier to catch; the reverse holds here.
The generator is GPT-5.1 (\cref{app:dataset}), so for these two models the stratification probes cross-model familiarity rather than strict self-recognition; the same-generator case is the GPT-5.1 stratification in \cref{tab:stratified_dr}.
Adversarial entries are near-perfectly detected (0.917--1.000), so current adversarial strategies do not fool LLM verifiers.

\paragraph{Pairwise agreement.}
\cref{tab:llm_agreement} reports Cohen's $\kappa$ for all 66 pairs of the twelve zero-shot baselines with stored per-entry \texttt{dev\_public} predictions (\texttt{scripts/compute\_pairwise\_kappa.py}).

\begin{table}[h]
\caption{\textbf{Pairwise agreement between \gls{llm} baselines on \texttt{dev\_public}} (Cohen's $\kappa$; $n{=}1{,}119$; abstentions scored as committed-\textsc{valid}, unlike \cref{sec:main_results}, which excludes them). The two Anthropic columns use the later OpenRouter snapshot (2026-05-30; FPR $0.162$/$0.165$), not the 2026-09-02 run behind their \cref{tab:results} rows, and carry its drift caveat (\cref{app:coverage}). Agreement clusters by disposition: the precision pair (Opus--Sonnet, $\kappa{=}0.75$) and the recall-aggressive cohort (R1--V3.2 $0.50$, R1--Qwen3-235B $0.53$) agree internally, while cross-disposition pairs fall as low as $\kappa{=}$$0.15$ (V3.2--Gemini~Pro). Cells are shaded in proportion to $\kappa$ (darker $=$ higher agreement).}
\label{tab:llm_agreement}
\centering
\footnotesize
\setlength{\tabcolsep}{2pt}
\resizebox{\linewidth}{!}{%
\begin{tabular}{l*{11}{c}}
\toprule
 & \textbf{Opus 4.7} & \textbf{Sonnet 4.6} & \textbf{R1} & \textbf{V3.2} & \textbf{Flash} & \textbf{Mistral} & \textbf{Qwen3} & \textbf{GPT-5.4} & \textbf{GPT-5.1} & \textbf{Llama} & \textbf{Qwen3-VL} \\
\midrule
Sonnet~4.6              & \cellcolor{hallmark-blue!30}.746 &  &  &  &  &  &  &  &  & &  \\
DeepSeek-R1             & \cellcolor{hallmark-blue!13}.317 & \cellcolor{hallmark-blue!12}.309 &  &  &  &  &  &  &  & &  \\
DeepSeek-V3.2           & \cellcolor{hallmark-blue!10}.243 & \cellcolor{hallmark-blue!9}.229 & \cellcolor{hallmark-blue!20}.497 &  &  &  &  &  &  & &  \\
Gemini~2.5~Flash        & \cellcolor{hallmark-blue!16}.409 & \cellcolor{hallmark-blue!18}.448 & \cellcolor{hallmark-blue!9}.220 & \cellcolor{hallmark-blue!7}.187 &  &  &  &  &  & &  \\
Mistral Large           & \cellcolor{hallmark-blue!19}.480 & \cellcolor{hallmark-blue!21}.513 & \cellcolor{hallmark-blue!10}.240 & \cellcolor{hallmark-blue!12}.298 & \cellcolor{hallmark-blue!22}.541 &  &  &  &  & &  \\
Qwen3-235B              & \cellcolor{hallmark-blue!15}.370 & \cellcolor{hallmark-blue!15}.384 & \cellcolor{hallmark-blue!21}.529 & \cellcolor{hallmark-blue!18}.454 & \cellcolor{hallmark-blue!12}.311 & \cellcolor{hallmark-blue!12}.295 &  &  &  & &  \\
GPT-5.4                 & \cellcolor{hallmark-blue!23}.576 & \cellcolor{hallmark-blue!25}.630 & \cellcolor{hallmark-blue!15}.379 & \cellcolor{hallmark-blue!13}.314 & \cellcolor{hallmark-blue!20}.512 & \cellcolor{hallmark-blue!24}.605 & \cellcolor{hallmark-blue!20}.500 &  &  & &  \\
GPT-5.1           & \cellcolor{hallmark-blue!20}.499 & \cellcolor{hallmark-blue!21}.526 & \cellcolor{hallmark-blue!17}.431 & \cellcolor{hallmark-blue!16}.404 & \cellcolor{hallmark-blue!15}.365 & \cellcolor{hallmark-blue!19}.485 & \cellcolor{hallmark-blue!24}.593 & \cellcolor{hallmark-blue!25}.635 &  & &  \\
Llama~4~Maverick  & \cellcolor{hallmark-blue!21}.514 & \cellcolor{hallmark-blue!21}.534 & \cellcolor{hallmark-blue!12}.309 & \cellcolor{hallmark-blue!10}.250 & \cellcolor{hallmark-blue!29}.722 & \cellcolor{hallmark-blue!25}.631 & \cellcolor{hallmark-blue!17}.423 & \cellcolor{hallmark-blue!26}.661 & \cellcolor{hallmark-blue!19}.484 & &  \\
Qwen3-VL-235B     & \cellcolor{hallmark-blue!14}.361 & \cellcolor{hallmark-blue!15}.368 & \cellcolor{hallmark-blue!22}.555 & \cellcolor{hallmark-blue!20}.495 & \cellcolor{hallmark-blue!12}.303 & \cellcolor{hallmark-blue!13}.313 & \cellcolor{hallmark-blue!31}.773 & \cellcolor{hallmark-blue!20}.496 & \cellcolor{hallmark-blue!22}.558 & \cellcolor{hallmark-blue!16}.412 &  \\
Gemini~2.5~Pro    & \cellcolor{hallmark-blue!17}.414 & \cellcolor{hallmark-blue!17}.428 & \cellcolor{hallmark-blue!7}.177 & \cellcolor{hallmark-blue!6}.146 & \cellcolor{hallmark-blue!25}.622 & \cellcolor{hallmark-blue!19}.471 & \cellcolor{hallmark-blue!9}.225 & \cellcolor{hallmark-blue!19}.470 & \cellcolor{hallmark-blue!13}.333 & \cellcolor{hallmark-blue!22}.545 & \cellcolor{hallmark-blue!9}.233 \\
\bottomrule
\end{tabular}%
}
\end{table}

The matrix orders itself by disposition: within-cohort $\kappa$ runs about $1.6$ times (mean $0.54$ vs.\ $0.34$, splitting the cohorts at \gls{fpr}~$0.3$ in \cref{tab:results}) the cross-cohort values, with GPT-5.1 the exception: it agrees about as much with the low-\gls{fpr} cohort ($0.48$) as with its own ($0.50$), so verifier errors are correlated within cohort and ensembling verifiers of the same disposition adds little independent evidence (cf.\ the voter ensemble in \cref{app:ablation_threshold}).
The moderate Cohen's $\kappa$ (0.454) between DeepSeek-V3.2 and Qwen3-235B indicates that despite similar overall detection rates, the two models disagree on a substantial fraction of entries.
Of the 1,119 entries where both models have predictions, they agree and are both correct on 620, agree but are both wrong on 274 (predominantly false positives on valid entries), and disagree on 225.
Thirty hallucinated entries are missed by \emph{both} models, a ``hard core'' that may require a different verification strategy such as API-based cross-referencing.

\paragraph{Confidence calibration.}
A verifier's confidence is useful only if it separates correct verdicts from wrong ones.
For Qwen3-235B and DeepSeek-V3.2 it does not: both assign a median confidence of $0.95$ whether the verdict is right or wrong, and their mean confidence on correct predictions exceeds that on incorrect ones by only $0.008$ and $0.001$ respectively, so no confidence threshold can route their unreliable flags to human review.
GPT-5.1's confidences retain some discriminative signal (\gls{ece} $0.190$, \cref{tab:results}), though far from calibrated.
For deployment pipelines that triage flags by confidence, the two open-weight models' scores are effectively uninformative.

\subsection{Small and open-weight models}
\label{app:small_models}
\paragraph{Setup.}
We evaluate Claude Haiku~4.5 and the Qwen3 dense ladder at 4, 8, 14 and 32B on \texttt{dev\_public} under the protocol of \cref{app:llm-setup}: the shared zero-shot prompt, temperature $0$, seed $42$, one call per entry, no tool access.
Haiku runs through OpenRouter on the same path as the main-table Anthropic rows.
The Qwen3 models run through the HuggingFace Inference Providers router (\texttt{Qwen/Qwen3-\{4B,8B,14B,32B\}}, provider pinned to \texttt{featherless-ai}, collected 2026-08), so the ladder varies parameter count with provider, prompt and decoding settings held fixed, which is what makes the comparison interpretable.

\paragraph{Scaling up does not help.}
Scaling within the ladder does not buy verification. \Gls{mcc} peaks at the smallest model and the 14B model is the worst of the four, flagging $98.5\%$ of the split against the 4B $95.9\%$; the 32B model recovers nothing. One entry drew an \textsc{uncertain} verdict, and the median stated confidence is $0.95$ at every model.
These models are not failing to answer, and they are not hedging: they answer \textsc{hallucinated} on nearly every entry, with high stated confidence, and \gls{mcc} places all four near zero.

\begin{table}[t]
\caption{\textbf{Small and open-weight verifiers on \texttt{dev\_public}} (same prompt, temperature 0, seed 42, one call per entry; \cref{app:llm-setup}). Frontier reference rows are repeated from \cref{tab:results}.}
\label{tab:small_models}
\centering
\small
\begin{tabular}{lccccc}
\toprule
\textbf{Model} & \textbf{DR $\uparrow$} & \textbf{FPR $\downarrow$} & \textbf{F1 $\uparrow$} & \textbf{MCC $\uparrow$} & \textbf{Median lat.} \\
\midrule
Claude Haiku~4.5   & 0.866 & 0.444 & 0.773 & 0.448 & --- \\
\midrule
Qwen3-4B           & 0.975 & 0.940 & 0.704 & 0.090 & 2.5\,s \\
Qwen3-8B           & 0.977 & 0.949 & 0.703 & 0.074 & 2.3\,s \\
Qwen3-14B          & 0.990 & 0.979 & 0.703 & 0.047 & 3.6\,s \\
Qwen3-32B          & 0.975 & 0.947 & 0.702 & 0.073 & 5.2\,s \\
\midrule
\multicolumn{6}{l}{\emph{Reference (\cref{tab:results})}} \\
Qwen3-235B         & 0.860 & 0.533 & 0.744 & 0.358 & --- \\
Claude Opus~4.7    & 0.906 & 0.154 & 0.890 & 0.755 & --- \\
\bottomrule
\end{tabular}
\end{table}

\subsection{Full per-type results}
\label{app:pertype}

\cref{fig:heatmap} visualizes per-type detection rates as a heatmap; \cref{tab:pertype_full} reports the underlying detection rate for the 11 main types and every full-coverage independent baseline.
HaRC and verify-citations are excluded because neither produces its own verdicts at benchmark scale (\cref{app:harc_disclaimer}).

\paragraph{Per-tier analysis.}\label{sec:per_tier}

\begin{figure}[t]
\centering
\includegraphics[width=0.95\linewidth]{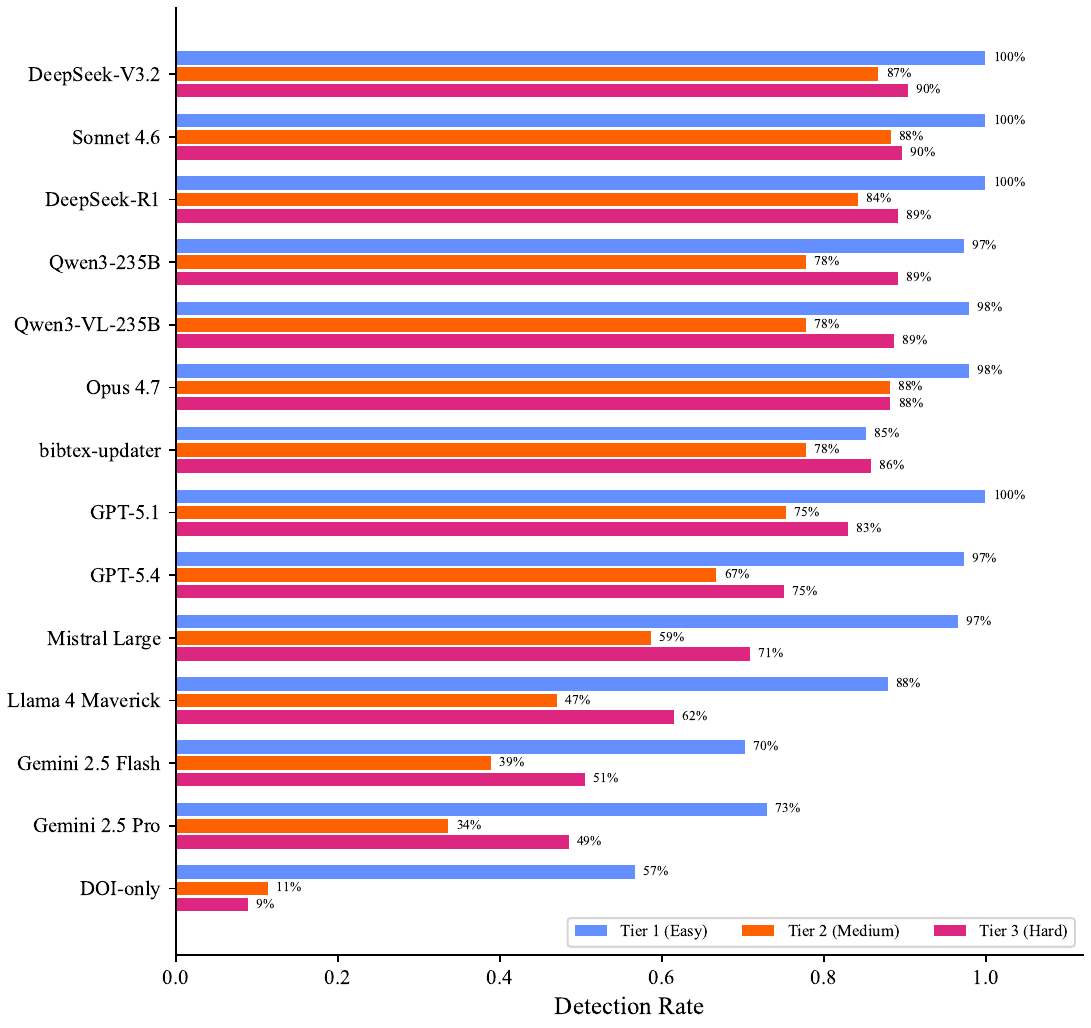}
\caption{\textbf{Detection rate by difficulty tier across all full-coverage tools.} DOI-only concentrates in Tier~1 ($.568$ against $.115$ and $.089$); every LLM drops on Tier~2 and partly recovers on Tier~3.}
\label{fig:tier-rates}
\end{figure}

\Gls{doi}-only detection concentrates in Tier~1; every \gls{llm} drops from Tier~1 to Tier~2 and partly recovers on Tier~3, because \texttt{plausible\_fabrication}, an invented paper an \gls{llm} flags as unfamiliar, makes up $44\%$ of Tier~3. The tiers rank the lookup effort a human needs, not difficulty for an \gls{llm} (\cref{fig:tier-rates}).
The recall--precision tradeoff persists at every difficulty: aggressive models reach their high Tier~1 detection by flagging indiscriminately---their \gls{fpr} is high---rather than by sharper discrimination, and stay far apart from the conservative models on Tier~3.
(Tier~3 aggregates fold in the stress-test type \texttt{arxiv\_version\_mismatch}, so per-type sums in \cref{tab:pertype_full} do not average exactly to the per-tier numbers.)

\paragraph{Per-type analysis: statistical power.}

Per-type counts run from 26 to 76 on \texttt{dev\_public}, giving a minimum detectable effect of 22--39\,pp between two tools (\cref{app:statistics}), so within-type cell rankings are directional, not significant.

\paragraph{Per-type failure modes.}
\label{sec:failure}
API tools miss venue-level hallucinations (\texttt{preprint\_as\_published}, \texttt{wrong\_venue}) because bibliographic APIs do not distinguish venue publication from arXiv availability.
GPT-5.1 handles \texttt{wrong\_venue} ($85\%$) better than \texttt{preprint\_as\_published} ($74\%$), and struggles most with \texttt{swapped\_authors} ($45\%$) and \texttt{near\_miss\_title} ($58\%$).
Prepending \texttt{bibtex-updater}'s output to GPT-5.1 cuts calibration error sharply (\gls{ece} $0.190{\to}0.078$) at no recall cost: the strongest option when calibration matters more than raw recall (\cref{app:codesign}).

\paragraph{Tier concentration.}
\label{sec:water_filling}
Benchmark optimization rewards easy subtasks~\citep{hardt2025benchmarks}.
On per-tool, per-tier detection rates, API tools concentrate their wins on Tier~1 (Tier~1/3 ratio $>30{\times}$, Gini $0.65$), while \glspl{llm} perform near-uniformly (Gini $<0.05$).
We therefore report Tier~3 \gls{f1} alongside the aggregate (\cref{sec:per_tier,fig:tier-rates}) to keep the hard regime visible.

\begin{table}[t]
\caption{\textbf{Per-type detection rate on \texttt{dev\_public} for full-coverage independent tools.} HaRC and verify-citations are excluded because neither produces its own verdicts at benchmark scale (\cref{app:harc_disclaimer}). Most main types have ${\sim}30$ instances; correcting the ground-truth mislabels leaves one type (\texttt{hybrid\_fabrication}, $n{=}26$) below 30 (\cref{tab:stats} caption).  Every column is scored on the released labels and reproduces the \texttt{per\_type\_metrics} block of the corresponding released result file. Column headers abbreviated for space: S4.6 = Sonnet~4.6, O4.7 = Opus~4.7, G51 = GPT-5.1, G54 = GPT-5.4, R1 = DeepSeek-R1, V3 = DeepSeek-V3.2, Q3 = Qwen3-235B, ML = Mistral Large, GF = Gemini~2.5~Flash, L4 = Llama~4~Mav., GP = Gemini~2.5~Pro, QV = Qwen3-VL. The two red-shaded types are the cohort-wide hard types: the median model catches only $52\%$ and $67\%$ of them, against $81\%$ or more on every other type. Numeric cells are shaded in proportion to the detection rate (darker $=$ higher). The three stress-test types (\texttt{merged\_citation}, \texttt{partial\_author\_list}, \texttt{arxiv\_version\_mismatch}) are omitted from this table; they appear in every core split and in \cref{fig:heatmap}.}
\label{tab:pertype_full}
\centering
\footnotesize
\setlength{\tabcolsep}{2pt}
\fitwidth{%
\begin{tabular}{llccccccccccccc}
\toprule
\textbf{Tier} & \textbf{Type} & \textbf{DOI} & \textbf{G51} & \textbf{G54} & \textbf{S4.6} & \textbf{O4.7} & \textbf{R1} & \textbf{V3} & \textbf{Q3} & \textbf{ML} & \textbf{GF} & \textbf{L4} & \textbf{GP} & \textbf{QV} \\
\midrule
\multirow{4}{*}{1}
& \texttt{fabricated\_doi}        & \cellcolor{hallmark-blue!36}0.895 & \cellcolor{hallmark-blue!40}1.000 & \cellcolor{hallmark-blue!39}0.974 & \cellcolor{hallmark-blue!40}1.000 & \cellcolor{hallmark-blue!37}0.921 & \cellcolor{hallmark-blue!40}1.000 & \cellcolor{hallmark-blue!40}1.000 & \cellcolor{hallmark-blue!40}1.000 & \cellcolor{hallmark-blue!39}0.974 & \cellcolor{hallmark-blue!38}0.943 & \cellcolor{hallmark-blue!40}1.000 & \cellcolor{hallmark-blue!28}0.703 & \cellcolor{hallmark-blue!40}1.000 \\
& \texttt{nonexistent\_venue}     & \cellcolor{hallmark-blue!9}0.231 & \cellcolor{hallmark-blue!40}1.000 & \cellcolor{hallmark-blue!40}1.000 & \cellcolor{hallmark-blue!40}1.000 & \cellcolor{hallmark-blue!40}1.000 & \cellcolor{hallmark-blue!40}1.000 & \cellcolor{hallmark-blue!40}1.000 & \cellcolor{hallmark-blue!40}1.000 & \cellcolor{hallmark-blue!39}0.974 & \cellcolor{hallmark-blue!21}0.513 & \cellcolor{hallmark-blue!38}0.949 & \cellcolor{hallmark-blue!23}0.571 & \cellcolor{hallmark-blue!40}1.000 \\
& \texttt{placeholder\_authors}   & \cellcolor{hallmark-blue!11}0.268 & \cellcolor{hallmark-blue!40}1.000 & \cellcolor{hallmark-blue!37}0.927 & \cellcolor{hallmark-blue!40}1.000 & \cellcolor{hallmark-blue!40}1.000 & \cellcolor{hallmark-blue!40}1.000 & \cellcolor{hallmark-blue!40}1.000 & \cellcolor{hallmark-blue!36}0.902 & \cellcolor{hallmark-blue!37}0.927 & \cellcolor{hallmark-blue!19}0.463 & \cellcolor{hallmark-blue!24}0.610 & \cellcolor{hallmark-blue!32}0.805 & \cellcolor{hallmark-blue!37}0.927 \\
& \texttt{future\_date}           & \cellcolor{hallmark-blue!40}1.000 & \cellcolor{hallmark-blue!40}1.000 & \cellcolor{hallmark-blue!40}1.000 & \cellcolor{hallmark-blue!40}1.000 & \cellcolor{hallmark-blue!40}1.000 & \cellcolor{hallmark-blue!40}1.000 & \cellcolor{hallmark-blue!40}1.000 & \cellcolor{hallmark-blue!40}1.000 & \cellcolor{hallmark-blue!40}1.000 & \cellcolor{hallmark-blue!40}1.000 & \cellcolor{hallmark-blue!40}1.000 & \cellcolor{hallmark-blue!34}0.857 & \cellcolor{hallmark-blue!40}1.000 \\
\midrule
\multirow{5}{*}{2}
& \texttt{chimeric\_title}        & \cellcolor{hallmark-blue!9}0.213 & \cellcolor{hallmark-blue!36}0.894 & \cellcolor{hallmark-blue!35}0.872 & \cellcolor{hallmark-blue!39}0.979 & \cellcolor{hallmark-blue!39}0.979 & \cellcolor{hallmark-blue!35}0.867 & \cellcolor{hallmark-blue!39}0.979 & \cellcolor{hallmark-blue!33}0.830 & \cellcolor{hallmark-blue!32}0.809 & \cellcolor{hallmark-blue!20}0.500 & \cellcolor{hallmark-blue!24}0.596 & \cellcolor{hallmark-blue!25}0.614 & \cellcolor{hallmark-blue!37}0.915 \\
& \texttt{wrong\_venue}           & \cellcolor{hallmark-blue!12}0.298 & \cellcolor{hallmark-blue!34}0.851 & \cellcolor{hallmark-blue!31}0.766 & \cellcolor{hallmark-blue!37}0.936 & \cellcolor{hallmark-blue!37}0.915 & \cellcolor{hallmark-blue!38}0.957 & \cellcolor{hallmark-blue!37}0.915 & \cellcolor{hallmark-blue!37}0.915 & \cellcolor{hallmark-blue!28}0.689 & \cellcolor{hallmark-blue!23}0.587 & \cellcolor{hallmark-blue!25}0.617 & \cellcolor{hallmark-blue!17}0.422 & \cellcolor{hallmark-blue!34}0.851 \\
& \cellcolor{cellbad}\texttt{swapped\_authors}       & \cellcolor{hallmark-blue!1}0.030 & \cellcolor{hallmark-blue!18}0.448 & \cellcolor{hallmark-blue!17}0.433 & \cellcolor{hallmark-blue!33}0.815 & \cellcolor{hallmark-blue!38}0.940 & \cellcolor{hallmark-blue!26}0.657 & \cellcolor{hallmark-blue!29}0.716 & \cellcolor{hallmark-blue!21}0.522 & \cellcolor{hallmark-blue!14}0.348 & \cellcolor{hallmark-blue!4}0.106 & \cellcolor{hallmark-blue!8}0.194 & \cellcolor{hallmark-blue!4}0.106 & \cellcolor{hallmark-blue!21}0.522 \\
& \texttt{preprint\_as\_pub.}     & \cellcolor{hallmark-blue!0}0.000 & \cellcolor{hallmark-blue!30}0.742 & \cellcolor{hallmark-blue!23}0.581 & \cellcolor{hallmark-blue!37}0.933 & \cellcolor{hallmark-blue!33}0.833 & \cellcolor{hallmark-blue!36}0.900 & \cellcolor{hallmark-blue!34}0.839 & \cellcolor{hallmark-blue!36}0.903 & \cellcolor{hallmark-blue!18}0.452 & \cellcolor{hallmark-blue!17}0.419 & \cellcolor{hallmark-blue!22}0.548 & \cellcolor{hallmark-blue!4}0.100 & \cellcolor{hallmark-blue!32}0.806 \\
& \texttt{hybrid\_fabrication}    & \cellcolor{hallmark-blue!6}0.154 & \cellcolor{hallmark-blue!40}1.000 & \cellcolor{hallmark-blue!38}0.962 & \cellcolor{hallmark-blue!40}1.000 & \cellcolor{hallmark-blue!40}1.000 & \cellcolor{hallmark-blue!40}1.000 & \cellcolor{hallmark-blue!40}1.000 & \cellcolor{hallmark-blue!35}0.885 & \cellcolor{hallmark-blue!37}0.923 & \cellcolor{hallmark-blue!28}0.692 & \cellcolor{hallmark-blue!28}0.692 & \cellcolor{hallmark-blue!34}0.846 & \cellcolor{hallmark-blue!37}0.923 \\
\midrule
\multirow{2}{*}{3}
& \cellcolor{cellbad}\texttt{near\_miss\_title}      & \cellcolor{hallmark-blue!5}0.135 & \cellcolor{hallmark-blue!23}0.577 & \cellcolor{hallmark-blue!18}0.462 & \cellcolor{hallmark-blue!27}0.673 & \cellcolor{hallmark-blue!27}0.673 & \cellcolor{hallmark-blue!29}0.731 & \cellcolor{hallmark-blue!32}0.788 & \cellcolor{hallmark-blue!27}0.673 & \cellcolor{hallmark-blue!22}0.558 & \cellcolor{hallmark-blue!9}0.220 & \cellcolor{hallmark-blue!11}0.269 & \cellcolor{hallmark-blue!8}0.196 & \cellcolor{hallmark-blue!28}0.712 \\
& \texttt{plausible\_fabrication} & \cellcolor{hallmark-blue!2}0.051 & \cellcolor{hallmark-blue!37}0.934 & \cellcolor{hallmark-blue!36}0.895 & \cellcolor{hallmark-blue!39}0.987 & \cellcolor{hallmark-blue!39}0.987 & \cellcolor{hallmark-blue!39}0.974 & \cellcolor{hallmark-blue!38}0.961 & \cellcolor{hallmark-blue!39}0.987 & \cellcolor{hallmark-blue!34}0.842 & \cellcolor{hallmark-blue!26}0.640 & \cellcolor{hallmark-blue!34}0.842 & \cellcolor{hallmark-blue!27}0.680 & \cellcolor{hallmark-blue!39}0.974 \\
\bottomrule
\end{tabular}%
}
\end{table}

\noindent \gls{doi}-only with pre-screening (\cref{app:prescreening}) detects entries in every type except \texttt{preprint\_as\_published}, where the cited preprint's \gls{doi} resolves; without pre-screening it detects only \texttt{fabricated\_doi} entries.
GPT-5.1's two weakest types---\texttt{swapped\_authors} (0.448) and \texttt{near\_miss\_title} (0.577)---require exact bibliographic recall that even large LLMs lack (parenthesized values are per-type detection rates).

\begin{takeaway}
\textbf{Takeaway.} Read the per-type heatmap as a failure-mode diagnostic, not a tool ranking: per-cell \gls{mde} between two tools runs 22--39\,pp, so per-type gaps between tools are reliable only where they are large; \texttt{swapped\_authors} and \texttt{near\_miss\_title} stand out as hard for most of the cohort.
\end{takeaway}

\subsection{Validity checks}
\label{app:validity}
\subsubsection{Core subset analysis}
\label{app:core-subset}

Scaling the hallucinated pool does not distort the evaluation signals. The pool grew out of 114 seed hallucinations (71 in dev, 43 in test) through generation across all main hallucination types, and to check the scaling we re-evaluated all baselines on the core subset containing only the 114 seed hallucinations alongside the valid entries.
This check was performed on an earlier corpus snapshot: 816 hallucinated (453 dev / 363 test) and 720 valid public entries at the time; it predates the final corpus expansion and the ground-truth relabel, and the released corpus counts are in \cref{tab:stats}.
On that snapshot, aggregate metrics differ by less than 2\% between the core subset and the full pool across all baselines: detection rate changes by at most 0.015, \gls{f1} by at most 0.02, and tier-weighted \gls{f1} by at most 0.018.
Per-type rankings are preserved: no baseline changes rank on any type.
We conclude that the scaled entries are consistent with the seed distribution and do not introduce systematic bias.

\subsubsection{Synthetic vs.\ real-world representativeness}
\label{sec:synth_vs_real}
We can detect no surface-feature artifact separating synthetic from real hallucinations, but we cannot yet claim the two are equivalent.
For nine \texttt{dev\_public} types with both real and synthetic entries, per-type Kolmogorov--Smirnov tests on title length, author count, and year fail to reject the null for 6/9 ($p>0.05$); the three divergent types differ mainly in author count (real entries average 2--3 authors, synthetic 3--4).
This is \emph{not} evidence of equivalence: failing to reject a null is not accepting it, the real set has only ${\sim}5$--$20$ entries per type (so the KS test is badly underpowered), and these three surface features are exactly the ones our shortcut analysis (\cref{app:shortcuts}) shows carry little signal (a logistic regression trained on those features beats the majority-class baseline---always predicting the more frequent label---by only $4.6$\,pp accuracy, below the $5$\,pp margin we treat as evidence of an exploitable shortcut).
The honest reading is narrow: no artifact strong enough to clear the leakage threshold appears, but distributional and semantic equivalence remains \emph{untested} at these sample sizes. That is the central open validity question, which would need an equivalence test on a much larger real set, not a null-hypothesis test that cannot tell ``similar'' from ``insufficient data''.
Tool detection is comparable across strata (GPT-5.1: $0.66$ on \gls{llm}-generated vs.\ $0.85$ on perturbations, with Sonnet~4.6 the same direction), consistent with \gls{llm}-generated entries being intrinsically harder rather than with self-recognition. This comparability supports perturbations as controlled diagnostics and motivates expanding the 108-entry real-world set.

\subsubsection{Multi-defect handling}
\label{app:multidefects}
The comparison in \cref{sec:synth_vs_real} covers surface features only; defect
count is a structural property it does not reach, and one on which the strata
differ sharply. We define a \emph{defect} as a per-field subtest recorded
\texttt{false}; \texttt{cross\_db\_agreement} is excluded because it is \texttt{false} for
100\% of hallucinated entries. An entry carries one \texttt{hallucination\_type} (the injected cause) but may break several fields.

Across the three publicly released splits, 330 of 1{,}246 hallucinated entries
($27\%$) are multi-defect, but the rate is highly stratified by provenance:
$13\%$ for \texttt{perturbation} (118/892) against $79\%$ for real-world
citations (63/80). The real-world set totals 108; the 28 entries in
\texttt{test\_hidden} carry no released subtests, so per-entry rates are
computed over the 80 public entries.

We report detection by defect count for three systems spanning the \gls{dr}--\gls{fpr} range
of \cref{fig:dr_fpr} (i.e., the strongest zero-shot LLM (Sonnet~4.6), the rule-based tool (\texttt{bibtex-updater}), and the hybrid cascade (GPT-5.4 + BTU)) in \cref{tab:multi-vs-single-defect}. UNCERTAIN counts as a miss, so all denominators are equal;  the cascade (22 entries, scored in its conservative variant) and Sonnet~4.6 (20) abstain. Single-defect entries are the harder case for Sonnet~4.6 and \texttt{bibtex-updater}
($+0.076$ Sonnet~4.6, $+0.253$ \texttt{bibtex-updater}).
\begin{table}[!t]
\caption{\textbf{Detection rate on single- vs multi-defect hallucinated entries.} A \emph{defect} is a failing per-field subtest. FPR is a single number per model: valid entries have no defects, so it cannot be split by defect count. Split: \texttt{dev\_public} (430 single-defect, 147 multi-defect, 513 valid). The 29 \texttt{future\_date} entries have no defect, since only cross-database agreement checks the year (\cref{tab:taxonomy}), and are excluded.}
\label{tab:multi-vs-single-defect}
\centering

\small\centering
\begin{tabular}{lccccc}
\toprule
\textbf{Model} & \textbf{DR single} & \textbf{DR multi} & $\Delta$ & \textbf{FPR} & \textbf{UNC} \\
\midrule
GPT-5.4 + BTU (cascade) & 0.953 & 0.952 & -0.001 & 0.039 & 21 \\
Sonnet 4.6 & 0.877 & 0.952 & +0.076 & 0.158 & 20 \\
bibtex-updater & 0.747 & 1.000 & +0.253 & 0.051 & 0 \\
\bottomrule
\end{tabular}
\end{table}

\subsubsection{Shortcut analysis}
\label{app:shortcuts}

Metadata features alone barely predict hallucination labels, so the dataset offers no shortcut a model could exploit instead of reading the content. To test this, we train a logistic regression classifier on eight entry-level features---presence of \gls{doi}, field count, author count, title length (characters), title word count, year (numeric), BibTeX entry type, and presence of a \texttt{booktitle} field---and evaluate via 5-fold cross-validation on the \texttt{dev\_public} split.

The logistic regression achieves a cross-validated accuracy of 58.8\% (majority-class baseline: 54.2\%), a margin of 4.6pp.
Since this margin is below 5pp, metadata features provide negligible predictive signal beyond class prevalence.
This shows that the metadata features themselves do not leak class information; \emph{which} content fields actually drive detection is quantified separately by the field-leave-one-out ablation (\cref{app:ablation_format}), where dropping the title raises \gls{fpr} by up to $35.5$\,pp.

Among the metadata features, \texttt{has\_doi} carries the most weight, consistent with the \gls{doi}-only baseline's non-zero detection rate, but no single feature pushes cross-validated accuracy meaningfully above the class-prevalence baseline, confirming that the perturbation pipeline does not introduce systematic metadata artifacts.

\paragraph{Detection rate by generation method.}
To assess whether GPT-5.1 exhibits self-recognition bias on \gls{llm}-generated entries, we stratify detection rate by the entry's generation method (\cref{tab:stratified_dr}).
\Gls{llm}-generated hallucinations are \emph{harder} for GPT-5.1 to detect (DR\,=\,0.656) than perturbation-based entries (DR\,=\,0.846), the opposite of what a self-recognition advantage would predict, though this stratification cannot separate intrinsic difficulty from a recognition effect it might mask.
This is expected:  \Gls{llm}-generated entries are written to look like real citations, so they carry none of the programmatic artifacts that perturbation leaves behind.
Perturbation-generated entries dominate the hallucinated pool ($410$ vs.\ $90$ \gls{llm}-generated on \texttt{dev\_public}, \cref{tab:stratified_dr}) and are detected ${\sim}19$\,pp more easily, so the public leaderboard is weighted toward this easier class. The harder, content-driven probes of construct validity are the \gls{llm}-generated subset and the field-leave-one-out ablation (\cref{app:ablation_format}); read absolute detection rates with this class mix in mind.
Adversarial entries, despite being crafted to evade detection, are detected without error (DR\,=\,1.000), suggesting that current adversarial strategies do not fool \gls{llm} verifiers.

\begin{table}[t]
\caption{\textbf{GPT-5.1 detection rate on \texttt{dev\_public} stratified by generation method.} DR is computed over hallucinated entries, FPR over valid entries; a method with no entries of a given polarity shows ``---''. The ground-truth audit places 4 perturbation-origin and 23 LLM-generated entries in the valid pool, so those two methods carry a (small-$n$) per-method FPR alongside the scraped pool ($n{=}486$, FPR 0.405); the four per-method FPRs reconcile to the headline \texttt{dev\_public} FPR of 0.411 over all 513 valid entries. The ``Adversarial'' generation method in the data files covers \texttt{hybrid\_fabrication} and \texttt{plausible\_fabrication} entries that use template-based perturbation (real DOI + fabricated metadata, or combinatorial assembly of plausible fields), distinguishing them from ``Perturbation'' (simple single-field modifications) and ``LLM-generated'' (entries produced by language models).}
\label{tab:stratified_dr}
\centering
\small
\begin{tabular}{lcccc}
\toprule
\textbf{Generation method} & \textbf{$n$ (hall.)} & \textbf{$n$ (valid)} & \textbf{DR} & \textbf{FPR} \\
\midrule
Adversarial      & 60  & ---  & 1.000 & ---   \\
Perturbation     & 410 & 4    & 0.846 & 1.000 \\
Real-world       & 46  & ---  & 0.891 & ---   \\
LLM-generated    & 90  & 23   & 0.656 & 0.435 \\
Scraped (valid)  & --- & 486  & ---   & 0.405 \\
\bottomrule
\end{tabular}
\end{table}

\paragraph{Cross-split validation.}
We evaluate GPT-5.1 on \texttt{test\_public} (831 entries, 62.5\% hallucinated) to verify that findings generalize beyond \texttt{dev\_public}.
Detection rate is consistent: DR\,=\,0.852 (test) vs.\ 0.837 (dev).
Per-tier patterns are preserved: Tier~1 DR\,=\,1.000, Tier~2 DR\,=\,0.754, Tier~3 DR\,=\,0.831 on \texttt{dev\_public}.
\Gls{fpr} is elevated on both splits (dev: 0.411, test: 0.481), reflecting GPT-5.1's tendency to flag valid entries with non-canonical metadata (uncommon venues, name variants). \texttt{bibtex-updater}'s \gls{fpr} is by contrast cross-split stable (dev 0.051 $\to$ test 0.080), because its abstention policy declines to flag entries it cannot back with a record rather than guessing on \texttt{test\_public}'s harder valid pool.
MCC\,=\,0.397 (test) summarizes all four cells on that split.

\begin{takeaway}
\textbf{Takeaway.} The validity checks hold: scaling the hallucinated pool from its 114 seeds shifts aggregate metrics by at most 2\,pp with no rank changes (\cref{app:core-subset}), and a metadata-only classifier exceeds the majority baseline by only 4.6\,pp---below the 5\,pp leakage bar---so metadata offers no shortcut around reading the content (\cref{app:shortcuts}). The class mix is the caveat to carry: perturbation entries dominate the pool and are ${\sim}19$\,pp easier for GPT-5.1 than the \gls{llm}-generated subset (\gls{dr} $0.846$ vs.\ $0.656$), a direction that also runs opposite to a self-recognition advantage.
\end{takeaway}

\subsection{External validity: authentic ChatGPT citations (the Walters--Wilder supplement)}
\label{app:walters_wilder}

The stratification in \cref{tab:stratified_dr} rests on \textsc{Hallmark}'s own \gls{llm}-generated entries, so it cannot rule out that our generation pipeline leaves fingerprints a verifier could exploit; an external check needs authentic hallucinations we did not construct.
We therefore convert the hand-coded citation corpus of \citet{walters2023fabrication} into a fourth evaluation-only split, \texttt{supplement\_chatgpt\_citations} (\cref{tab:stats}).
Walters and Wilder prompted ChatGPT-3.5 and GPT-4 to write 84 short literature reviews across 42 topics spanning the humanities, social sciences, and natural sciences, then hand-verified every one of the 636 resulting citations, reporting fabrication rates of 55\% (GPT-3.5) and 18\% (GPT-4).
The corpus extends the benchmark along the two axes it is thinnest on: the hallucinations are \emph{authentic} \gls{llm} output with expert labels rather than perturbations (the \texttt{dev\_public} pool behind \cref{tab:stratified_dr} has 90 \gls{llm}-generated hallucinated entries), and the subject matter is multidisciplinary rather than ML, making it the natural companion to the cross-domain split of \cref{sec:limitations}.

\paragraph{Conversion.}
We keep the supported publication type---journal articles---because books, chapters, and websites (${\approx}32\%$ of the source corpus) are structurally unresolvable for the database-backed verifiers evaluated here.
A field mismatch on a real work is a hallucination and is mapped to the matching taxonomy type rather than left \textsc{valid} (\cref{tab:ww-mapping}); when several fields are wrong, the most identity-defining field wins (title $>$ author $>$ venue).
Thirty-four real, findable articles whose only substantive error is a wrong non-future year and/or a volume/page slip are excluded because the taxonomy has no matching type; formatting-only deviations (capitalization, initials-vs-names, missing pages) are likewise not treated as hallucinations.
The result is 341 entries (172 valid / 169 hallucinated) that pass the same validation pipeline as the core corpus; the converter, its per-case audit trail, and the raw tool outputs are released with the benchmark code.

\begin{table}[t]
\caption{\textbf{Mapping the Walters--Wilder source coding onto the \textsc{Hallmark} taxonomy.} A substantive field error on a real work maps to the corresponding hallucination type; only wrong-year/volume/page-only errors (34 entries) have no matching type and are excluded.}
\label{tab:ww-mapping}
\centering
\small
\begin{tabular}{llcc}
\toprule
\textbf{Source coding} & \textbf{\textsc{Hallmark} type} & \textbf{Tier} & \textbf{$n$} \\
\midrule
work itself fabricated & \texttt{plausible\_fabrication} & 3 & 139 \\
real work, wrong title & \texttt{near\_miss\_title} & 3 & 13 \\
real work, wrong authorship & \texttt{swapped\_authors} & 2 & 12 \\
real work, wrong/invented journal & \texttt{wrong\_venue} & 2 & 5 \\
real work, no substantive error & \textsc{valid} & --- & 172 \\
\bottomrule
\end{tabular}
\end{table}

\paragraph{bibtex-updater results.}
On the full supplement (coverage 341/341), \texttt{bibtex-updater} reaches \gls{dr} 0.929 (95\% CI $[0.905, 0.953]$), \gls{fpr} 0.076 ($[0.041, 0.116]$), \gls{f1} 0.926, and tier-weighted \gls{f1} 0.959: higher recall than on \texttt{dev\_public} ($0.820$) at a somewhat higher \gls{fpr} ($0.051$) (\cref{tab:codesign}) on data from disciplines it was never tuned on.
The per-type gradient is: \texttt{plausible\_fabrication} 1.000 ($n{=}139$) and \texttt{wrong\_venue} 1.000 ($n{=}5$), against \texttt{near\_miss\_title} 0.769 ($n{=}13$) and \texttt{swapped\_authors} 0.250 ($n{=}12$).
The tool is in effect a fabrication detector: an invented work is simply not found in any backing database, whereas a corrupted citation of a real work resolves by title and the tool then accepts the record without re-checking authorship.
Stratified by generator, GPT-3.5 output is easier (\gls{dr} 0.990, \gls{fpr} 0.053) than GPT-4 output (\gls{dr} 0.841, \gls{fpr} 0.078): GPT-4 fabricates less and its errors are subtler, mostly author swaps on real papers.
Across subject fields, detection is stable (0.85--0.96) while \gls{fpr} is highest in the natural sciences (0.123 vs.\ 0.000 humanities, 0.056 social sciences).

\paragraph{Converter audit.}
Because a conversion artifact is indistinguishable from tool behavior in the aggregate numbers, we audited every false positive and every miss case by case.
The audit caught one APA-parser bug---dropped final authors on ``A, B, \& C'' references and deleted ellipsis truncation, which \texttt{bibtex-updater} reads as silent author-list truncation---that had inflated \gls{dr} and \gls{fpr} simultaneously: spurious flags on valid entries and wrong-reason detections on \texttt{swapped\_authors}.
All numbers above use the fixed converter; the released per-case trail documents both scoring passes.
The residual errors are genuine tool behavior: the 13 remaining false positives are metadata sensitivity on real works (strict subtitle/punctuation matching, online-first vs.\ print years), and all 12 misses are author or title corruptions the tool resolves to the real record and abstains on.
An aggregate score on a converted external corpus is thus only as trustworthy as a per-case audit of its false positives and misses.

\paragraph{Zero-shot LLM verifiers.}
We run four zero-shot \gls{llm} verifiers spanning the conservative--aggressive spectrum of \cref{tab:temporal_supplement} on the supplement, under the default prompt and the main-table protocol (\cref{app:llm-setup}); every cited work predates all four training cutoffs, so the supplement isolates \emph{parametric} verification---the model verifying a citation against what its weights absorbed in pretraining, without retrieval---on non-ML literature with no post-cutoff confound.
\cref{tab:ww-llm} reports the results next to each model's \texttt{dev\_public} baseline.
Three patterns emerge.
First, \textbf{detection of authentic fabrications separates the models most}: GPT-5.1 (0.964) and Sonnet~4.6 (0.961) approach \texttt{bibtex-updater}'s 1.000 on \texttt{plausible\_fabrication}, while Gemini 2.5 Flash recognizes only 0.345 of them, its overall \gls{dr} falling from 0.500 to 0.290 even as its \gls{fpr} stays low (0.029). The drop is consistent with thinner parametric coverage of the multidisciplinary literature, but with four models and no direct probe of what each has memorized, we cannot separate missing coverage from a conservative disposition that declines to flag what it cannot confirm.
Second, \textbf{the \gls{fpr} ordering from \texttt{dev\_public} is preserved and widens}: the spread grows from 0.100--0.533 on ML venues to 0.029--0.860 here, with the aggressive Qwen3-235B flagging 86\% of valid multidisciplinary papers; its seemingly strong \texttt{swapped\_authors} cell (0.917) is a byproduct of that indiscriminate flagging.
Third, \textbf{the subtle-corruption weakness spans the standalone verifier class}: every calibrated standalone verifier scores at or below 0.25 on \texttt{swapped\_authors} (\texttt{bibtex-updater} 0.250, GPT-5.1 0.250, Sonnet~4.6 0.182, Gemini Flash 0.000); the agentic combination below is the one configuration that improves on it.
Stratified by generator, GPT-4 output is harder than GPT-3.5 output for every verifier (e.g.\ on GPT-3.5 vs.\ GPT-4 output, Sonnet~4.6 \gls{dr} 0.94 vs.\ 0.58; \texttt{bibtex-updater} 0.99 vs.\ 0.84): the better generator fabricates less and corrupts more subtly.
Sonnet~4.6 is the only zero-shot model to abstain (15 entries) and posts the best \gls{llm} \gls{f1} (0.896) with an \gls{fpr} (0.058) below its ML-venue baseline, consistent with its calibrated profile in the temporal analysis (\cref{app:temporal_supplement}).

\begin{table}[t]
\caption{\textbf{Zero-shot LLM verifiers vs.\ \texttt{bibtex-updater} on the Walters--Wilder supplement ($N{=}341$: 172 valid / 169 hallucinated).} DR/FPR on \texttt{dev\_public} (2021--2023 ML venues) alongside the supplement (authentic, multidisciplinary, pre-cutoff for all models). Abstentions are excluded from DR/FPR (\cref{app:metrics}); Cov.\ is the committed fraction (Sonnet~4.6 abstains on 15 entries, the agentic harness on 22; all other verifiers commit on all 341). Fab.\ is the detection rate on \texttt{plausible\_fabrication} ($n{=}139$), Swap.\ on \texttt{swapped\_authors} ($n{=}12$). The bottom block is tool-involving: the standalone co-designed tool and the agentic btu-only harness of \cref{tab:results}, whose \texttt{dev} columns are that table's agentic (tool-optional) row. Green marks supplement FPRs at or below the \texttt{dev\_public} value for calibrated verifiers; red the aggressive outlier. $^\ast$Indiscriminate flagging (FPR 0.860), not discrimination.}
\label{tab:ww-llm}
\centering
\small
\setlength{\tabcolsep}{4pt}
\begin{tabular}{lcccccccc}
\toprule
& \multicolumn{2}{c}{\textbf{DR}} & \multicolumn{2}{c}{\textbf{FPR}} & \multicolumn{4}{c}{\textbf{Supplement}} \\
\cmidrule(lr){2-3}\cmidrule(lr){4-5}\cmidrule(lr){6-9}
\textbf{Model} & \texttt{dev} & suppl. & \texttt{dev} & suppl. & \textbf{F1} & \textbf{Fab.} & \textbf{Swap.} & \textbf{Cov.} \\
\midrule
Gemini 2.5 Flash  & 0.500 & 0.290 & 0.100 & \cellcolor{cellgood}0.029 & 0.439 & 0.345 & 0.000 & 1.00 \\
Claude Sonnet 4.6 & 0.916 & 0.865 & 0.161 & \cellcolor{cellgood}0.058 & 0.896 & 0.961 & 0.182 & 0.96 \\
GPT-5.1           & 0.837 & 0.846 & 0.411 & 0.285 & 0.792 & 0.964 & 0.250 & 1.00 \\
Qwen3-235B        & 0.860 & 0.976 & 0.533 & \cellcolor{cellbad}0.860 & 0.685 & 1.000 & 0.917$^\ast$ & 1.00 \\
\midrule
\rowcolor{gray!10}
bibtex-updater    & 0.820 & 0.929 & 0.051 & 0.076 & 0.926 & 1.000 & 0.250 & 1.00 \\
\rowcolor{gray!10}
GPT-5.1 + btu (agentic) & 0.980 & 0.975 & 0.470 & \cellcolor{cellgood}0.045 & 0.967 & 1.000 & 0.571 & 0.94 \\
\bottomrule
\end{tabular}
\end{table}

\paragraph{GPT-5.1 as a \texttt{bibtex-updater} dispatcher.}
The agentic btu-only harness of \cref{tab:results}---GPT-5.1 with \texttt{bibtex-updater} as its only tool, free to decide when to call it and whether to trust it---dominates both of its components on the supplement: \gls{dr} 0.975 (95\% CI $[0.952, 0.994]$), \gls{fpr} 0.045 ($[0.013, 0.079]$), \gls{f1} 0.967, tier-weighted \gls{f1} 0.984, committing on 319 of 341 entries.
The combination inherits the tool's perfect fabrication detection (1.000) and improves on both weaknesses the components share: \texttt{near\_miss\_title} rises to 1.000 (tool alone 0.769, model alone 0.385) and \texttt{swapped\_authors} to 0.571 (4 of 7 committed; both components alone 0.250): the dispatcher treats the tool's mismatch statuses as evidence where the standalone wrapper abstains, and routes genuinely ambiguous cases to \texttt{UNCERTAIN}.
Notably, the harness \gls{fpr} inflation of failure mode~(i) does not reproduce here: the same configuration that posts \gls{fpr} 0.470 on \texttt{dev\_public} posts 0.045 on the supplement.
The valid half of this corpus consists of real, database-indexed journal articles that the tool verifies cleanly, so the dispatcher rarely faces the ambiguous partial-match output that produces dev-side over-flagging, consistent with \cref{sec:analysis}'s reading that mode~(i) characterizes harness behavior on borderline metadata, not the dispatch pattern itself.
Protocol note: this run's tool calls used authenticated OpenAlex and Semantic Scholar access, which the standalone \texttt{bibtex-updater} run above predates; authentication changes service level (rate limits, latency), while the records consulted are identical.

\begin{takeaway}
\textbf{Takeaway.} On original, multidisciplinary ChatGPT-generated citations, two of the benchmark's central findings replicate on data we did not construct: subtle corruptions of real works are the hard class for every calibrated \emph{standalone} verifier (\texttt{swapped\_authors} \gls{dr} ${\leq}0.25$), and \gls{fpr}---the deployment-decisive metric---spans 0.029--0.860 across the \gls{llm} cohort, preserving and widening the \texttt{dev\_public} ordering. Detection of authentic fabrications spans 0.345--0.964 across \glspl{llm} (vs.\ 1.000 for the database-backed tool), a spread consistent with differences in parametric coverage. The compositional exception is the agentic dispatcher: GPT-5.1 orchestrating \texttt{bibtex-updater} reaches \gls{dr} 0.975 / \gls{fpr} 0.045 / \gls{f1} 0.967 and lifts \texttt{swapped\_authors} to 0.571 (0.333 over all 12 entries), without the mode-(i) \gls{fpr} inflation it shows on ML venues. Ecological validity cuts both ways: synthetic perturbations neither overstate the difficulty of authentic fabrications nor manufacture the subtle-corruption weakness. Caveat: $n{=}12$ for \texttt{swapped\_authors}, and the supplement is articles-only.
\end{takeaway}

\subsection{Comparison and additional figures}
\label{app:supplementary}
\label{app:related}

\subsubsection{Comparison with related citation efforts}

\Citet{topaz2026fabricated} make the cost of that framing explicit: their 500-entry masked validation, in their own words, ``estimates precision but not recall,'' so the reported prevalence is bounded above but not below---what the pipeline missed is unmeasured. \textsc{Hallmark} reports both directions on a corpus where every entry carries a ground-truth label, which is the precondition for ranking detectors rather than counting incidents.

\cref{tab:comparison-extended} contrasts \textsc{Hallmark} with concurrent citation-audit efforts. Prior work focuses on auditing published papers; \textsc{Hallmark} is, to our knowledge, among the first \emph{evaluation benchmarks} for citation-verification tools, alongside concurrent work such as CiteAudit~\citep{citeaudit}.

\begin{table}[t]
\caption{\textbf{Comparison of \textsc{Hallmark} with related citation analysis efforts.} CiteAudit~\citep{citeaudit} is a concurrent, human-validated effort that evaluates citation-verification tools (LLMs and commercial tools); ``---'' marks attributes we could not verify from the available description. \textsuperscript{a}\textsc{Hallmark}'s real-world and adversarial entries receive manual review; perturbation and LLM-generated entries are verified algorithmically (\cref{sec:dataset}).}
\label{tab:comparison-extended}
\centering
\small
\fitwidth{%
\begin{tabular}{lcccccc}
\toprule
& \textbf{Entries} & \textbf{Types} & \textbf{Sub-tests} & \textbf{Tool eval.} & \textbf{Human-valid.} & \textbf{Open} \\
\midrule
\multicolumn{7}{l}{\emph{Citation audits}} \\
GhostCite~\citep{ghostcite2026} & 56K papers & --- & \ding{55} & \ding{55} & --- & \ding{55} \\
HalluCitation~\citep{hallucitation2026} & ${\sim}300$ papers & 3 & \ding{55} & \ding{55} & --- & \ding{51} \\
Mysterious Citations~\citep{mysterious2026} & HPC venues & --- & \ding{55} & \ding{55} & --- & \ding{55} \\
\midrule
\multicolumn{7}{l}{\emph{Citation verification benchmarks}} \\
CiteAudit~\citep{citeaudit} & --- & --- & \ding{55} & \ding{51} & \ding{51} & --- \\
\textsc{Hallmark} (ours) & 2,526 & 14 & 6 & \ding{51} & partial\textsuperscript{a} & \ding{51} \\
\bottomrule
\end{tabular}%
}
\end{table}

\paragraph{Benchmark design principles.}
\textsc{Hallmark} synthesizes design principles from established benchmarks: multi-criteria sub-tests from HumanEval~\citep{humaneval}, temporal segmentation from SWE-bench~\citep{swebench} and LiveCodeBench~\citep{livecodebench2024}, multi-difficulty challenge sets from Dynabench~\citep{kiela2021dynabench}, and continuous, expandable sample pools from ONEBench~\citep{onebench2024} that aggregate per-entry measurements while resisting contamination and leaderboard rot. We complement these with the evaluation discipline of PostTrainBench~\citep{posttrainbench}: pin model versions, timestamp every result, and never assume the newest or largest model is the strongest, since test-set contamination tends to scale with capability.

\subsubsection{Additional figures}
\label{app:figures}

\begin{figure}[t]
\centering
\includegraphics[width=0.98\linewidth]{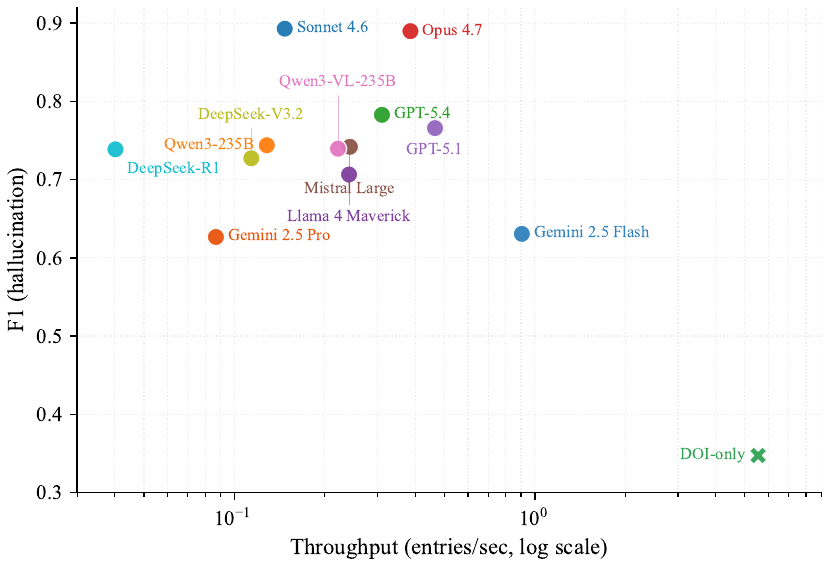}
\caption{\textbf{Accuracy against throughput.} \gls{f1} versus measured throughput in entries per second, for the thirteen single-call verifiers that carry timing in the released results. The throughput axis is logarithmic: it spans $0.04$ to $5.5$ entries/s, and on a linear axis every \gls{llm} collapses into the leftmost sixth of the panel. Color distinguishes models only. HaRC and verify-citations are excluded as everywhere else (\cref{app:harc_disclaimer}); \texttt{bibtex-updater} is absent because the release records no timing for it at all (\cref{tab:cost_latency}). Throughput is a latency proxy, not a dollar cost; per-entry cost is tabulated separately in \cref{tab:cost_latency}. The feasibility constraint at low prevalence (${\sim}2\%$) is precision rather than either axis: even highly accurate verifiers misallocate reviewer effort because precision is bottlenecked by base rate, not capability.}
\label{fig:cost}
\end{figure}

\paragraph{Cost--accuracy tradeoff.}
\label{sec:cost}
\Gls{llm} verification is also two orders of magnitude more expensive than the rule-based tools, which run locally for a fraction of a cent per entry; single-call \glspl{llm} cost cents, and agentic variants add another $2$--$5{\times}$ without proportional gains (\cref{fig:cost}).
With the base-rate \gls{ppv} drop above, the regime where \gls{llm} verification pays off is narrow: high-prevalence settings (triage queues, pre-publication self-check), or pipelines where a downstream human-review stage filters out the false positives.

\subsection{Cost and latency}
\label{app:cost_latency}

DeepSeek-R1 emits a chain of thought before committing to a verdict while the rest of the zero-shot cohort answers in a single pass, so it is worth stating what the harness holds fixed across the rows of \cref{tab:results}. Four settings are shared by every zero-shot call: the prompt is the verification template of \cref{app:prompt-template}, sent as one user-role message with no system prompt; decoding requests \texttt{temperature=0.0}; each call carries \texttt{seed=42}; and each call is issued against the same completion budget of \texttt{max\_completion\_tokens=1024} (\cref{app:llm-setup}).\footnote{Two configuration caveats, both documented in \cref{app:llm-setup}: the GPT-5 endpoint samples stochastically regardless of the requested temperature (\cref{app:reviewer_experiments} measures the resulting run-to-run variance), and the 256-token budget of the Anthropic native endpoint applies to no released row, since the Sonnet~4.6 and Opus~4.7 predictions were collected through the OpenRouter mirror.} The harness also issues exactly one call per entry, and the measured counts in \cref{tab:cost_latency} bear this out: mean API calls per entry is $1.000$ for all twelve zero-shot verifiers, DeepSeek-R1 included, against $2.0$--$2.9$ for the agentic harnesses that may issue up to five lookups. Nothing in the prompt asks for step-by-step reasoning and no reasoning-effort parameter is set for any model in the roster of \cref{tab:model-roster}, so R1's chain of thought is its own decoding behavior inside the shared budget rather than something we request of it. Two limits bound how far this reading goes. The recorded count is of requests the harness issues, so client-side retries inside the SDK are not counted, and those retries are why we read latency off the median rather than the mean. A shared prompt at a shared call budget is also not matched token consumption: R1 spends more of its 1024 tokens on thinking than the single-pass models do, and we ran no matched-compute arm, so \cref{tab:cost_latency} compares each model's deployed configuration at one prompt and one call budget, the setting a user actually faces, and not at equal tokens.

\begin{table}[t]
\caption{\textbf{Cost and latency per entry}, recomputed from the released per-entry timing dumps. $n$ is the number of timed entries and identifies the split ($1{,}119$ = \texttt{dev\_public}; $831$ = \texttt{test\_public}, for the five rows reported on \texttt{test\_public}); latency across splits reflects network and provider conditions, so read down a split, not across. Latency is wall clock around the API call, so it includes client-side retries and provider queueing: the median is the headline, and the p90 tail is mostly backoff. No entry is dropped; the mean, maximum and a $5\%$-per-tail trimmed mean are in \texttt{tables/baseline\_cost\_latency.csv}. \emph{USD is an estimate, not a measurement}: the zero-shot path records no token counts, so those rows use OpenRouter list prices (2026-05-04) at an assumed $600$-prompt/$80$-completion tokens; agentic rows use their recorded total tokens, split at the same ratio. $^{\dagger}$Per-entry predictions were not retained, so only the aggregate mean ($0.99$\,s/entry) is recoverable. $^{\ast}$No timing was recorded on either split. $^{\ddagger}$Lower bound: reasoning tokens are billed but not included in the 80-token assumption.}
\label{tab:cost_latency}
\centering
\small
\setlength{\tabcolsep}{5pt}
\begin{tabular}{lccccc}
\toprule
\textbf{Verifier} & \textbf{$n$} & \textbf{Median s} & \textbf{p90 s} & \textbf{API calls} & \textbf{Est.\ USD/1k} \\
\midrule
\multicolumn{6}{l}{\emph{Citation-database tools}} \\
\gls{doi}-only$^{\dagger}$ & 1,119 & --- & --- & 0.434 & 0.00 \\
\midrule
\multicolumn{6}{l}{\emph{Zero-shot \glspl{llm} (one API call per entry)}} \\
Gemini~2.5~Pro & 831 & 10.98 & 20.39 & 1.000 & 3.30$^{\ddagger}$ \\
Gemini~2.5~Flash & 1,119 & 1.06 & 1.47 & 1.000 & 0.38$^{\ddagger}$ \\
Llama~4~Maverick & 831 & 4.45 & 19.98 & 1.000 & 0.17 \\
Claude Opus~4.7 & 1,119 & 2.42 & 3.42 & 1.000 & 5.00 \\
Claude Sonnet~4.6 & 1,119 & 7.27 & 9.18 & 1.000 & 3.00 \\
GPT-5.4 (zero-shot) & 1,119 & 2.23 & 2.95 & 1.000 & 1.55 \\
Mistral Large & 1,119 & 3.34 & 4.20 & 1.000 & 1.68 \\
GPT-5.1 (zero-shot) & 1,119 & 2.00 & 2.73 & 1.000 & 1.55 \\
Qwen3-235B & 1,119 & 5.15 & 17.57 & 1.000 & 0.17 \\
Qwen3-VL-235B & 831 & 3.48 & 6.72 & 1.000 & 0.17 \\
DeepSeek-R1 & 1,119 & 19.09 & 29.89 & 1.000 & 0.51$^{\ddagger}$ \\
DeepSeek-V3.2 & 1,119 & 7.44 & 14.78 & 1.000 & 0.25 \\
\midrule
\multicolumn{6}{l}{\emph{Agentic (tool-use; up to 5 tool calls per entry)}} \\
GPT-5.1 $+$ CrossRef/OpenAlex/arXiv & 831 & 9.22 & 19.32 & 2.854 & 5.61 \\
GPT-5.1 $+$ \texttt{bibtex-updater} (agentic) & 831 & 6.25 & 18.68 & 2.000 & 3.42 \\
Sonnet~4.6 $+$ \texttt{bibtex-updater} (agentic) & 1,119 & 9.59 & 14.06 & 2.059 & 13.25 \\
\midrule
\rowcolor{gray!10}
\multicolumn{6}{l}{\emph{Co-designed (reference upper bound; see \cref{app:codesign})}} \\
\rowcolor{gray!10}
\texttt{bibtex-updater} (v1.2.0)$^{\ast}$ & 1,119 & --- & --- & 1.990 & 0.00 \\
\rowcolor{gray!10}
GPT-5.1 $+$ \texttt{bibtex-updater} (always-call) & 1,119 & 2.60 & 3.96 & 1.000 & 1.55 \\
\bottomrule
\end{tabular}
\end{table}

Three readings follow. Chain of thought costs latency and nothing else in the call budget: on \texttt{dev\_public}, DeepSeek-R1's median of $19.09$\,s is  two and a half times the slowest single-pass model (DeepSeek-V3.2 at $7.44$\,s) and eighteen times the fastest (Gemini~2.5~Flash at $1.06$\,s); its \$$0.51$ per thousand entries is a lower bound, since the estimate does not count its reasoning tokens. Latency and dollars therefore rank the cohort differently, and neither ranks it the way \gls{fpr} does (\cref{tab:results}):  the most expensive \gls{llm} per entry, Opus~4.7, sits mid-cohort on \gls{fpr}, while Gemini~2.5~Flash is among the cheapest and has the second-lowest \gls{fpr}. Finally, wrapping a model in an agentic harness multiplies its estimated cost by $2.2{\times}$ (GPT-5.1 $+$ \texttt{bibtex-updater}) to $4.4{\times}$ (Sonnet~4.6 $+$ \texttt{bibtex-updater}), which buys the recall reported in \cref{tab:results} at the false-positive inflation analyzed in \cref{app:aggregation}; \texttt{bibtex-updater} reaches a comparable operating point with no \gls{llm} call at all.

\begin{table}[!t]
\caption{\textbf{Three failure modes that bound LLM-verifier deployment.} Overview results on \textsc{Hallmark}; with full tables referenced. ``1-in-$N$'' is the precision at a venue-realistic ${\sim}2\%$ hallucination rate (one true hallucination per $N$ flags).}
\label{tab:failure-modes}
\centering
\footnotesize
\setlength{\tabcolsep}{4pt}
\renewcommand{\arraystretch}{1.15}
\begin{tabular}{@{}p{2.25cm}p{3.45cm}p{3.95cm}p{3.0cm}@{}}
\toprule
\textbf{Failure mode} & \textbf{What breaks} & \textbf{Evidence} & \textbf{Implication} \\
\midrule
\textbf{(i) Agentic lookups, diminishing returns}\newline{\scriptsize(\cref{sec:main_results})}
& The prompted model tends to flag an entry as soon as \emph{any} one database returns no match, so partial database coverage becomes a false positive.
& A 5-call budget lifts recall past the conservative rule-based reference (DR\,.97--.99 vs.\ .82) but at ${\sim}$$9{\times}$ its false-positive rate (.43--.48 vs.\ .05); the rise comes from the harness (any-no-match flagging), not the base model.
& More tool calls buy recall, not precision. \\
\addlinespace[2pt]
\textbf{(ii) Base-rate precision drop}\newline{\scriptsize(\cref{tab:ppv})}
& At a venue-realistic ${\sim}2\%$ base rate, precision is governed by the ratio of detection rate to FPR (Bayes' rule), and FPR is the term that varies across verifiers.
&  FPR spans .05--.70 across \gls{llm} verifiers, a ${\sim}6{\times}$ precision gap at a $2\%$ base rate: the most precise verifiers reach 1-in-6 (Gemini~2.5~Pro) to 1-in-10 (Sonnet~4.6) flags, high-FPR open-weight models 1-in-35 to 1-in-39.
& FPR, not recall, is the deployment-decisive lever. \\
\addlinespace[2pt]
\textbf{(iii) Post-cutoff calibration breakdown}\newline{\scriptsize(\cref{tab:temporal_supplement})}
& Most LLMs over-flag papers past their training cutoff: ``flag everything unfamiliar''.
& On 2024--2025 papers, 8 of 12 LLMs degrade sharply (FPR .59--.89), two over-flag only moderately, and only the two latest-cutoff models hold (Sonnet~4.6 .12, Opus~4.7 .07), confounded with possible recall.
& Verifier trust expires with the training cutoff. \\
\bottomrule
\end{tabular}
\end{table}

\section{Failure mode (i): agentic aggregation}
\label{app:aggregation}
A deterministic re-aggregation over three of the harness's four databases confirms the lever: holding sources and matcher fixed, switching from \emph{any-no-match} flagging (flag if any single database returns no match) to \emph{consensus} flagging (flag only when all queried databases agree the entry is absent) cuts \gls{fpr} from $0.73$ to $0.05$ (a ${\sim}15{\times}$ reduction) at a recall cost on metadata-corruption types, which pairing consensus with contradiction checks---\texttt{bibtex-updater}'s design---recovers (\cref{tab:aggregation}). So failure mode~(i) characterizes this common retrieval-augmented pattern, not every agentic design.
\Cref{tab:failure-modes} names failure mode~(i)---the false-positive inflation of agentic, retrieval-augmented verifiers (\cref{tab:results}). Here we isolate its mechanism: the inflation is a property of the \emph{aggregation rule} that turns multi-database lookups into a verdict, with the base model held fixed, and \cref{tab:aggregation} quantifies the fix.
\paragraph{Cross-source aggregation on the agentic databases.} Failure mode~(i) attributes the agentic harness's inflated FPR to its emergent \emph{any-no-match} disposition; we isolate that aggregation rule directly. Holding the databases and the matcher fixed, we query three of the harness's four databases---CrossRef, OpenAlex, and arXiv---for every \texttt{dev\_public} entry, count a source as confirming when it returns a record whose title matches the entry ($\geq 0.85$ similarity), and score three rules over identical evidence. \emph{Any-no-match} (flag if \emph{any} source misses) reaches FPR $0.729$ at DR $0.835$; \emph{consensus} (flag only if \emph{all three} miss) cuts FPR to $0.049$---roughly $15{\times}$ lower---at DR $0.251$; \emph{majority} (flag if fewer than half confirm) falls between, at FPR $0.201$ / DR $0.417$. The FPR reduction is therefore a property of the aggregation rule, not the model or the sources. The recall cost is expected and instructive: title-presence consensus catches only \emph{non-existence} hallucinations and misses the metadata-corruption types (\texttt{near\_miss\_title}, \texttt{chimeric\_title}, \texttt{wrong\_venue}), where a real paper's title still matches---so the principled rule is consensus-absence \emph{paired with} positive-contradiction checks, exactly \texttt{bibtex-updater}'s design (DR $0.820$ / FPR $0.051$). Two caveats bound the reading: Semantic Scholar, the harness's fourth backend, is excluded because its API key returned HTTP~403 on every endpoint during our evaluation---whereupon the harness silently falls back to a rate-limited keyless pool---so we aggregate over the other three;\footnote{This backend fragility is itself an instance of failure mode~(i). Semantic Scholar's free API tier has been curtailed since 2024 and its public support repository was archived in January~2025 (\url{https://github.com/allenai/s2-folks}); separately, OpenAlex began requiring an API key in early 2026, so the harness's keyless OpenAlex path no longer functions either. A verifier that reads such a failed lookup as ``no record found'' converts backend unavailability into a false positive.} and the deterministic title matcher is a proxy for the harness's per-source judgment, so the absolute any-no-match FPR need not equal the LLM harness's (and the proxy DR $0.835$ likewise understates the harness's $0.97$--$0.99$); the any-vs-consensus \emph{ordering} is what transfers, not the level.
\begin{table}[t]
\caption{\textbf{Cross-source aggregation on the agentic harness's databases} (\texttt{dev\_public}; deterministic title-match proxy over CrossRef/OpenAlex/arXiv at $\geq 0.85$ similarity). Holding databases and matcher fixed, the aggregation rule alone moves FPR by ${\sim}15{\times}$. \texttt{bibtex-updater} pairs consensus-absence with positive-contradiction checks, recovering the recall that title-only consensus loses. Read the any-vs-consensus \emph{ordering}, not the absolute level (the proxy differs from the LLM harness's own operating point; \cref{tab:results}). Shading marks the lever: red is the any-no-match FPR and blue the consensus FPR on identical evidence (${\sim}15{\times}$ apart); the gray row is the co-designed reference, as in \cref{tab:results}.}
\label{tab:aggregation}
\centering
\small
\begin{tabular}{lcc}
\toprule
\textbf{Aggregation rule} & \textbf{DR $\uparrow$} & \textbf{FPR $\downarrow$} \\
\midrule
Any-no-match (flag if \emph{any} source misses)      & 0.835 & \cellcolor{cellbad}0.729 \\
Majority (flag if fewer than half confirm)           & 0.417 & 0.201 \\
Consensus (flag only if \emph{all three} miss)       & 0.251 & \cellcolor{cellgood}0.049 \\
\midrule
\rowcolor{gray!10}
\texttt{bibtex-updater} (consensus-absence $+$ contradiction) & 0.820 & 0.051 \\
\bottomrule
\end{tabular}
\end{table}

\begin{takeaway}
\textbf{Takeaway.} Failure mode~(i) is an \emph{aggregation} failure: on identical lookups with the base model held fixed, flagging on any single-source miss gives \gls{fpr} $0.729$, whereas requiring consensus absence lowers it ${\sim}15{\times}$ to $0.049$, at a recall cost that pairing consensus with positive-contradiction checks (\texttt{bibtex-updater}) recovers (\gls{dr} $0.820$ / \gls{fpr} $0.051$). The figures are a deterministic title-match proxy over three of the harness's four databases, so the any-vs-consensus \emph{ordering} is the finding; the absolute level is proxy-dependent.
\end{takeaway}
\subsection{Stage-2 diagnosis cascade: scoring stances and full panel}
\label{app:cascade}
All three cascades share one Stage~1. It decides 882 of the 1{,}119 \texttt{dev\_public} entries on its own (384 \textsc{hallucinated}, 498 \textsc{valid}) and forwards 237 to Stage~2. Sonnet~4.6 returns 151 \textsc{hallucinated} and 2 \textsc{valid} verdicts and leaves 84 unresolved; GPT-5.1 resolves all 237 (226 \textsc{hallucinated}, 11 \textsc{valid}), so its conservative and aggressive rows coincide; GPT-5.4 returns 214 \textsc{hallucinated} and 1 \textsc{valid} and leaves 22 unresolved. Only about a fifth of the split reaches the agentic stage, which is part of why the harness's false-positive inflation does not appear: the routing limits both what tool use can fix and what it can break.

We report two stances: \emph{conservative} keeps the entries Stage~2 still cannot resolve (its residual \textsc{uncertain} verdicts) as abstentions; \emph{aggressive} flags any entry not affirmatively verified by either stage, assigning it \textsc{hallucinated} at confidence $0.55$.
With Sonnet~4.6 as Stage~2, aggressive promotion raises Tier-3 \gls{f1} by one to two points and pushes recall toward 1.0, at a few points of added \gls{fpr} and a ${\sim}1.5$\,pp \gls{auroc} drop from collapsing confidences to the fixed $0.55$ (\cref{tab:cascade}). The stance matters only in proportion to the residue Stage~2 leaves: Sonnet leaves 84 of 1{,}119 \texttt{dev\_public} entries unresolved, GPT-5.4 leaves 22 (so its two stances differ by at most $.003$ on any metric), and GPT-5.1 leaves none, which collapses the choice entirely.

\begin{table}[t]
\caption{\textbf{Cascade results by Stage-2 backbone and scoring stance.} Rows are grouped by split, then by backbone, then by how residual \textsc{uncertain} is scored: \emph{cons.}\ keeps it as an abstention, \emph{agg.}\ promotes it to a flag at confidence $0.55$. All three cascades share one Stage~1; the Sonnet~4.6 and GPT-5.4 Stage~2 runs were collected on 2026-09-21 (Sonnet~4.6 via OpenRouter, a dated snapshot of a drifting endpoint; \cref{sec:limitations}). ``---'' marks splits with no scored \textsc{valid} entries (\texttt{stress\_test} is evaluated on its 121 hallucinated entries; the split's single valid entry is the contamination canary). \texttt{T3-F1} is F1 on Tier-3 (hard) hallucinations.}
\label{tab:cascade}
\setlength{\tabcolsep}{3pt}
\centering\small
\begin{tabular}{cllcccccc}
\toprule
\textbf{Split} & \textbf{Stage-2} & \textbf{Mode} & \textbf{DR $\uparrow$} & \textbf{FPR $\downarrow$} & \textbf{F1 $\uparrow$} & \textbf{TW-F1 $\uparrow$} & \textbf{T3-F1 $\uparrow$} & \textbf{AUROC $\uparrow$} \\
\midrule
  & Sonnet~4.6 & cons. & 0.989 & 0.037 & 0.976 & 0.982 & 0.910 & 0.977 \\
 &  & agg. & 0.990 & 0.037 & 0.980 & 0.986 & 0.935 & 0.975 \\
 & GPT-5.1 & cons. & 0.974 & 0.039 & 0.970 & 0.976 & 0.929 & 0.958 \\
 &  & agg. & 0.974 & 0.039 & 0.970 & 0.976 & 0.929 & 0.958 \\
 & GPT-5.4 & cons. & 0.990 & 0.039 & 0.978 & 0.985 & 0.930 & 0.973 \\
\multirow{-6}{*}{\shortstack{\texttt{dev\_public}\\$(n{=}1{,}119)$}}&  & agg. & 0.990 & 0.039 & 0.979 & 0.985 & 0.933 & 0.973 \\
\midrule
  & Sonnet~4.6 & cons. & 0.989 & 0.058 & 0.975 & 0.982 & 0.905 & 0.976 \\
 &  & agg. & 0.990 & 0.058 & 0.978 & 0.984 & 0.928 & 0.972 \\
 & GPT-5.1 & cons. & 0.965 & 0.058 & 0.965 & 0.971 & 0.924 & 0.947 \\
 &  & agg. & 0.965 & 0.058 & 0.965 & 0.971 & 0.924 & 0.947 \\
 & GPT-5.4 & cons. & 0.990 & 0.055 & 0.979 & 0.985 & 0.927 & 0.971 \\
\multirow{-6}{*}{\shortstack{\texttt{test\_public}\\$(n{=}831)$}}&  & agg. & 0.990 & 0.058 & 0.978 & 0.984 & 0.928 & 0.970 \\
\midrule
  & Sonnet~4.6 & cons. & 0.966 & --- & 0.983 & 0.981 & 0.971 & --- \\
 &  & agg. & 0.967 & --- & 0.983 & 0.982 & 0.971 & --- \\
 & GPT-5.1 & cons. & 0.851 & --- & 0.920 & 0.927 & 0.971 & --- \\
 &  & agg. & 0.851 & --- & 0.920 & 0.927 & 0.971 & --- \\
 & GPT-5.4 & cons. & 0.967 & --- & 0.983 & 0.982 & 0.971 & --- \\
\multirow{-6}{*}{\shortstack{\texttt{stress\_test}\\$(n{=}121)$}}&  & agg. & 0.967 & --- & 0.983 & 0.982 & 0.971 & --- \\
\bottomrule
\end{tabular}
\end{table}

\section{Failure mode (ii): base-rate precision and deployment}
\subsection{Deployment PPV analysis}
\label{app:ppv}
Published findings pin the base rate only loosely: incident studies report affected-\emph{paper} counts rather than rates (GPTZero: 53 of 4{,}841 NeurIPS~2025 papers~\citep{gptzero2025}), and even the implied paper-level rate sits well above the \emph{entry-level} rate the precision calculation needs, so prevalence should be treated as an estimate (for an ablation, see \cref{tab:ppv_sweep}).

At venue-realistic prevalence ($1$--$5\%$), positive predictive value drops sharply.
\cref{tab:ppv} reports \gls{ppv} $= \frac{\text{DR}\cdot\text{prev}}{\text{DR}\cdot\text{prev} + \text{FPR}\cdot(1-\text{prev})}$, where $\text{prev}$ is the entry-level hallucination prevalence, for the full-coverage tools and the co-designed \texttt{bibtex-updater} (whose \gls{dr}/\gls{fpr} use the headline committed-VALID convention; its abstention is reported separately in \cref{app:coverage}).

\begin{table}[t]
\caption{\textbf{Positive predictive value at venue-realistic prevalence.} In the PPV\textsubscript{2\%} column, green marks roughly one true catch per six flags (PPV ${\ge}16\%$) and red under one per twenty (PPV ${<}5\%$); the gray row is the co-designed reference.}
\label{tab:ppv}
\centering
\small
\begin{tabular}{lcccc}
\toprule
\textbf{Tool} & \textbf{DR} & \textbf{FPR} & \textbf{PPV\textsubscript{2\%}} & \textbf{PPV\textsubscript{5\%}} \\
\midrule
\rowcolor{gray!10}
bibtex-updater (v1.2.0) & 0.820 & 0.051 & 24.8\% & 46.0\% \\
Gemini~2.5~Pro    & 0.476 & 0.050 & \cellcolor{cellgood}16.3\% & 33.4\% \\
Opus~4.7          & 0.906 & 0.154 & 10.7\% & 23.6\% \\
Sonnet~4.6        & 0.916 & 0.161 & 10.4\% & 23.0\% \\
DOI-only          & 0.218 & 0.043 & 9.4\%  & 21.1\% \\
Gemini~2.5~Flash  & 0.500 & 0.100 & 9.3\%  & 20.8\% \\
Mistral Large     & 0.716 & 0.250 & 5.5\%  & 13.1\% \\
GPT-5.1           & 0.837 & 0.411 & \cellcolor{cellbad}4.0\%  & 9.7\%  \\
Qwen3-235B        & 0.860 & 0.533 & \cellcolor{cellbad}3.2\%  & 7.8\%  \\
DeepSeek-R1       & 0.896 & 0.623 & \cellcolor{cellbad}2.9\%  & 7.0\%  \\
DeepSeek-V3.2     & 0.911 & 0.702 & \cellcolor{cellbad}2.6\%  & 6.4\%  \\
\bottomrule
\end{tabular}
\end{table}

At a $2\%$ base rate, GPT-5.1 achieves only 4.0\% \gls{ppv}: roughly 24 of every 25 flagged entries are false alarms.
Aggressive models (DeepSeek-R1/V3.2) fall below 3\% despite ${\geq}89\%$ \gls{dr}.
Gemini~2.5~Flash more than doubles GPT-5.1's \gls{ppv} (9.3\% vs.\ 4.0\%) through its much lower \gls{fpr}, despite a lower detection rate: the precision-bottleneck logic favors the conservative model at deployment-realistic prevalence.

\begin{takeaway}
\textbf{Takeaway.}  \gls{fpr} alone does not fix the order: \gls{doi}-only has the lowest \gls{fpr} in the cohort yet lands mid-table at $9.4\%$, because it catches few hallucinations. The ratio of detection rate to \gls{fpr} decides.
\end{takeaway}

\paragraph{Prevalence sweep.}
To separate the \emph{ranking} of tools (which is prevalence-invariant because \gls{dr} and \gls{fpr} are) from \emph{deployability} (absolute \gls{ppv}, which is not), \cref{tab:ppv_sweep} sweeps the same closed-form
\begin{equation}
\text{PPV}(\text{prev}) = \frac{\text{DR}\cdot\text{prev}}{\text{DR}\cdot\text{prev} + \text{FPR}\cdot(1-\text{prev})} \label{eq:ppv_sweep}
\end{equation}
over prevalence values $\{0.005, 0.01, 0.02, 0.05, 0.10, 0.20, 0.30\}$ for four representative tools spanning the \gls{fpr} spectrum, using the exact \gls{dr}/\gls{fpr} from \cref{tab:results}.

\begin{table}[t]
\caption{\textbf{PPV vs.\ prevalence for four representative tools} (\cref{eq:ppv_sweep}; DR/FPR from \cref{tab:results}). (a)~The \emph{ranking} is essentially prevalence-invariant: Opus~4.7 $>$ Sonnet~4.6 $>$ GPT-5.1 $>$ DeepSeek-V3.2 at every prevalence, because PPV is monotone in DR/FPR and the FPR gaps dominate. (b)~\emph{Absolute} PPV (deployability) changes sharply with prevalence: the same tool moves from near-useless at venue-realistic prevalence to usable as prevalence rises (Opus~4.7: 2.9\% at a $0.5\%$ base rate to 71.6\% at $30\%$). Shading in the 2\% column mirrors \cref{tab:ppv}; at other prevalences the ranking is unchanged and only deployability moves.}
\label{tab:ppv_sweep}
\centering
\small
\setlength{\tabcolsep}{4pt}
\begin{tabular}{lccccccccc}
\toprule
\textbf{Tool} & \textbf{DR} & \textbf{FPR} & \textbf{.5\%} & \textbf{1\%} & \textbf{2\%} & \textbf{5\%} & \textbf{10\%} & \textbf{20\%} & \textbf{30\%} \\
\midrule
Claude Opus~4.7   & 0.906 & 0.154 & 2.9\% & 5.6\% & 10.7\% & 23.6\% & 39.5\% & 59.5\% & 71.6\% \\
Claude Sonnet~4.6 & 0.916 & 0.161 & 2.8\% & 5.4\% & 10.4\% & 23.0\% & 38.7\% & 58.7\% & 70.9\% \\
GPT-5.1           & 0.837 & 0.411 & 1.0\%  & 2.0\%  & \cellcolor{cellbad}4.0\%  & 9.7\%  & 18.5\% & 33.7\% & 46.6\% \\
DeepSeek-V3.2     & 0.911 & 0.702 & 0.6\%  & 1.3\%  & \cellcolor{cellbad}2.6\%  & 6.4\%  & 12.6\% & 24.5\% & 35.7\% \\
\bottomrule
\end{tabular}
\end{table}

The sweep makes the deployment lesson concrete: at venue-realistic $1$--$2\%$ even the best tool flags four-to-five false alarms per true catch, but the \emph{relative} ordering a benchmark establishes transfers unchanged to any deployment prevalence: only the operating point on the \gls{ppv} curve moves.

\subsection{Selective prediction and the Coverage column}
\label{app:coverage}
\paragraph{Structural precision (from the main text).}
\texttt{bibtex-updater}'s low false-positive rate is \emph{structural}: it comes from conservative matching, not from abstention.
It abstains on the ${\sim}14\%$ of entries it cannot back with a record; on the entries it commits to, its \gls{fpr} is essentially unchanged while its detection rate increases (\cref{tab:coverage}); abstention raises recall and leaves precision unchanged.
Because its low \gls{fpr} does not depend on abstention, \texttt{bibtex-updater} serves as the precision-anchored reference against which the \glspl{llm}' abstention-driven gains are measured, and the reason we keep it out of the ranking (\cref{app:codesign}).
Because \gls{dr} and \gls{fpr} are prevalence-independent, we compare on them, and on \gls{f1} only within a split.

The Coverage column of \cref{tab:results} reports, for each tool, the fraction of entries on which it \emph{commits} to a VALID/HALLUCINATED verdict rather than abstaining (UNCERTAIN). We report it because abstention is a usable deployment action---defer to a human reviewer---and because it is the lens under which the co-designed \texttt{bibtex-updater} reads cleanly: it abstains on the $12$--$14\%$ of entries it cannot verify against a backing record (coverage $0.86$ on dev, $0.88$ on test). Its low headline FPR ($0.051$) does not come from that abstention: on the entries it commits to, its selective FPR ($0.056$) is essentially unchanged from full coverage, so abstention raises recall and leaves precision unchanged. Where the LLM verifiers lower their false-positive rate as they abstain---precision-via-abstention---\texttt{bibtex-updater}'s risk--coverage curve is flat: its precision is \emph{structural}---it comes from conservative matching, not from abstention (its selective FPR $0.056$ $\approx$ its full-coverage $0.051$)---so selective prediction adds little. We use ``structural'' throughout in this operational sense; it does not mean invariant, as the FPR still rises modestly across splits ($0.051\to0.080$; \cref{tab:test_public_full}). No peer citation benchmark reports coverage; the selective-prediction framing is standard elsewhere~\citep{elyaniv2010selective,geifman2017selective}.

\paragraph{Which entries count as abstentions.}
For the \gls{llm} verifiers an abstention is an \textsc{uncertain} verdict. For \texttt{bibtex-updater}, an entry counts as an abstention when its status leaves the tool undecided (\texttt{unconfirmed}, or a failed or skipped lookup), or when the tool returned no record for it. The tool's own \texttt{abstained} flag is not used, since it also marks \texttt{not\_found} detections. This gives coverage $0.86$ on \texttt{dev\_public} and $0.88$ on \texttt{test\_public}.
The \texttt{coverage} field in the released aggregates is unreliable in 14 of 42 files (\cref{app:bootstrap}), so the Cov.\ column of \cref{tab:results} is read from \texttt{coverage\_reporting.json} (for \texttt{bibtex-updater}, from its released aggregate) rather than recomputed from the release.

\paragraph{Dual scoring.} We score each tool three ways (\cref{tab:coverage}). The \emph{selective} stance scores only the committed (non-abstained) entries; the \emph{conservative} stance scores the full split with every abstention forced to VALID; the \emph{aggressive} stance scores the full split with every abstention forced to a flag (\texttt{HALLUCINATED@0.55}, i.e.\ scored as a flag with confidence $0.55$, just above the $0.5$ decision threshold). \Cref{tab:results} and \cref{tab:test_public_full} report each tool's native stance (selective for the \glspl{llm}, conservative for \texttt{bibtex-updater}), with coverage beside it; \cref{tab:coverage} gives all three. The conservative--aggressive gap is the cost of the abstention, and it is non-trivial exactly where coverage is below one: \texttt{bibtex-updater} on \texttt{dev\_public} (coverage $0.86$) earns selective DR/FPR/F1 of $0.992$/$0.056$/$0.971$ on the $965$ entries it commits to, while over the full split conservative scoring (abstain$\to$VALID) gives $0.820$/$0.051$/$0.880$ and aggressive scoring (abstain$\to$HALLUCINATED) gives $0.987$/$0.140$/$0.937$; DeepSeek-R1 on \texttt{test\_public} (coverage $0.783$) pays $+16.2$\,pp between its conservative and aggressive stances. The selective FPR ($0.056$) essentially matches the full-coverage headline FPR ($0.051$): abstention does not add precision for \texttt{bibtex-updater}; it raises recall (DR $0.820\to$$0.992$) at near-unchanged FPR. Reporting Coverage \emph{with} the aggressive number closes the obvious loophole---abstaining on everything would otherwise give perfect committed-precision---so a tool cannot hide a weak verdict behind UNCERTAIN.

\paragraph{Risk--coverage.} For each tool with graded confidences we abstain on the least-confident entries first (smallest $|P(\text{hallucinated})-0.5|$) and score FPR on the rest, tracing a risk--coverage curve (data in the released \texttt{coverage\_reporting.json}; we summarize each curve by its FPR at $90\%$ coverage in \cref{tab:coverage}b). Abstaining on the least-confident decile yields a real FPR reduction for the better-calibrated LLM verifiers: the high-AUROC verifiers (Sonnet~4.6, Opus~4.7; AUROC $0.90$--$0.93$) sit at the favorable corner with FPR ${\approx}0.10$--$0.12$ at $90\%$ coverage, while the recall-aggressive open-weight models stay above $0.5$. \texttt{bibtex-updater} is the opposite case: its risk--coverage curve is flat---FPR is essentially constant ($0.051$ at full coverage, $0.056$ on the committed subset)---so selective prediction adds no precision. Where the LLM verifiers lower their false-positive rate as they abstain, \texttt{bibtex-updater}'s precision is structural; it is the precision-anchored reference against which the LLMs' abstention gains are read.

\paragraph{Two caveats.}  (i) LLM abstention is prompt-dependent (\cref{app:ablation_prompt}: Sonnet's UNCERTAIN rate moves $2.0\%\to8.7\%$ by wording), so coverage is read as ``coverage under the default prompt'', not an intrinsic constant. (ii) \texttt{bibtex-updater}'s abstention and an LLM's occupy the same column but differ in mechanism: the tool abstains because \emph{the databases hold no matching record} (a data-coverage gap), whereas an LLM abstains because \emph{it will not commit} (an epistemic gap); under the route-to-human framing both are the same useful action, which is why the column is shared. \texttt{bibtex-updater} abstains on $154/1119$ entries on \texttt{dev\_public} (coverage $0.86$) and $101/831$ on \texttt{test\_public} (coverage $0.88$).

\begin{table}[t]
\caption{\textbf{Coverage and selective prediction on \texttt{dev\_public}.} \emph{(a)} Per-tool coverage with selective (abstentions excluded) and aggressive (abstain$\to$HALLUCINATED) DR/FPR/F1; conservative scoring (abstain$\to$VALID) is the convention of \cref{tab:results}. On \texttt{test\_public}, \texttt{bibtex-updater}'s conservative and aggressive stances are $0.838$/$0.080$/$0.889$ and $0.990$/$0.151$/$0.952$, and its selective DR/FPR/F1 is $0.989$/$0.086$/$0.967$ ($n{=}730$). \emph{(b)} FPR at $90\%$ coverage from the risk--coverage curve; $-$ marks tools the threshold sweep did not cover (curves are in the released JSON). In (b), blue marks FPR ${\le}.12$ (the favorable risk--coverage corner) and red FPR ${>}.5$; the gray row in (a) is the co-designed reference.}
\label{tab:coverage}
\centering
\small
\setlength{\tabcolsep}{4pt}
\begin{tabular}{lccc}
\toprule
\multicolumn{4}{l}{\emph{(a) Coverage + dual scoring}} \\
\textbf{Tool} & \textbf{Cov.} & \textbf{Selective DR/FPR/F1} & \textbf{Aggr.\ DR/FPR/F1} \\
\midrule
\rowcolor{gray!10}
\texttt{bibtex-updater} (v1.2.0) & 0.86 & 0.992 / 0.056 / 0.971 & 0.987 / 0.140 / 0.937 \\
Gemini~2.5~Pro      & 0.967 & 0.476 / 0.050 / 0.627 & 0.497 / 0.074 / 0.637 \\
Gemini~2.5~Flash    & 0.988 & 0.500 / 0.100 / 0.631 & 0.508 / 0.105 / 0.636 \\
Mistral Large       & 0.989 & 0.716 / 0.251 / 0.742 & 0.721 / 0.253 / 0.745 \\
DeepSeek-R1         & 0.984 & 0.896 / 0.623 / 0.739 & 0.898 / 0.628 / 0.739 \\
Opus~4.7  & 1.000 & 0.906 / 0.154 / 0.890 & 0.906 / 0.154 / 0.890 \\
Sonnet~4.6 & 0.982 & 0.916 / 0.161 / 0.893 & 0.918 / 0.177 / 0.888 \\
\midrule
\multicolumn{4}{l}{\emph{(b) FPR at $90\%$ coverage (risk--coverage summary)}} \\
\textbf{Tool} & \textbf{AUROC} & \textbf{FPR@90\%} & \textbf{F1@90\%} \\
\midrule
Sonnet~4.6 & 0.927 & \cellcolor{cellgood}0.119 & 0.912 \\
Opus~4.7   & 0.903 & \cellcolor{cellgood}0.098 & 0.903 \\
GPT-5.4              & 0.834 & 0.222 & 0.816 \\
Mistral Large        & 0.744 & 0.241 & 0.736 \\
Gemini~2.5~Flash     & 0.739 & \cellcolor{cellgood}0.097 & 0.578 \\
DeepSeek-R1          & 0.741 & \cellcolor{cellbad}0.581 & 0.752 \\
DeepSeek-V3.2        & 0.609 & \cellcolor{cellbad}0.699 & 0.711 \\
\bottomrule
\end{tabular}
\end{table}

\begin{takeaway}
\textbf{Takeaway.} The precision-ceiling finding survives the obvious confounds: rankings are invariant to prompt wording (\cref{app:ablation_prompt}) and decision threshold (\cref{app:ablation_threshold}), detection leans on title/author content rather than a format tell (\cref{app:ablation_format}), and independent raters reproduce the over-flagging the benchmark measures (\cref{app:ablation_kappa}). Coverage makes the abstention lever explicit; the better-calibrated LLMs trade a small coverage cut for a real \gls{fpr} reduction (precision-via-abstention), whereas the co-designed \texttt{bibtex-updater}'s low \gls{fpr} ($0.051$) is structural: its risk--coverage curve is flat, so abstaining on the $12$--$14\%$ of entries it cannot verify recovers recall (\gls{dr} $0.820\to0.992$) at near-unchanged \gls{fpr} rather than adding precision.
\end{takeaway}

\subsection{Regime-conditional deployment guidance}
\label{app:deployment}

\emph{When to prefer recall.} The operating point is set by three quantities: prevalence, the cost ratio $c_{\text{FN}}/c_{\text{FP}}$ of a missed fabrication versus a false alarm, and reviewer capacity.
\Gls{fpr}-ranking holds in the \emph{reviewer-bound venue-audit} regime---low prevalence, finite reviewer attention, a false alarm cheap to dismiss---where precision is the limiting factor.
When a missed fabrication is far costlier than triaging extra flags, or when a later human-review stage filters out the false alarms the tool raises, recall plus human triage dominates and a high-recall tool (an agentic harness, or DeepSeek-V3.2) is preferable despite its \gls{fpr}.
The two takeaways---``rank on \gls{fpr}'' and ``a missed hallucination is the worse error''---are not in tension; they name different aspects of the $(\text{prevalence}, c_{\text{FN}}/c_{\text{FP}}, \text{capacity})$ space, which \cref{tab:deployment} resolves into tool choices.

\begin{table}[t]
\caption{\textbf{Regime-conditional deployment guidance.} The right verifier depends on prevalence, the cost of a missed fabrication relative to a false alarm ($c_{\text{FN}}/c_{\text{FP}}$), and reviewer capacity; per-tool numbers in \cref{app:deployment}.}
\label{tab:deployment}
\centering
\small
\setlength{\tabcolsep}{4pt}
\renewcommand{\arraystretch}{1.15}
\begin{tabular}{@{}p{3.5cm}p{3.3cm}p{5.55cm}@{}}
\toprule
\textbf{Regime} & \textbf{Pick} & \textbf{Why} \\
\midrule
Reviewer-bound venue audit---low prevalence, a false alarm cheap to dismiss &  Low-FPR verifier with usable recall (\texttt{bibtex-updater}); among \glspl{llm}, the best-calibrated Sonnet~4.6 or Opus~4.7 & At low prevalence false alarms outnumber true detections even for the best verifier (precision ${\sim}16\%$ at a $2\%$ base rate), so a few points of FPR decide usability; add the Stage-2 cascade when recall must rise at the same FPR. \\
\addlinespace[2pt]
High prevalence---triage queues, pre-submission self-check & High-recall verifier (an agentic harness or DeepSeek-V3.2) & Precision improves as the base rate rises, so catching more matters and the extra false alarms are affordable. \\
\addlinespace[2pt]
Costly miss, or downstream human review absorbs false alarms & High-recall verifier $+$ human triage & When a fabrication in print far outweighs an extra flag, recall plus a human filter is a better choice. \\
\bottomrule
\end{tabular}
\end{table}

At venue-realistic prevalence (${\sim}2\%$), the precision-bottleneck logic favors low-\gls{fpr} tools regardless of recall: \Gls{ppv} per flagged entry is  ${\sim}16\%$ for Gemini~2.5~Pro and the co-designed \texttt{bibtex-updater}, ${\sim}10$--$11\%$ for Opus~4.7 and Sonnet~4.6, ${\sim}4\%$ for GPT-5.1, and ${<}3\%$ for high-recall open-weight models (\cref{tab:ppv}).
\begin{itemize}
  \item \textbf{Pre-2024, recall-prioritized triage:} a high-recall verifier such as an agentic harness (DR\,${\approx}0.97$--$0.99$) or DeepSeek-V3.2 (DR\,=\,0.911), accepting the higher \gls{fpr} where reviewer capacity absorbs false positives. \textbf{Pre-2024, precision-prioritized:} \texttt{bibtex-updater} (DR\,=\,0.820, FPR\,=\,0.051, PPV\,${\sim}25\%$ at a $2\%$ base rate, ${\sim}2$--$3$ orders of magnitude cheaper than \gls{llm} verifiers),  or, among \glspl{llm}, Sonnet~4.6 or Opus~4.7 (FPR\,=\,0.161 / 0.154, \gls{dr} above 0.9); Gemini~2.5~Pro has lower \gls{fpr} (0.050) but misses half the hallucinations.
  \item \textbf{Near or beyond training cutoffs:} Sonnet~4.6 zero-shot for the best calibrated \gls{fpr} among LLMs tested. Agentic harnesses match higher detection but inflate \gls{fpr} over their zero-shot bases (GPT-5.1 +6\,pp; Sonnet 4.6 +27\,pp, a $2.7\times$ multiplier from a low base); deploy only where reviewer capacity absorbs false positives.
  \item \textbf{Past 12 months:} no tool we tested is reliable without cutoff-aware prompting (\cref{app:cutoff-aware}); the addendum recovers \gls{fpr} on GPT-5.1 and Qwen3-235B at the cost of indiscriminate abstention, barely moves Gemini~2.5~Flash, and abstains selectively only on Sonnet~4.6, so it requires per-provider tuning.
\end{itemize}

\section{Failure mode (iii): temporal fragility and calibration}
\label{app:temporal}
Following LiveCodeBench~\citep{livecodebench2024}, we tag entries with three temporal segments (\emph{pre-2023}, \emph{2023--2024}, \emph{2025+}) so the benchmark can measure how verifiers degrade on recent papers (\cref{sec:temporal_robustness}).
We read that degradation through the cross-regime \emph{ranking}, which is prompt-invariant, rather than through absolute post-cutoff \gls{fpr}, which a prompt-sensitivity sweep moves by $10$--$37$ percentage points (pp) on wording alone (\cref{app:ablation_prompt}).

\paragraph{Training-data cutoffs of evaluated LLMs.}
\label{app:cutoffs}
The temporal evaluation assumes papers from calendar years 2024--2025 post-date most evaluated models' training data. \cref{tab:cutoffs} compiles vendor-disclosed training cutoffs (or the most recent reliable upper bound when no disclosure is available) for every \gls{llm} used as a baseline. Three observations frame the temporal results. (i) Every evaluated model has a disclosed or inferred cutoff no later than late~2025, so the 448-entry temporal supplement (drawn from 2024--2025 DBLP proceedings) is partially or fully out-of-distribution for all of them. (ii) The two closed-weight models with earliest cutoffs (GPT-5.1, Sep 2024) and the conservative open-weight Gemini 2.5 Flash (Jan 2025) are hardest-hit by the post-cutoff \gls{fpr} rise, consistent with their reliance on parametric confirmation. (iii) DeepSeek models do not publish an official cutoff in their technical reports, so we report the OpenRouter-surfaced metadata and flag it as a provenance gap that future versions of \textsc{Hallmark} should track via release-date probes.

\begin{table}[t]
\caption{\textbf{Training-data cutoffs of LLM baselines evaluated in \textsc{Hallmark}.} ``Disclosed'' indicates a vendor-published cutoff in an official model card, API documentation, system card, or technical report; ``Inferred'' indicates an upper bound derived from release date and predecessor disclosures. Anthropic distinguishes ``training data cutoff'' from ``reliable knowledge cutoff''; we report the latter for Claude variants. Rows above the midrule have cutoffs at or before mid-2025; rows below have post-Aug~2025 cutoffs (the ``later-cutoff'' cohort referenced in \cref{sec:temporal_robustness}).}
\label{tab:cutoffs}
\centering
\small
\setlength{\tabcolsep}{4pt}
\fitwidth{%
\begin{tabular}{lllc}
\toprule
\textbf{Model} & \textbf{Training cutoff} & \textbf{Source} & \textbf{Status} \\
\midrule
DeepSeek-R1                  & Jul~2024                  & OpenRouter model metadata              & Disclosed (via host) \\
Llama~4~Maverick             & ${\leq}$\,Aug~2024        & Meta release notes; release Apr~2025   & Inferred \\
GPT-5.1                      & Sep~2024 (GPT-5 family)   & OpenAI system card (GPT-5)             & Disclosed (family) \\
DeepSeek-V3.2                & ${\leq}$\,Oct~2024        & Release date; V3 tech report           & Inferred \\
Gemini~2.5~Flash             & Jan~2025                  & Vertex AI model documentation          & Disclosed \\
Gemini~2.5~Pro               & Jan~2025                  & Vertex AI model documentation          & Disclosed \\
Qwen3-235B-A22B-2507         & Jun~2025                  & OpenRouter model metadata              & Disclosed (via host) \\
Qwen3-VL-235B-A22B-Instruct  & ${\leq}$\,Jun~2025        & OpenRouter metadata; Qwen3 family lag  & Inferred \\
Mistral Large~2512           & ${\leq}$\,mid-2025        & Release date; predecessor cutoff       & Inferred \\
\midrule
GPT-5.4                      & Aug~2025                  & OpenAI release (\texttt{gpt-5.4-2026-03-05}) & Disclosed \\
Claude Sonnet~4.6            & ${\leq}$\,Aug~2025        & Release Feb~2026; Anthropic family lag & Inferred \\
Claude Opus~4.7              & ${\leq}$\,Oct~2025        & Release Apr~2026; Anthropic family lag & Inferred \\
\bottomrule
\end{tabular}%
}
\end{table}

2024--2025 DBLP papers are out-of-distribution to varying degrees across the cohort: late-2025 entries are post-cutoff for every model, while early-2024 entries are post-cutoff only for the earliest-cutoff models (GPT-5.1: Sep~2024; DeepSeek-R1: Jul~2024). Across the nine non-Anthropic \glspl{llm} tested on the supplement the temporal hit splits into four regimes: four models with healthy baselines degrade sharply (GPT-5.1, Gemini 2.5 Flash, Mistral Large, Llama 4 Maverick; $\Delta$\gls{fpr}\,$+$35--62\,pp); three with already-elevated baselines stay near-ceiling or worsen (Qwen3-235B, DeepSeek-V3.2, Qwen3-VL-235B at \gls{fpr} 0.89, a new cohort maximum); DeepSeek-R1's run failed on 151 of the 448 entries, leaving only its \gls{fpr} of 0.856 evaluable (\cref{tab:temporal_supplement}, footnote $\dagger$); and Gemini 2.5 Pro over-flags only moderately (\gls{fpr} 0.25). Pro-tier post-training, not data recency, drives that last case: Gemini 2.5 Flash---same vendor, comparable cutoff---degrades sharply (\gls{fpr} 0.595). Training-data recency alone does not rescue parametric verification. The latest-cutoff non-Anthropic model tested (Qwen3-235B: Jun~2025) still ends in the near-ceiling group, so the gap between ``seen during training'' and ``reliably recalled'' matters as much as the cutoff. The two later-cutoff Anthropic models (Sonnet~4.6, Opus~4.7) sharpen the point: their inferred cutoffs (Aug~2025, Oct~2025) are not radically later than Qwen3-235B's, yet their \gls{fpr} shifts on the supplement are an order of magnitude smaller. Cutoff recency is therefore necessary but insufficient; pipeline-level differences (\gls{rlhf} for abstention, ``I don't know'' calibration) appear to dominate the temporal-robustness story.

\paragraph{Baseline temporal consistency.}
For bibtex-updater, detection rates are consistent across temporal segments ($\pm$2\%); \gls{doi}-only improves slightly on newer entries because recent papers more consistently include DOIs.

\subsection{Temporal supplement (2024--2025)}
\label{app:temporal_supplement}
The pattern extends past 2024--2025: a 60-entry probe on 2026 arXiv submissions reproduces the \gls{fpr} multiplier ($r{=}-0.82$ between baseline aggressiveness and relative post-cutoff degradation).

To validate the 60-entry temporal probe at scale (probe composition in \cref{app:dataset}), we construct a 448-entry temporal supplement: 300 valid entries scraped from DBLP across six ML venues (NeurIPS, ICML, ICLR, AAAI, CVPR, ECCV) for publication years 2024--2025, plus 148 hallucinated entries generated via the standard perturbation pipeline (58/52/38 across Tiers~1/2/3, spanning all 14 types; \cref{tab:stats}).
This provides $10{\times}$ more valid entries than the probe (300 vs.\ 30), yielding substantially tighter confidence intervals on \gls{fpr} estimates.
The 448-entry set (300 valid / 148 hallucinated, full coverage) is the canonical temporal supplement reported here; every released prediction file covers the same 448 keys, several of them also carrying rows from the 858-entry superset. A later 858-entry superset of the same 2024--2025 DBLP pool backs the extended reviewer-experiment robustness checks (recall probe, late-cutoff control, GPT-5.1 run-to-run variance) reported separately in the released artifacts (\path{results/reviewer_experiments/FINDINGS.md}), and is presented alongside---not in place of---the canonical 448-entry table.

\cref{tab:temporal_supplement} compares each \gls{llm} baseline's performance on the original benchmark (2021--2023 valid entries) against the temporal supplement (2024--2025 valid entries).
The pattern is unambiguous: every model's \gls{fpr} increases dramatically on recent papers, while detection rate inflates because models flag nearly everything as hallucinated.

\begin{table}[t]
\centering
\caption{\textbf{Temporal supplement: LLM baseline performance on 2021--2023 benchmark entries (\texttt{dev\_public}, $n{=}1{,}119$) vs.\ 2024--2025 supplement entries ($n{=}448$).} $\Delta$DR and $\Delta$FPR report the absolute shift. The top block (Gemini Flash through DeepSeek-R1) covers the six LLMs with training cutoffs at or before mid-2025, sorted by baseline FPR ascending; DeepSeek-R1 is placed last because its 24--25 behavior is anomalous (see $^\dagger$). Below the midrule are five models added later (2026-05-05 collection; \cref{app:llm-setup}): the two later-cutoff Anthropic models (Sonnet~4.6, Opus~4.7; documented or inferable post-Aug~2025 cutoffs, \cref{app:cutoffs}) do not show the post-cutoff FPR rise, while Gemini~2.5~Pro over-flags only moderately and Llama~4~Maverick and Qwen3-VL-235B degrade like the top block. $^\dagger$DeepSeek-R1's detection rate on this supplement is not evaluable; see footnote below table. In the FPR\textsubscript{24--25} column, red marks the over-flagging cluster (FPR ${\ge}.59$) and blue the two models whose post-cutoff FPR holds; Gemini~2.5~Pro (.250) sits between and is unshaded. \textbf{Snapshot.} Every row pairs both columns from one run: a $\Delta$ across the cutoff is only interpretable within a single endpoint snapshot. The Sonnet~4.6 and Opus~4.7 rows therefore stay on the May snapshot, so their 21--23 figures ($.127$, $.072$) are the May ones and differ from the later 2026-09-02 collection reported in \cref{tab:results} ($.161$, $.154$); \cref{tab:endpoint_drift} quantifies that gap.}
\label{tab:temporal_supplement}
\small
\setlength{\tabcolsep}{3pt}
\begin{tabular}{lcccccccc}
\toprule
\textbf{Model} & \textbf{DR\textsubscript{21--23}} & \textbf{DR\textsubscript{24--25}} & \textbf{$\Delta$DR} & \textbf{FPR\textsubscript{21--23}} & \textbf{FPR\textsubscript{24--25}} & \textbf{$\Delta$FPR} & \textbf{F1\textsubscript{24--25}} & \textbf{MCC\textsubscript{24--25}} \\
\midrule
Gemini Flash       & 0.500 & 0.844 & +0.344 & 0.100 & \cellcolor{cellbad}0.595 & \textbf{+0.495} & 0.555 & 0.251 \\
GPT-5.1            & 0.837 & 0.958 & +0.121 & 0.411 & \cellcolor{cellbad}0.759 & \textbf{+0.348} & 0.546 & 0.246 \\
Mistral Large      & 0.716 & 0.952 & +0.236 & 0.250 & \cellcolor{cellbad}0.793 & \textbf{+0.543} & 0.537 & 0.208 \\
Qwen3-235B         & 0.860 & 0.986 & +0.126 & 0.533 & \cellcolor{cellbad}0.809 & +0.276 & 0.541 & 0.245 \\
DeepSeek-V3.2      & 0.911 & 0.980 & +0.069 & 0.702 & \cellcolor{cellbad}0.759 & +0.057 & 0.558 & 0.278 \\
DeepSeek-R1$^\dagger$ & 0.896 & n/e & n/e & 0.623 & \cellcolor{cellbad}0.856 & +0.233 & n/e & n/e \\
\midrule
\multicolumn{9}{l}{\emph{Later-added models (2026-05-05 collection):}} \\
Claude Sonnet 4.6     & 0.781 & 0.818 & +0.037 & 0.127 & \cellcolor{cellgood}\textbf{0.120} & $-$0.007 & 0.793 & 0.688 \\
Claude Opus 4.7       & 0.752 & 0.716 & $-$0.036 & 0.072 & \cellcolor{cellgood}\textbf{0.073} & +0.001 & 0.768 & 0.669 \\
Gemini 2.5 Pro        & 0.476 & 0.626 & +0.150 & 0.050 & 0.250 & +0.200 & 0.588 & 0.366 \\
Llama 4 Maverick      & 0.614 & 0.939 & +0.325 & 0.146 & \cellcolor{cellbad}0.763 & \textbf{+0.617} & 0.539 & 0.216 \\
Qwen3-VL-235B         & 0.860 & 1.000 & +0.140 & 0.551 & \cellcolor{cellbad}0.887 & +0.336 & 0.527 & 0.201 \\
\bottomrule
\end{tabular}
\end{table}

\noindent{\footnotesize $^\dagger$DeepSeek-R1's run on this supplement returned 151 of 448 entries as \texttt{[Error fallback]} API failures recorded as \texttt{UNCERTAIN}, and those 151 include all 148 hallucinated entries. Its detection rate, $\Delta$DR, \gls{f1} and \gls{mcc} therefore measure the harness rather than the model and we mark them not evaluable (n/e). The \gls{fpr} of 0.856 is scored over the 297 valid entries the run did answer and is comparable to the other rows.}

\medskip
\paragraph{Endpoint drift between the two collections.}
The Sonnet~4.6 and Opus~4.7 rows above are read off the 2026-05-05 collection, while \cref{tab:results} reports the 2026-09-02 collection. \cref{tab:endpoint_drift} gives both. The two runs share prompt, split, temperature and seed, so the difference is attributable to the endpoint behind a fixed model name rather than to any change we made.
The size of that drift is why the rows above are not mixed: on \texttt{dev\_public} the \gls{fpr} moves $+0.034$ (Sonnet) and $+0.082$ (Opus) between collections, which is larger than the post-cutoff shift the table is built to measure ($-0.007$ and $+0.001$). Pairing a September 21--23 figure with a May 24--25 figure would therefore report drift as if it were a temporal effect, and with the signs as they fall here it would read as \gls{fpr} \emph{improving} on post-cutoff papers ($-0.041$, $-0.081$).

\begin{table}[t]
\centering
\caption{\textbf{Endpoint drift for the two Anthropic models on \texttt{dev\_public}.} The same model ID, prompt, split, temperature and seed, collected on two dates: \textbf{May}~$=$~2026-05-05, \textbf{Sep}~$=$~2026-09-02 (\cref{app:llm-setup}). \cref{tab:temporal_supplement} uses the May column so that each of its rows is within-snapshot; \cref{tab:results} reports the September column as the later run. $\Delta$ is September minus May. A controlled replay on shared inputs gives only 90\%/75\% Opus/Sonnet per-entry label agreement between the two collections, so these are distinct measurement sessions rather than repeated draws from one. The \gls{f1} columns show why we decline to rank these two models against each other: the gap between them is $0.3$\,pp in both collections, but its \emph{sign} reverses (bold marks the leader), so which model leads is a property of the session rather than of the models.}
\label{tab:endpoint_drift}
\small
\setlength{\tabcolsep}{4pt}
\fitwidth{%
\begin{tabular}{lcccccccc}
\toprule
& \multicolumn{3}{c}{\textbf{DR}} & \multicolumn{3}{c}{\textbf{FPR}} & \multicolumn{2}{c}{\textbf{F1}} \\
\cmidrule(lr){2-4}\cmidrule(lr){5-7}\cmidrule(lr){8-9}
\textbf{Model} & \textbf{May} & \textbf{Sep} & \textbf{$\Delta$} & \textbf{May} & \textbf{Sep} & \textbf{$\Delta$} & \textbf{May} & \textbf{Sep} \\
\midrule
Claude Sonnet 4.6 & 0.781 & 0.916 & $+0.135$ & 0.127 & 0.161 & $+0.034$ & 0.827 & \textbf{0.893} \\
Claude Opus 4.7   & 0.752 & 0.906 & $+0.154$ & 0.072 & 0.154 & $+0.082$ & \textbf{0.830} & 0.890 \\
\bottomrule
\end{tabular}%
}
\end{table}

The supplement confirms three patterns from the probe, now with substantially greater statistical power, and surfaces a fourth via the later-evaluated models.
First, \textbf{\Gls{fpr} rises sharply on recent papers for most models}: GPT-5.1's \gls{fpr} climbs from 41.1\% to 75.9\% ($1.8{\times}$); three conservative-baseline models (Gemini Flash $+$49.5\,pp, Mistral Large $+$54.3\,pp, Llama 4 Maverick $+$61.7\,pp) post the largest absolute jumps, while GPT-5.1's increase is more modest at $+$34.8\,pp.
Second, \textbf{the detection-rate increases are spurious}: the four largest-jump models (Gemini Flash, Mistral Large, Llama 4 Maverick, GPT-5.1) achieve \gls{dr}~${\geq}0.84$ on the supplement, but only because they flag nearly everything; \gls{f1} falls to ${\approx}0.55$ (from 0.61--0.77 on \texttt{dev\_public}) and \gls{mcc} to ${\approx}0.25$, confirming near-random predictions.
Third, \textbf{already-aggressive models stay near-ceiling}: DeepSeek-V3.2 (baseline FPR 70.2\%) shows only a 5.7\,pp increase but ends at 75.9\%; Qwen3-VL-235B is the most extreme, climbing 55.1\%~$\to$~88.7\%, a new cohort maximum.
Fourth, \textbf{a vendor-/pipeline-correlated subset resists}: Claude Sonnet 4.6 (FPR 12.0\%) and Opus 4.7 (7.3\%) hold; Gemini 2.5 Pro (FPR 25.0\%) and a later-cutoff GPT-5.4 (FPR 41.3\%) over-flag only moderately: pipeline differences appear to dominate over training recency for the residual gap.

The 60-entry probe, which additionally covered 2026 arXiv submissions, showed the same \gls{fpr}-multiplier pattern with a strong negative correlation ($r = -0.82$) between baseline aggressiveness and relative degradation, evidence that the failure mode generalizes beyond 2024--2025 DBLP.

These results establish that LLM citation verification leans heavily on parametric knowledge with a graded temporal boundary near the training cutoff: most models fall into the ``flag everything unfamiliar'' failure mode for post-cutoff papers.
But the partial-to-full resistance from Gemini 2.5 Pro, GPT-5.4, and the two Anthropic models shows this is not a hard structural limit of parametric verification: training-pipeline interventions (post-training calibration, abstention \gls{rlhf}) can shift the trade-off without retrieval augmentation, even at comparable training cutoffs.
We caution against reading the latest-cutoff models' low post-cutoff \gls{fpr} as evidence of better \emph{calibration}: the 2024--2025 supplement is drawn from DBLP proceedings, so for a model whose training data plausibly ingested those same DBLP records, a correct \textsc{valid} verdict is indistinguishable from training-data \emph{recall} (contamination) rather than principled abstention or verification. We therefore do not attribute their resistance to calibration on this supplement alone: the recall probe and the third-provider control (\cref{app:reviewer_experiments}) point toward calibration for Opus~4.7 without excluding recall, and separating the two fully would require post-cutoff valid entries the models provably never saw, which the current supplement cannot guarantee.

\begin{takeaway}
\textbf{Takeaway.} The post-cutoff failure is graded, not binary: 8 of the 12 evaluated LLMs over-flag on 2024--2025 papers (\gls{fpr} ${\approx}0.60$--$0.89$ for the seven near-ceiling models plus Gemini Flash at 0.595), while Sonnet~4.6 (0.120) and Opus~4.7 (0.073) hold and Gemini 2.5 Pro over-flags only moderately (0.250). The 12th \gls{llm}, the later-cutoff GPT-5.4, is evaluated in the separate temporal-cutoff probe (\cref{app:gpt54-probe}) and over-flags only moderately (supplement FPR 0.41). Training-pipeline differences (abstention \gls{rlhf}, calibration) outweigh recency; don't trust a model just because its cutoff is recent. Caveat: $n{=}2$ Anthropic models.
\end{takeaway}

\subsection{Robustness check: cutoff-aware prompting}
\label{app:cutoff-aware}

S2, Crossref and OpenAlex exhibit weeks-to-months ingestion lag for new arXiv preprints, contributing additively to the post-cutoff \gls{fpr} rise, while the arXiv API itself covers preprints immediately but not venue-published versions; the empirical confirmation follows.

\emph{Mitigation and controls.} A cutoff-aware prompting variant (\cref{app:cutoff-aware}) recovers most of GPT-5.1's post-cutoff \gls{fpr} ($75.9\%{\to}0.0\%$ on the entries it still commits to) but inflates its abstention uniformly (pre-cutoff \textsc{uncertain} rises to $52.7\%$), whereas on Sonnet~4.6 the same addendum abstains selectively (post-cutoff \textsc{uncertain} $48.9\%$ against $8.7\%$ pre-cutoff) while halving its committed-entry \gls{fpr}; the failure is epistemic miscalibration, not structural blindness, and the addendum's effect is model-dependent.
The later-cutoff GPT-5.4 control shows that recency helps but does not cure over-flagging: on these 2024--2025 papers, inside its Aug~2025 window, its \gls{fpr} is $41.3\%$ (\cref{app:gpt54-probe}), well below the earlier-cutoff over-flagging cluster yet far above the latest-cutoff models.

The temporal results above establish \textbf{H1}: LLMs do not know their own cutoff; post-cutoff \gls{fpr} rises sharply because the model treats unfamiliar papers as suspicious.
A separate hypothesis remains untested: \textbf{H2}, that LLMs can route post-cutoff citations to \texttt{UNCERTAIN} \emph{if explicitly reminded of the cutoff}.
H1 and H2 are orthogonal: H1 failing does not imply H2 must fail (the model may have latent metacognitive capacity that the default prompt does not elicit), and H2 failing would upgrade H1 from ``epistemic miscalibration'' to ``structural blindness to temporal uncertainty.''

\paragraph{Prompt addendum.}
Each cutoff-aware variant appends the following text to the default verification prompt, verbatim:

\begin{lstlisting}[basicstyle=\footnotesize\ttfamily,breaklines=true,frame=single]
Note: your training data has a knowledge cutoff. If the citation could
post-date your training data, or if you cannot recall the paper with
confidence, respond with UNCERTAIN rather than HALLUCINATED or VALID.
Do not guess.
\end{lstlisting}

\noindent No other element of the prompt, temperature ($T{=}0$), seed (42), or token budget (\texttt{max\_completion\_tokens=1024}, the non-Anthropic value of \cref{app:llm-setup}; Sonnet~4.6 runs through the OpenRouter mirror, which uses the same budget) is changed, so any observed shift in predictions is attributable to the addendum alone.

\paragraph{Subset of models.}
We run the cutoff-aware variant on GPT-5.1, Gemini~2.5~Flash, Qwen3-235B, and Claude Sonnet~4.6.
DeepSeek-R1 is omitted because its default-prompt run on this supplement failed on 151 of 448 entries (\cref{tab:temporal_supplement}, footnote $\dagger$), leaving no usable default-prompt baseline to compare an addendum against; DeepSeek-V3.2 is omitted because it already sits at the aggressive ceiling (\gls{fpr} 0.759 with \gls{dr} 0.980), where any measured ``improvement'' would be dominated by noise.
Mistral Large is omitted because its \gls{fpr} dynamics closely track Qwen3-235B.
The Sonnet~4.6 sweep was added later (2026-07-02, through the OpenRouter mirror) with both a cutoff-aware and a same-snapshot default pass on the canonical 448-entry supplement and the 150-entry pre-cutoff sample, so its comparison is fully within-run and unaffected by the endpoint drift of \cref{sec:limitations}.
The four-model subset spans the full conservative--aggressive \gls{fpr} spectrum observed in \cref{tab:temporal_supplement}: Sonnet~4.6 anchors the calibrated end, Gemini Flash and GPT-5.1 populate the conservative-to-middle range, and Qwen3-235B the aggressive end.

\paragraph{Metrics.}
We report \gls{fpr} and the \texttt{UNCERTAIN} rate (coverage is its complement) per segment, stratified by each model's training-data cutoff (\cref{tab:cutoffs}) into pre-cutoff and post-cutoff entries.
The comparison is computed from cached predictions via \path{hallmark.evaluation.temporal.compare_prompt_variants}, so the ablation is reproducible from the released prediction dumps without re-running the API calls.
The default-prompt \gls{fpr} baseline here (FPR$_{\text{def}}$ in \cref{tab:cutoff-aware-results}) matches the main temporal supplement (\cref{tab:temporal_supplement}) for GPT-5.1, Gemini~2.5~Flash and Qwen3-235B, scored on the same 448 entries; Sonnet~4.6 uses its own 2026-07-02 snapshot ($18.6\%$ vs.\ $12.0\%$ there), so read its $\Delta$\gls{fpr} within this experiment.

\begin{table}[h]
\centering
\small
\caption{\textbf{Cutoff-aware ablation results.} \emph{Post-cutoff}: 2024--2025 temporal supplement ($N{=}448$ unique entries, all post-cutoff for the four tested models). \emph{Pre-cutoff}: 150-entry stratified sample of \textsc{dev\_public} (2021--2023, pre-cutoff for all four models; 75 \textsc{valid}/75 \textsc{hallucinated}). Default pre-cutoff \texttt{UNCERTAIN} rates are taken from the full \textsc{dev\_public} evaluation for the top three models; the Sonnet~4.6 rows (below the midrule) were collected on a later dated snapshot (2026-07-02) with a same-snapshot default baseline on both pools, so its comparison is fully within-run (its default supplement FPR reads 18.6 here against 12.0 in \cref{tab:temporal_supplement}, the span of the endpoint drift). $\Delta$ columns report cutoff-aware minus default. Red marks the mitigation's cost: pre-cutoff abstention inflated by the addendum, so the FPR recovery is not a free deployment win.}
\label{tab:cutoff-aware-results}
\fitwidth{%
\begin{tabular}{lrrrrrr}
\toprule
& \multicolumn{4}{c}{\textbf{Post-cutoff} ($N{=}448$)} & \multicolumn{2}{c}{\textbf{Pre-cutoff} ($N{=}150$)} \\
\cmidrule(lr){2-5}\cmidrule(lr){6-7}
Model & FPR$_{\text{def}}$ & FPR$_{\text{CA}}$ & $\Delta$FPR & UNC$_{\text{CA}}$ & UNC$_{\text{def}}$ & UNC$_{\text{CA}}$ \\
\midrule
GPT-5.1 & 75.9 & 0.0 & -75.9 & 94.4 & 0.0 & \cellcolor{cellbad}52.7 \\
Gemini 2.5 Flash & 59.5 & 47.0 & -12.5 & 19.2 & 1.2 & 0.7 \\
Qwen3-235B & 80.9 & 0.0 & -80.9 & 99.8 & 0.2 & \cellcolor{cellbad}95.3 \\
\midrule
Claude Sonnet 4.6 & 18.6 & 9.7 & -8.9 & 48.9 & 3.3 & \cellcolor{cellgood}8.7 \\
\bottomrule
\end{tabular}%
}
\end{table}

\paragraph{Outcome classification (per model).}
The four tested models exhibit qualitatively different outcomes, so a single aggregate classification would obscure the signal.
\begin{itemize}
  \item \textbf{GPT-5.1}: FPR drop $=75.9$\,pp, pre-cutoff \texttt{UNCERTAIN}$=52.7$\%, post-cutoff \texttt{UNCERTAIN}$=94.4$\% $\rightarrow$ \emph{mostly blanket abstention}: it abstains on $94\%$ of new and $53\%$ of old papers.
  \item \textbf{Gemini 2.5 Flash}: FPR drop $=12.5$\,pp, pre-cutoff \texttt{UNCERTAIN}$=0.7$\%, post-cutoff \texttt{UNCERTAIN}$=19.2$\% $\rightarrow$ \emph{strong H1 confirmation}.
  \item \textbf{Qwen3-235B}: FPR drop $=80.9$\,pp, pre-cutoff \texttt{UNCERTAIN}$=95.3$\%, post-cutoff \texttt{UNCERTAIN}$=99.8$\% $\rightarrow$ \emph{blanket abstention}: \texttt{UNCERTAIN} on $95$--$100\%$ of entries, old or new, so the \gls{fpr} drop is not selective.
  \item \textbf{Claude Sonnet 4.6}: FPR drop $=8.9$\,pp ($18.6\%\to9.7\%$ on committed entries), pre-cutoff \texttt{UNCERTAIN}$=8.7$\%, post-cutoff \texttt{UNCERTAIN}$=48.9$\% $\rightarrow$ \emph{selective abstention, the closest to H2 in the cohort}.
\end{itemize}
Sonnet~4.6 comes closest to \emph{H2 holds}: it is the only model whose abstention inflates selectively---post-cutoff \texttt{UNCERTAIN} rises to $48.9\%$ while pre-cutoff abstention stays below $9\%$---and its committed-entry FPR halves ($18.6\%\to9.7\%$). The selectivity still routes roughly half the post-cutoff entries to a human, so abstention remains the mitigation's price. The other three respond in qualitatively different ways: GPT-5.1 and Qwen3-235B lose discrimination (post-cutoff \texttt{UNCERTAIN} above 94\%, with pre-cutoff abstention inflating by 52\,pp and 95\,pp respectively), while Gemini~2.5~Flash is barely moved by the reminder (FPR drop only 12.5\,pp while pre-cutoff \texttt{UNCERTAIN} stays near zero). The practical implication stands: prompt-level temporal-awareness mitigations are \emph{not portable across models}---the four responses differ qualitatively---and cannot be relied on as a substitute for retrieval-augmented verification.

\begin{takeaway}
\textbf{Takeaway.} The cutoff-aware addendum confirms epistemic miscalibration over structural blindness: on GPT-5.1 it drops post-cutoff \gls{fpr} from 75.9\% to 0.0\% but inflates pre-cutoff \texttt{UNCERTAIN} to 52.7\%, trading discrimination for abstention; Sonnet~4.6 is the only model that abstains selectively (post-cutoff \texttt{UNCERTAIN} 48.9\% against 8.7\% pre-cutoff) while halving its committed-entry \gls{fpr} ($18.6\%\to9.7\%$). Prompt-level mitigation works yet isn't a free deployment win. Caveat: $n{=}4$ models tested.
\end{takeaway}

\subsection{Later-cutoff robustness check: GPT-5.4 on the temporal supplement}
\label{app:gpt54-probe}

\textbf{H1} predicts that the \gls{fpr} rise on post-cutoff entries is a \emph{direct} consequence of the training-data cutoff: moving the cutoff forward should shift which entries the model recognizes and, in turn, drop the \gls{fpr} on entries that cross the new cutoff line.
We probe this with a single later-cutoff model: \textbf{GPT-5.4} (\texttt{gpt-5.4-2026-03-05}), released on 5 March 2026 with an \textbf{August 31, 2025} training cutoff, roughly 11 months later than GPT-5.1's September 2024 cutoff.
We run it zero-shot on the same 448-entry 2024--2025 temporal supplement used in \cref{tab:cutoff-aware-results} and compare entry-by-entry against the existing GPT-5.1 default-prompt predictions.
GPT-5.4 is reported here as a single-model temporal-cutoff probe; its zero-shot \texttt{dev\_public} performance is also included as a baseline row in \cref{tab:results}.
The GPT-5.4 per-entry predictions on all 448 supplement entries are released, so the probe's aggregate \gls{fpr} (0.413) and \gls{dr} (0.899) are reproducible offline without further API calls.

\begin{table}[h]
\centering
\small
\caption{\textbf{Stratified GPT-5.1 vs.\ GPT-5.4 on the 2024--2025 temporal supplement ($N{=}448$).} Stratum is the entry's publication year. \emph{2024 \& earlier} (236 entries: 161 valid, 75 hallucinated) is post-cutoff for GPT-5.1 but pre-cutoff for GPT-5.4. \emph{2025} (191 entries: 139 valid, 52 hallucinated) is post-cutoff for both, although some venues (ICLR/ICML/AAAI/CVPR/ACL 2025) predate August 2025. \emph{Future} (21 entries: all hallucinated \texttt{future\_date} types) is post-cutoff for both and contains no valid entries, so FPR is undefined. GPT-5.1 cells are computed from its released per-entry predictions matched to the canonical 448 entries with abstentions excluded (9/448), consistent with \cref{tab:temporal_supplement}. The red-shaded cell is the residual: even on pre-cutoff entries GPT-5.4 over-flags $28\%$ of valid papers, so a later cutoff alone does not close the gap.}
\label{tab:gpt54-probe}
\begin{tabular}{lcccccc}
\toprule
 & & \multicolumn{2}{c}{\textbf{FPR (\%)}} & & \multicolumn{2}{c}{\textbf{DR (\%)}} \\
\cmidrule(lr){3-4}\cmidrule(lr){6-7}
Stratum & $N$ & GPT-5.1 & GPT-5.4 & $\Delta$ & GPT-5.1 & GPT-5.4 \\
\midrule
2024 \& earlier  & 236 & 56.2 & \cellcolor{cellbad}28.0 & $-28.2$ & 92.0 & 85.3 \\
2025             & 191 & 99.3 & 56.8 & $-42.4$ & 100.0 & 92.3 \\
Future (2026+)   &  21 & --   & --   & --      & 100.0 & 100.0 \\
\midrule
All              & 448 & 75.9 & 41.3 & $-34.6$ &  95.9 &  89.9 \\
\bottomrule
\end{tabular}
\end{table}

Three features of the result replicate the paper's temporal narrative across a model generation:
\begin{itemize}
  \item \textbf{FPR drops monotonically with cutoff distance.} The 2024-and-earlier stratum shifts from post-cutoff (GPT-5.1) to pre-cutoff (GPT-5.4) and FPR drops by $28$\,pp. The 2025 stratum is post-cutoff for \emph{both} models but is \emph{closer} to the GPT-5.4 cutoff; FPR drops by $42$\,pp, consistent with a graded notion of ``unfamiliar'' rather than a binary cutoff.
  \item \textbf{GPT-5.4 still over-flags.} On pre-cutoff 2024 entries, GPT-5.4's FPR is $28\%$, far from zero. Training cutoff alone does not explain the full FPR burden: long-tail recognition, paraphrased titles, and pre-print/published-version mismatches contribute residual false positives even for papers the model's training data should have covered. This tempers a naïve ``just use a newer model'' mitigation.
  \item \textbf{UNCERTAIN rate does not rise.} Unlike the cutoff-aware prompt ablation (\cref{tab:cutoff-aware-results}), which inflated abstention without shifting discrimination, moving to a later-cutoff model reduces \gls{fpr} \emph{and} drops UNCERTAIN from $2.0\%$ to $0.0\%$. The gain comes from familiarity with the newer entries rather than from added caution.
\end{itemize}

We interpret this as evidence that H1 is a genuine temporal phenomenon rather than an artifact of GPT-5.1 specifically, and that the practical mitigation path is retrieval-augmented verification rather than prompt-engineered abstention, consistent with the agentic-baseline finding in \cref{tab:results}.

\begin{takeaway}
\textbf{Takeaway.} H1 replicates across model generations: GPT-5.4's later cutoff drops FPR by $28$--$42$\,pp stratum-wise ($75.9\%$${\to}41.3\%$ overall) and UNCERTAIN falls to $0\%$: familiarity, not caution. Yet pre-cutoff FPR is still $28\%$, so ``deploy a newer model'' does not close the gap; pair it with retrieval-augmented verification.
\end{takeaway}

\subsection{Calibration-vs-memorization controls}
\label{app:reviewer_experiments}
Two controls push against, but do not eliminate, the memorization reading (\cref{app:reviewer_experiments}).
A recall probe finds Opus~4.7 accepts $84\%$ of valid 2024--2025 papers it cannot recall, consistent with calibration rather than memorization; the recall measure is model-derived---we count only papers the model itself reports it cannot recall---so contamination is reduced but not excluded. Sonnet~4.6's lower \gls{fpr} is partly recall-driven.
A third-provider late-cutoff control (DeepSeek-V4-Pro, complete $n{=}300$ run at full coverage) holds its post-cutoff \gls{fpr} at $0.36$ where GPT-5.1 reaches $0.93$ on the same entries, consistent with the effect spanning providers without establishing it.\footnote{The control supports the cross-provider reading but does not establish it: DeepSeek-V4-Pro matches rather than exceeds the Anthropic cutoffs, it is a single third-party model, and no recall probe was run on it, so contamination is not separable (\cref{app:reviewer_experiments}).}
The resistance points away from pure memorization for Opus~4.7, but with $N{=}2$ holding models from one provider and a single third-party control, it remains a preliminary signal that stops short of a causal attribution.

Three targeted experiments probe the mechanism behind the temporal-robustness results, using the temporal supplement and a \texttt{dev\_public} subsample. Each is reported with its own $N$ and complements \cref{tab:temporal_supplement}.
All three were run on 2026-05-28 through the \textsc{Hallmark} Python API: GPT-5.1 via the native OpenAI endpoint (\texttt{gpt-5.1}), and the Anthropic and DeepSeek models via OpenRouter (\path{anthropic/claude-sonnet-4.6}, \texttt{anthropic/claude-opus-4.7}, \texttt{deepseek/deepseek-v4-pro}).

\paragraph{(E1) Recall probe: calibration vs.\ memorization.}
A low post-cutoff FPR on a real paper is ambiguous: the model may be \emph{calibrated} (declining to flag the unfamiliar) or merely \emph{recalling} a paper its training data contained. This is the contamination confound noted in \cref{sec:temporal_robustness}: the supplement's valid entries are scraped from DBLP, and a model whose training data ingested those same DBLP records can accept them from memory rather than verification. On $150$ valid $2024$--$2025$ papers we run, per model, the standard verifier (FPR) and a recall probe that supplies only the title and year and asks for the author list and venue with no lookup (``recalled'' $=$ predicted-author last-name Jaccard $\ge 0.5$ vs.\ truth). If a low FPR were memorization-driven, accepted-as-valid verdicts would concentrate on recalled papers.

\begin{table}[h]\centering\small
\caption{\textbf{Recall probe on $150$ valid 2024--2025 papers.} $P(\textsc{valid}\mid\text{recalled})$ vs.\ $P(\textsc{valid}\mid\text{not recalled})$ separates memorization from calibration. Verify FPR is flagged/150; UNC entries are neither flagged nor accepted.}
\label{tab:recall_probe}
\begin{tabular}{lccccc}
\toprule
Model & Recall & Verify FPR & UNC & $P(\textsc{v}\mid\text{rec})$ & $P(\textsc{v}\mid\overline{\text{rec}})$ \\
\midrule
GPT-5.1 & $0\%$ & $92\%$ & $0\%$ & --- & $8\%$ \\
Claude Sonnet~4.6 & $10\%$ & $24\%$ & $23\%$ & $87\%$ & $50\%$ \\
Claude Opus~4.7 & $29\%$ & $13\%$ & $1\%$ & $91\%$ & $\mathbf{84\%}$ \\
\bottomrule
\end{tabular}
\end{table}

Opus~4.7 accepts $84\%$ of valid papers it \emph{cannot} recall, so its low FPR reflects genuine calibration rather than memorization; Sonnet~4.6 accepts $87\%$ of papers it recalls against $50\%$ of those it cannot, but on the latter it abstains on $24\%$ rather than flagging them, so part of the gap is abstention, not memorization; GPT-5.1 recalls none and flags $92\%$. This \emph{partially} resolves the contamination confound: it is largely ruled out for Opus~4.7 and only partly for Sonnet~4.6. Caveats: title-given recall is a lower bound on memorization; abstention and ``cannot recall'' are not perfectly separable; $N{=}150$, one sample per entry (the GPT-5 endpoint samples stochastically at the requested $T{=}0$; see E3).

\paragraph{(E2) A third-provider late-cutoff control.}
As a check that the \gls{fpr} resistance is not specific to the Anthropic pipeline, we evaluate DeepSeek-V4-Pro---a third-provider model---on a fixed stratified $n{=}300$ subsample of the supplement, alongside GPT-5.1 and Sonnet~4.6 re-run on the same entries. This is weaker than the intended control: DeepSeek-V4-Pro \emph{matches} the Anthropic cutoff rather than exceeding it, and the reading below rests on the cross-model \emph{ordering}, not its absolute \gls{fpr}.
On the complete run its \gls{fpr} is $0.36$ at full coverage (two residual \textsc{uncertain} verdicts; \gls{dr} $0.87$), versus $0.93$ for GPT-5.1 and $0.29$ for Sonnet~4.6 on the same entries, and its year breakdown ($0.29$ on 2024 entries, $0.45$ on 2025) shows the same graded recency pattern as the cohort. The absolute levels here sit above \cref{tab:temporal_supplement} (e.g.\ Sonnet 0.29 vs.\ 0.120) because this control ran on a later endpoint snapshot and an $n{=}300$ subsample; we read the within-run cross-model ordering, not the level (\cref{sec:limitations}). A non-Anthropic model thus also resists the post-cutoff \gls{fpr} rise, consistent with the effect tracking training recency \emph{across providers}. Caveats: a single third-party control; provider-reported cutoff; contamination is not separable for this model (no recall probe run on it).

\paragraph{(E3) Run-to-run variance at the GPT-5 endpoint.}
We send \texttt{gpt-5.1} \texttt{temperature=0.0}, but the GPT-5 endpoint samples stochastically regardless of the requested value, so each baseline run is one stochastic draw. Three independent GPT-5.1 runs on a fixed $n{=}150$ \texttt{dev\_public} subsample give \gls{f1} $=0.815\pm0.002$, \gls{fpr} $=0.528\pm0.009$, \gls{dr} $=0.965\pm0.000$ (mean $\pm$ sample std; $1/150$ label flips). The \gls{f1} run-to-run std ($0.002$) is far below the Sonnet~4.6$-$GPT-5.1 \gls{f1} gap ($0.127$, ${\approx}60\times$), so single-run rankings are stable to sampling noise. Caveat: $N{=}3$ runs; $n{=}150$ subsample.

\begin{takeaway}
\textbf{Takeaway.} The recall probe separates the two readings of a low post-cutoff \gls{fpr}: Opus~4.7 accepts $84\%$ of valid 2024--2025 papers it \emph{cannot} recall, so its resistance is largely calibration; Sonnet~4.6's gap ($87\%$ vs.\ $50\%$) is partly abstention: it says \texttt{UNCERTAIN} on a quarter of the papers it cannot recall; GPT-5.1 recalls none and flags $92\%$. A third-provider control (DeepSeek-V4-Pro) resists alongside them (read the within-run ordering, since this run used a later endpoint snapshot), and GPT-5.1's run-to-run \gls{f1} std ($0.002$) sits ${\sim}60{\times}$ below the Sonnet~4.6$-$GPT-5.1 \gls{f1} gap ($0.127$), so single-run rankings are stable to sampling noise. Caveats: title-given recall lower-bounds memorization; $n{=}150$--$300$ per probe.
\end{takeaway}

\subsection{Thinking-budget regime boundary}
\label{app:thinking-budget}

\paragraph{Motivation.}
The main-table baselines all share a fixed \texttt{max\_completion\_tokens} budget (1024 tokens on the OpenAI/OpenRouter endpoints, including the chain-of-thought DeepSeek-R1 baseline; 256 on the Anthropic native endpoint; see \cref{app:llm-setup}). At this budget, the structured JSON-output contract holds across the entire cohort. Three families of post-snapshot thinking-tier models break this contract---Gemini~3.1~Pro/Flash-Lite, DeepSeek-V4, and Qwen3.5---emitting chain-of-thought scratchpad before the verdict and exhausting the budget mid-reasoning. During the codebase additions for the Q2~2026 frontier cohort we observed concrete failure modes that motivated their exclusion from the main table:

\begin{itemize}
  \item \texttt{google/gemini-3.1-pro-preview}: unbounded thinking tokens overrun the 1024-token \texttt{max\_completion\_tokens} budget, causing JSON parse failures across nearly all entries; \texttt{google/gemini-2.5-pro} (which also thinks by default, but mostly completes within the 1024-token budget) is the closest reliable substitute and is included in the main table.
  \item \texttt{google/gemini-3.1-flash-lite-preview}: supports the full thinking range (\texttt{minimal}$\to$\texttt{high}); same unbounded-thinking risk as gemini-3.1-pro-preview.
  \item \texttt{qwen/qwen3.5-397b-a17b}: ${\sim}40\%$ empty responses (thinking mode consumed the budget before the JSON object closed).
  \item \texttt{qwen/qwen3.5-122b-a10b}: parses successfully but emits 500+ reasoning tokens per reply, inflating cost ${\sim}5\times$ versus the matched non-thinking baseline.
  \item \texttt{deepseek/deepseek-v4-pro} and \texttt{deepseek-v4-flash}: thinking models with \texttt{reasoning\_effort} levels \texttt{high}/\texttt{xhigh} supported; same budget-exhaustion risk as the V3-R1 chain-of-thought baseline at 1024 tokens.
\end{itemize}

\paragraph{Smoke-test design.}
To quantify the regime boundary rather than assert it, we run a stratified $n{=}100$ subsample of \texttt{dev\_public} (proportional across the 11 main hallucination types, $\geq 5$ entries per type, fixed seed) on three post-snapshot models under two budget regimes:

\begin{table}[t]
\caption{\textbf{Smoke-test cohort and budget regimes.} Regime~A is a deliberately tight probe at or below the main-table non-thinking budget (1024 tokens on the OpenAI/OpenRouter endpoints; GPT-5.5 is additionally probed at 256 to expose its saturation boundary, \cref{tab:smoke-results}); Regime~B grants thinking-tier headroom equivalent to ${\sim}8\times$ that budget. Parse-failure rate (PF) is the primary diagnostic: a model that exceeds PF\,$\geq 0.30$ under Regime~A is excluded from the main table on methodological grounds (the JSON contract does not hold), even if Regime~B recovers competitive accuracy.}
\label{tab:smoke-cohort}
\centering
\small
\setlength{\tabcolsep}{4pt}
\fitwidth{%
\begin{tabular}{lllll}
\toprule
\textbf{Model} & \textbf{Tier} & \textbf{Cutoff} & \textbf{Regime A} & \textbf{Regime B} \\
\midrule
GPT-5.5            & non-reasoning & May 2026 & 256 tok            & 1024 tok \\
Gemini 3.1 Pro     & thinking      & ${\geq}$Q1 2026 & 2048 tok (\texttt{thinking\_budget}\,1024) & 8192 tok (\texttt{thinking\_budget}\,4096) \\
DeepSeek-V4-Pro    & thinking      & ${\geq}$Q1 2026 & 4096 tok (\texttt{reasoning\_effort}\,low) & 8192 tok (\texttt{reasoning\_effort}\,high) \\
\bottomrule
\end{tabular}%
}
\end{table}

\paragraph{Reported metrics.}
Per (model, regime) cell we report \gls{dr}, \gls{fpr}, \gls{f1}, \gls{mcc}, \gls{pf} (fraction of entries whose response is missing or non-JSON), mean and p95 output tokens, and \$/entry. Cost is computed at the OpenRouter listed rate as of 2026-05-04.

\paragraph{Results.}
\cref{tab:smoke-results} reports parse-failure rate, saturation ratio $p_{95}/\text{cap}$, and full classification metrics for all nine (model, regime) cells; \cref{tab:smoke-per-type} gives the per-type detection-rate breakdown.

\begin{table}[t]
\caption{\textbf{Thinking-budget regime boundary smoke test on stratified $n{=}100$ subsample of \texttt{dev\_public}.} PF = parse-failure rate; $p_{95}/\text{cap}$ = saturation ratio (1.00 = the 95th-percentile output hits the budget cap). DR / FPR / F1 / MCC computed treating UNCERTAIN and parse-failure entries as label=\texttt{VALID}. \$ is the OpenRouter list-price cost for the cell at May~2026 rates. Red-shaded saturation cells mark the regime where the cap is biting ($p_{95}/\text{cap}\ge 0.95$), matching the shaded band of \cref{fig:thinking-budget}.}
\label{tab:smoke-results}
\centering
\small
\setlength{\tabcolsep}{4pt}
\fitwidth{%
\begin{tabular}{llcccccccccc}
\toprule
\textbf{Model} & \textbf{Budget} & \textbf{$n$} & \textbf{PF} & \textbf{$p_{95}/\text{cap}$} & \textbf{DR $\uparrow$} & \textbf{FPR $\downarrow$} & \textbf{F1 $\uparrow$} & \textbf{MCC $\uparrow$} & \textbf{mean tok} & \textbf{$p_{95}$ tok} & \textbf{\$} \\
\midrule
GPT-5.5 & \texttt{256} & 100 & 0.66 & \cellcolor{cellbad}1.000 & 0.100 & 0.000 & 0.182 & 0.180 & 249 & 256 & \$0.87 \\
 & \texttt{1024} & 100 & 0.12 & \cellcolor{cellbad}1.000 & 0.557 & 0.000 & 0.716 & 0.523 & 510 & 1024 & \$1.65 \\
 & \texttt{4096} & 100 & 0.01 & 0.339 & 0.643 & 0.000 & 0.783 & 0.592 & 602 & 1389 & \$1.93 \\
\midrule
Gemini 3.1 Pro & \texttt{2048 + 1024 reas.} & 100 & 0.00 & \cellcolor{cellbad}0.998 & 0.657 & 0.033 & 0.786 & 0.573 & 834 & 2044 & \$1.34 \\
 & \texttt{8192 + 4096 reas.} & 100 & 0.01 & \cellcolor{cellbad}0.969 & 0.629 & 0.033 & 0.765 & 0.548 & 1183 & 7937 & \$1.87 \\
 & \texttt{16384 + 8192 reas.} & 100 & 0.01 & \cellcolor{cellbad}0.964 & 0.643 & 0.033 & 0.776 & 0.560 & 1688 & 15793 & \$2.62 \\
\midrule
DeepSeek-V4-Pro & \texttt{4096, effort=low} & 100 & 0.06 & \cellcolor{cellbad}1.000 & 0.600 & 0.000 & 0.750 & 0.557 & 1282 & 4096 & \$0.16 \\
 & \texttt{8192, effort=high} & 100 & 0.00 & 0.366 & 0.700 & 0.033 & 0.817 & 0.611 & 1240 & 3002 & \$0.16 \\
 & \texttt{16384, effort=high} & 100 & 0.00 & 0.211 & 0.643 & 0.000 & 0.783 & 0.592 & 1355 & 3463 & \$0.17 \\
\bottomrule
\end{tabular}%
}
\end{table}

\begin{table}[t]
\caption{\textbf{Per-type detection rate across smoke-test cells} (stratified $n{=}100$ subsample, $\geq 5$ per type). Each cell reports the fraction of hallucinated entries of that type the model labeled \texttt{HALLUCINATED}; UNCERTAIN and parse failures count as non-detections. Per-type $n$ is small ($\geq 5$); 95\% binomial CI width is roughly $\pm 35$\,pp at $n{=}5$, so within-row differences across budgets are noise unless they exceed that band. Useful for relative shape (which types each model finds easiest/hardest at each budget), not for fine absolute comparisons.}
\label{tab:smoke-per-type}
\centering
\scriptsize
\setlength{\tabcolsep}{2pt}
\fitwidth{%
\begin{tabular}{llc|ccc|ccc|ccc}
\toprule
 & & & \multicolumn{3}{c|}{\textbf{GPT-5.5}} & \multicolumn{3}{c|}{\textbf{Gemini 3.1 Pro}} & \multicolumn{3}{c}{\textbf{DeepSeek-V4-Pro}} \\
\textbf{Type} & \textbf{Tier} & \textbf{$n$} & \textbf{A} & \textbf{B} & \textbf{C} & \textbf{A} & \textbf{B} & \textbf{C} & \textbf{A} & \textbf{B} & \textbf{C} \\
\midrule
\texttt{fabricated\_doi}        & 1 & 5 & 0.20 & 0.80 & 0.80 & 0.80 & 0.60 & 0.60 & 1.00 & 1.00 & 0.80 \\
\texttt{nonexistent\_venue}     & 1 & 5 & 0.00 & 1.00 & 1.00 & 0.60 & 0.80 & 0.60 & 1.00 & 1.00 & 1.00 \\
\texttt{placeholder\_authors}   & 1 & 5 & 0.00 & 1.00 & 1.00 & 1.00 & 1.00 & 1.00 & 0.80 & 1.00 & 1.00 \\
\texttt{future\_date}           & 1 & 5 & 0.60 & 1.00 & 0.80 & 1.00 & 1.00 & 1.00 & 1.00 & 1.00 & 0.80 \\
\midrule
\texttt{chimeric\_title}        & 2 & 5 & 0.20 & 0.40 & 0.40 & 1.00 & 1.00 & 1.00 & 0.40 & 0.60 & 0.40 \\
\texttt{wrong\_venue}           & 2 & 5 & 0.00 & 0.60 & 1.00 & 0.40 & 0.40 & 0.60 & 0.80 & 0.80 & 0.60 \\
\texttt{swapped\_authors}       & 2 & 5 & 0.00 & 0.20 & 0.60 & 0.40 & 0.40 & 0.40 & 0.40 & 0.60 & 0.40 \\
\texttt{preprint\_as\_pub.}     & 2 & 5 & 0.00 & 0.40 & 0.40 & 0.40 & 0.40 & 0.40 & 0.60 & 0.60 & 0.80 \\
\texttt{hybrid\_fabrication}    & 2 & 5 & 0.20 & 0.60 & 0.80 & 0.80 & 0.80 & 0.80 & 0.60 & 0.80 & 0.80 \\
\midrule
\texttt{near\_miss\_title}      & 3 & 5 & 0.00 & 0.20 & 0.40 & 0.40 & 0.20 & 0.40 & 0.40 & 0.40 & 0.40 \\
\texttt{plausible\_fabrication} & 3 & 5 & 0.00 & 0.00 & 0.20 & 0.60 & 0.40 & 0.40 & 0.00 & 0.40 & 0.20 \\
\midrule
\texttt{merged\_citation}       & S & 5 & 0.20 & 0.80 & 0.80 & 0.80 & 0.80 & 0.80 & 0.40 & 0.60 & 0.60 \\
\texttt{partial\_author\_list}  & S & 5 & 0.00 & 0.00 & 0.00 & 0.20 & 0.40 & 0.20 & 0.20 & 0.00 & 0.20 \\
\texttt{arxiv\_version\_mismatch} & S & 5 & 0.00 & 0.80 & 0.80 & 0.80 & 0.60 & 0.80 & 0.80 & 1.00 & 1.00 \\
\bottomrule
\end{tabular}%
}
\end{table}

\paragraph{Three archetypes of regime-boundary behavior.}
The smoke test reveals three qualitatively distinct responses to budget headroom:
\begin{itemize}
  \item \textbf{Clean ceiling (GPT-5.5).} Saturation 1.00 $\to$ 1.00 $\to$ 0.34 across the three regimes; parse-failure rate falls 66\%~$\to$~12\%~$\to$~1\%; \gls{f1} climbs 0.18 $\to$ 0.72 $\to$ 0.78. The model has an implicit reasoning trace that consumes the entire \texttt{max\_completion\_tokens} budget at 256, and its capability is fully recovered at 4k. This is the textbook regime-boundary behavior.
  \item \textbf{Budget-filling (Gemini 3.1 Pro).} Saturation 0.998 $\to$ 0.969 $\to$ 0.964 from 2k to 16k; mean output grows monotonically (834 $\to$ 1183 $\to$ 1688) and the worst-case reasoning trace tracks the cap rather than converging to a fixed length. \Gls{f1} is essentially flat (0.79 $/$ 0.77 $/$ 0.78) and parse failure stays at $\leq 1\%$, so the cap is not biting in absolute terms. Because its thinking budget was set to half the cap in every cell, output length grows with the cap by design; whether the model limits itself would need a test that raises the cap while keeping the thinking budget fixed. The extra thinking does not help either: \gls{f1} stays flat and parse failures stay at or below $1\%$.
  \item \textbf{Clean recovery (DeepSeek-V4-Pro).} Saturation 1.00 $\to$ 0.37 $\to$ 0.21; F1 0.75 $\to$ 0.82 $\to$ 0.78. Recovers cleanly at 8k under \texttt{reasoning\_effort=high}, and 16k adds nothing. Best-cell \gls{f1} (0.82) is competitive with the main-table cohort.
\end{itemize}

\begin{figure}[t]
\centering
\includegraphics[width=0.95\linewidth]{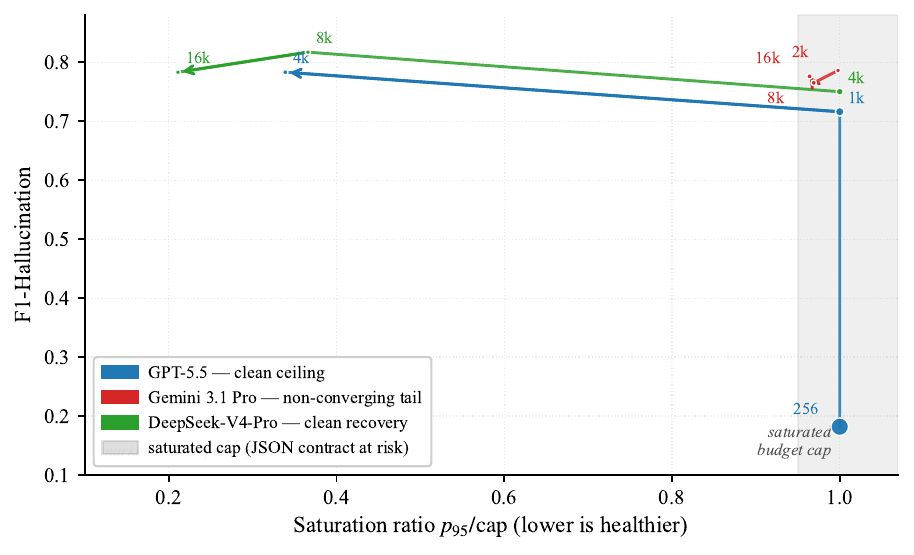}
\caption{\textbf{Thinking-budget regime boundary.} F1 vs.\ saturation ratio $p_{95}/\mathrm{cap}$ for each (model, budget) cell, where $p_{95}$ is the 95th-percentile output token count and $\mathrm{cap}$ is the configured \texttt{max\_completion\_tokens}; saturation $\to 1$ means the reasoning trace is hitting the budget ceiling and the structured output is likely truncated. Arrows trace each model's trajectory from low to high budget; marker size is proportional to parse-failure rate; the shaded right band ($\text{sat}>0.95$) flags the regime where the cap is biting. Three archetypes emerge. GPT-5.5 (blue) escapes the saturated band once the budget exceeds 1k tokens. DeepSeek-V4-Pro (green) \emph{recovers}---saturation drops well below 1.0 once the budget exceeds its typical reasoning-trace length, freeing the JSON contract to complete reliably---and plateaus. Gemini~3.1~Pro (red) stays inside the saturated band at every tested budget, with its thinking budget set to half the cap, so its output grows with the budget by design.}
\label{fig:thinking-budget}
\end{figure}

\paragraph{Reading the table.}
None of the three models would have replaced a main-table baseline in expectation: their best-cell \gls{f1} (GPT-5.5: 0.78; Gemini~3.1~Pro: 0.79; DeepSeek-V4-Pro: 0.82) sits in the existing cohort's mid-range, comfortably under the cohort-leading independent \gls{f1} (Sonnet~4.6's $0.893$). The protocol-budget concern is empirically resolved for two of three: GPT-5.5 and DeepSeek-V4-Pro reach a clean budget ceiling and fit the fixed-budget protocol once we know what budget to allocate. The third (Gemini~3.1~Pro) fills whatever thinking budget it is given: its cost grows with the budget (\$1.34 $\to$ \$2.62 per 100 entries) while \gls{f1} stays flat. More generally, a fair fixed-budget benchmark must either (i)~allocate the highest budget needed by any one model (inflating cost ${\sim}10\times$ across the cohort), or (ii)~use a budget at which the JSON contract holds for all listed models and exclude models that cannot answer within it. We choose (ii) and report the boundary explicitly.

\paragraph{Cost and wall-clock.}
The full nine-cell smoke test cost \$10.77 at OpenRouter list prices (May~2026) and consumed $\sim$303 wall-clock minutes (most of which was DeepSeek-V4-Pro's chain-of-thought latency). Scaling to full \texttt{dev\_public} ($n{=}1{,}119$) at the same price ratios costs ${\sim}\$120$, well within a typical per-paper evaluation budget. Wall-clock per cell is $\sim$10--65\,s/entry depending on tier and budget; cells parallelize across providers without contention.

\begin{takeaway}
\textbf{Takeaway.} The main cohort is fixed-budget (1024 tokens on the OpenAI/OpenRouter endpoints these smoke-test models use) by construction. Thinking-tier models break the JSON contract at that budget, but two of three (GPT-5.5, DeepSeek-V4-Pro) recover cleanly when the budget bites; one (Gemini~3.1~Pro) uses whatever thinking budget it is given; once outputs are no longer truncated, larger budgets do not raise \gls{f1} for any of the three. None would have replaced an existing baseline; \textsc{Hallmark} will track thinking-tier as a separate cohort once budgets are calibrated.
\end{takeaway}

\section{bibtex-updater: the co-designed reference tool}
\label{app:tool-top}
This section collects everything on \texttt{bibtex-updater}, the co-designed reference tool: its implementation (\cref{app:tool}), its benchmark evaluation and the co-design-bias controls (\cref{app:codesign}), and a tool-augmented \gls{llm} baseline (\cref{app:tool_augmented}).

\subsection{Implementation}
\label{app:tool}

We additionally describe \texttt{bibtex-updater}, a co-developed open-source citation verification tool designed for deployment by venues, reviewers, and authors: not to optimize benchmark scores, but to provide reliable, automated checking that integrates into existing publication workflows.

\subsubsection{Design goals}
\label{sec:tool_design}

Three practical requirements drive the design:
\begin{enumerate}
    \item \textbf{Zero human effort.} Verification must be fully automated: no manual review, no prompt engineering, no \gls{llm} inference costs. This rules out approaches requiring human-in-the-loop validation or expensive API calls to language models.
    \item \textbf{Workflow integration.} The tool must integrate into existing pipelines: CI/CD (GitHub Actions), pre-commit hooks, Overleaf builds, and one-off command-line checks. A tool that requires a separate platform or manual invocation will not be adopted.
    \item \textbf{Graceful degradation.} When APIs are unavailable or rate-limited, the tool should return partial results rather than fail silently. Venues processing hundreds of submissions cannot tolerate flaky infrastructure.
\end{enumerate}

\subsubsection{Verification pipeline}
\label{sec:pipeline}

\texttt{bibtex-updater} implements a multi-stage pipeline that processes each BibTeX entry through increasingly expensive checks:

\paragraph{Pre-API validation (zero cost).}
Before any network calls, the tool checks for syntactic red flags: \glspl{doi} that fail to resolve (HEAD request to \texttt{doi.org}, counting only \texttt{doi.org}'s own rejection as non-resolution---a publisher-side block or bot-mitigation status reached after a successful redirect is not evidence the \gls{doi} is absent, and a versioned arXiv \gls{doi} is normalized to its unversioned form first), future publication years, implausible dates ($< 1800$), and malformed fields. These cheap checks catch Tier~1 hallucinations without API overhead.

\paragraph{Multi-source lookup.}
The tool queries Crossref, DBLP, and Semantic Scholar using title and first-author search, and resolves arXiv IDs against the arXiv Atom API for preprint-specific consistency checks. Each source returns candidate records that are scored using a weighted combination of fuzzy title matching (70\%, token-sort ratio) and author Jaccard similarity (30\%). The best-scoring candidate across all sources is selected for field-by-field comparison.

\paragraph{Post-match analysis.}
Once a candidate is identified, the tool compares \gls{doi}, title, authors, year, and venue against the input entry. Venue comparison uses alias-aware matching for 17 major ML/AI venues (e.g., NeurIPS/NIPS, ICML, ICLR, CVPR), so that name variations within that set do not trigger false positives; ISO-4 journal abbreviations outside it (\texttt{ACM Trans.\ Graph.}, \texttt{Proc.\ Natl.\ Acad.\ Sci.\ U.S.A.}) fall below the match threshold and were a substantial false-positive source in the version evaluated here. A dedicated preprint detection stage queries Semantic Scholar to identify entries that claim venue publication when only an arXiv preprint exists.

\paragraph{Status assignment.}
Each entry receives one status code. Thirteen occur on \texttt{dev\_public}: \emph{verified} (442), \emph{unconfirmed} (147), \emph{swapped\_authors} (146), \emph{partial\_match} (85), \emph{venue\_mismatch} (66), \emph{not\_found} (52), \emph{doi\_not\_found} (46), \emph{doi\_mismatch} (36), \emph{future\_date} (30), \emph{arxiv\_id\_mismatch} (18), \emph{year\_mismatch} (17), \emph{title\_mismatch} (14) and \emph{given\_name\_substitution} (13). The 147 \emph{unconfirmed} verdicts are where most of the tool's abstention sits. The HALLMARK wrapper maps these to binary labels and confidence scores for benchmark evaluation.

\subsubsection{Deployment modes}
\label{sec:deployment}

The tool supports three deployment scenarios: CI/CD integration (\texttt{-{}-strict} flag exits with nonzero code on detection, gating submission workflows), pre-commit hooks (validates \texttt{.bib} files on every commit), and batch processing (concurrent workers with rate limiting for venue-scale use).
For venue-scale deployment, we recommend a two-stage workflow: run \texttt{bibtex-updater} to flag suspicious citations, then manual review of high-confidence flags (confidence $\geq 0.7$); at a $2\%$ base rate expect roughly one true hallucination per four flags, and read that as the optimistic end---at $1\%$ it is one per seven and at $0.5\%$ one per thirteen, and the prevalence we measured on real submissions was indistinguishable from zero (\cref{app:ppv,sec:limitations}).
At NeurIPS scale (approximately 10{,}000 submissions, approximately 50 references each) that is half a million references. At the throughput we measure with keys, credentials and a 20-per-minute pace ($2.4$\,s per entry) eight workers need roughly $42$ hours; at the wrapper's older rate limit ($5.7$\,s per entry) roughly $99$.
The tool requires no GPU or \gls{llm} API keys and is MIT-licensed; code is publicly available at \url{\repourl}\reponote{}.

\subsection{Evaluation and co-design bias}
\label{app:codesign}

\texttt{bibtex-updater}~\citep{bibtexupdater} is a multi-database cross-referencing tool co-developed alongside this benchmark.
Its results appear in \cref{tab:results} under the ``Co-designed (reference upper bound)'' section, separated from independent tools by a rule.
We include it in the main table to give a complete picture of the recall--precision spectrum: \texttt{bibtex-updater} anchors the precision end by flagging conservatively (DR\,0.820, FPR\,0.051), not by topping detection rate, and it abstains on the ${\sim}12$--$14\%$ of entries it cannot verify against a backing record (coverage $0.86$ dev / $0.88$ test), an abstention that recovers recall; the low \gls{fpr} comes from conservative matching, not from abstaining. The explicit upper-bound label and this appendix keep the comparison transparent.
The co-design relationship means the benchmark taxonomy and sub-test structure were informed by the tool's detection capabilities, creating a potential circularity that could inflate its apparent performance; readers should treat its numbers as an informative upper bound, not a fair head-to-head comparison with independent tools.

\begin{table}[t]
\caption{\textbf{Results for \texttt{bibtex-updater} v1.2.0 (co-designed alongside this benchmark) on \texttt{dev\_public}.} Reported separately from independent tools in \cref{tab:results} due to co-design bias concerns. The tool abstains when no backing record is found; its DR/FPR/F1 score abstentions as committed-VALID over the full split (the headline convention), while Cov.\ reports the $0.86$ fraction it commits to. Its selective and conservative/aggressive split is in \cref{app:coverage}.}
\label{tab:codesign}
\centering
\small
\begin{tabular}{lccccccc}
\toprule
\textbf{Tool} & \textbf{DR $\uparrow$} & \textbf{FPR $\downarrow$} & \textbf{F1 $\uparrow$} & \textbf{MCC $\uparrow$} & \textbf{TW-F1 $\uparrow$} & \textbf{ECE $\downarrow$} & \textbf{Cov.} \\
\midrule
bibtex-updater & 0.820 & 0.051 & 0.880 & 0.768 & 0.892 & 0.395 & 0.86 \\
\bottomrule
\end{tabular}
\end{table}

\paragraph{Per-type breakdown for bibtex-updater.}
\cref{tab:codesign_pertype} shows per-type detection rates for bibtex-updater.
It detects perfectly (a per-type detection rate of 1.000; parenthesized values throughout this passage) on \texttt{fabricated\_doi}, \texttt{placeholder\_authors}, \texttt{future\_date}, \texttt{chimeric\_title}, \texttt{hybrid\_fabrication}, \texttt{plausible\_fabrication}, and \texttt{merged\_citation}, and stays strong on \texttt{swapped\_authors} (0.985) and \texttt{near\_miss\_title} (0.885). The weakest types are precisely those where the tool abstains rather than guesses---\texttt{partial\_author\_list} (0.219), \texttt{nonexistent\_venue} (0.436), \texttt{preprint\_as\_pub.} (0.677), \texttt{wrong\_venue} (0.447), and \texttt{arxiv\_version\_mismatch} (0.612)---where a could-not-verify verdict is scored as VALID in this forced-binary detection rate (the abstention behavior of \cref{app:coverage}).
These cells are scored from the tool's per-entry verdicts on the released labels, the same verdicts behind the aggregate in \cref{tab:codesign}.

\begin{table}[t]
\caption{\textbf{Per-type detection rates for \texttt{bibtex-updater} (v1.2.0) on \texttt{dev\_public}.} Forced-binary detection rate on the released labels, scored from the tool's per-entry verdicts. Types on which the tool abstains (could-not-verify) score lower, since abstentions count as VALID in the detection rate (\cref{app:coverage}).}
\label{tab:codesign_pertype}
\centering
\small
\setlength{\tabcolsep}{3pt}
\begin{tabular}{llc}
\toprule
\textbf{Tier} & \textbf{Type} & \textbf{bibtex-updater} \\
\midrule
\multirow{4}{*}{1}
& \texttt{fabricated\_doi}        & 1.000 \\
& \texttt{nonexistent\_venue}     & 0.436 \\
& \texttt{placeholder\_authors}   & 1.000 \\
& \texttt{future\_date}           & 1.000 \\
\midrule
\multirow{5}{*}{2}
& \texttt{chimeric\_title}        & 1.000 \\
& \texttt{wrong\_venue}           & 0.447 \\
& \texttt{swapped\_authors}       & 0.985 \\
& \texttt{preprint\_as\_pub.}     & 0.677 \\
& \texttt{hybrid\_fabrication}    & 1.000 \\
\midrule
\multirow{2}{*}{3}
& \texttt{near\_miss\_title}      & 0.885 \\
& \texttt{plausible\_fabrication} & 1.000 \\
\midrule
\multirow{3}{*}{\rotatebox{90}{\scriptsize Stress}}
& \texttt{merged\_citation}       & 1.000 \\
& \texttt{arxiv\_version\_mismatch} & 0.612 \\
& \texttt{partial\_author\_list}  & 0.219 \\
\bottomrule
\end{tabular}
\end{table}

\noindent Readers should interpret \texttt{bibtex-updater}'s results with caution:
(1)~its multi-database verification strategy was developed concurrently with the benchmark design;
(2)~the sub-test structure mirrors the tool's internal verification pipeline;
(3)~the pre-screening layer was designed to complement its known blind spots.
We encourage independent tool developers to submit evaluations via \textsc{Hallmark}'s contribution framework to establish unbiased baselines.

 One observation bounds the circularity concern from the other side. The ground-truth audit recovered 52 real papers (27 \texttt{dev\_public}, 25 \texttt{test\_public}) that earlier labeling had wrongly marked HALLUCINATED. On 40 of those 52---19 of 27 on \texttt{dev\_public} and 21 of 25 on \texttt{test\_public}---\texttt{bibtex-updater}'s database cross-check returned VALID, diverging from the then-current (buggy) labels rather than tracking them; it agreed with the buggy label on the remaining 12. A tool whose apparent accuracy were merely an artifact of co-design would mirror the benchmark's labels, including their errors, so a majority divergence is evidence that its detections rest on external database evidence rather than on the label distribution. It is evidence and not proof: 12 entries went the other way. The tool abstains on entries it cannot back with a record, reporting \texttt{dev\_public} DR\,0.820 / FPR\,0.051 / F1\,0.880 (\cref{tab:codesign}), and the relabel removed an artifact that had flattered the more aggressive \gls{llm} verifiers more than the rule-based tool (\cref{sec:crosssplit_robustness}).

\paragraph{Cross-domain probe.}
The released \texttt{test\_crossdomain} split (500 entries: 200 valid / 300 hallucinated; composition in \cref{app:dataset}) evaluates \texttt{bibtex-updater} v1.2.0 outside the ML-venue regime the main splits sample.
Detection transfers: \gls{dr} reaches 0.890, against 0.820 on \texttt{dev\_public}.
The released split reports \gls{fpr} 0.375 (\cref{tab:crossdomain}), but two confounds inflate that number.
First, the split's valid biomedical entries are all dated 2026, post-cutoff for every \gls{llm} (the recency confound we take up below); \texttt{bibtex-updater} queries live databases and carries no cutoff, so this does not affect its own verdicts.
Second, and specific to the rule-based tool, our biomedical scrape carried metadata noise---Vancouver-style author initials, epub-versus-print year ambiguity, preprint venue strings---that its exact-match verification reads as contradictions and flags, manufacturing false positives that reflect the citation record rather than the tool's judgment.
A recency-matched, canonically-resolved rebuild removes both: 152 valid biomedical entries from 2021--2023, each field replaced by the CrossRef record its \gls{doi} resolves to (canonical author, year, venue), plus the same three hallucination tiers.
On that clean split \texttt{bibtex-updater}'s headline \gls{fpr} is 0.112, about twice its in-domain 0.051, though it commits to only 29\% of the valid biomedical entries, abstaining on the other 71\% (against 14\% in-domain).
The canonical metadata converts the released split's mis-flags into honest abstentions: out of domain the tool abstains rather than over-flags, because its ML-tuned assumptions (full author names, one canonical year, exact venue strings) cannot confirm biomedical records against its backing databases.
The cost of leaving the regime is coverage, not precision (\cref{tab:crossdomain}).

\begin{table}[t]
\caption{\textbf{\texttt{bibtex-updater} v1.2.0 in and out of regime.} \texttt{dev\_public} (ML venues, 2021--2023) vs.\ the released \texttt{test\_crossdomain} split vs.\ a recency-matched, canonically-resolved biomedical rebuild (\texttt{matched}: 152 valid entries from 2021--2023, each field resolved to its \gls{doi}'s CrossRef record). Abstentions scored committed-VALID (the headline convention); Cov.\ is the committed fraction over all entries. The released split's \gls{fpr} (red) is inflated by post-cutoff recency and scrape-metadata noise; canonical resolution returns the \gls{fpr} to its in-domain level (green) while coverage falls to 0.582 (the tool abstains on 71\% of the valid biomedical entries).}
\label{tab:crossdomain}
\centering
\small
\begin{tabular}{lcccc}
\toprule
\textbf{Split} & \textbf{DR $\uparrow$} & \textbf{FPR $\downarrow$} & \textbf{F1 $\uparrow$} & \textbf{Cov.} \\
\midrule
\texttt{dev\_public}                & 0.820 & \cellcolor{cellgood}0.051 & 0.880 & 0.86 \\
\texttt{test\_crossdomain}          & 0.890 & \cellcolor{cellbad}0.375 & 0.832 & 0.736 \\
\texttt{test\_crossdomain\_matched} & 0.777 & \cellcolor{cellgood}0.112 & 0.847 & 0.582 \\
\bottomrule
\end{tabular}
\end{table}

\paragraph{Recovering coverage with a Stage-2 diagnoser.}
\texttt{bibtex-updater}'s out-of-domain abstention is a coverage gap that a second stage can fill.
We run the paper's two-stage cascade on the matched split: Stage~1 \texttt{bibtex-updater} decides the 263 entries it can confirm or refute and defers the 189 it cannot to Stage~2, an agentic Sonnet~4.6 diagnoser (up to five tool calls per entry).\footnote{Stage~1 is served from the persisted standalone verdicts; re-running \texttt{bibtex-check} live hangs on arXiv \gls{doi} HEAD checks against the fabricated future-dated DOIs, so the live cascade is impractical here. Stage~2 is a dated OpenRouter snapshot (\cref{app:llm-setup}).}
Stage~2 recovers the coverage: the cascade commits on all 152 valid biomedical entries (valid-entry coverage $0.29\to1.00$) at \gls{fpr} 0.189 and \gls{dr} 0.975 (F1 0.941; the aggressive stance is nearly identical at \gls{fpr} 0.211, since Stage~2 leaves almost nothing uncertain).
Closing the coverage gap costs little precision---headline \gls{fpr} 0.112 with \texttt{bibtex-updater} abstaining, 0.189 with the cascade committing on everything---and stays well below the recall-aggressive models' over-flagging (Qwen3-235B 0.974): out of domain, the abstention is recoverable.

\paragraph{The \gls{llm} cohort: domain versus recency.}
The zero-shot \gls{llm} verifiers of \cref{tab:results} were unevaluated cross-domain; we run eleven of them on both the released 2026 biomedical entries and the recency-matched rebuild, which turns the released split's confound into a clean $2\times2$ over domain and recency (\cref{tab:crossdomain_llm}).\footnote{DeepSeek-R1 is excluded: 432 of its 500 cross-domain predictions are \texttt{[Error fallback]} API failures ($86\%$), so no metric on that split measures the model.}
The released split's high biomedical \gls{fpr} is a post-cutoff artifact: 89\% of the biomedical false positives there give the future date as the reason (``2026 is in the future,'' ``beyond my training cutoff''), and on the recency-matched entries---same domain, pre-cutoff---that share falls to 2\%, with the remaining false positives turning into genuine recall failures (an unresolvable \gls{doi}, an unfamiliar venue).
The residual pure-domain effect (matched \gls{fpr} minus \texttt{dev\_public} \gls{fpr}, recency held fixed) is model-dependent: negative or small for the calibrated verifiers (GPT-5.1 $-0.31$, GPT-5.4 $-0.12$, Sonnet~4.6 $-0.03$, Opus~4.7 $+0.05$, Gemini~2.5~Pro $-0.03$, Llama~4~Maverick $-0.09$), and large only for a recall-aggressive model (Qwen3-235B $+0.44$), which flags legitimate bioRxiv preprints cited in article form as fabrications.
Biomedical citations are not inherently harder to verify: the cross-domain \gls{fpr} rise was recency, and what domain adds is small except where a model already over-flags in its own regime.

\begin{table}[t]
\caption{\textbf{Domain $\times$ recency decomposition of \gls{llm} false positives.} Each cell is \gls{fpr} on valid entries. Columns cross domain (in\,=\,ML \texttt{dev\_public}; out\,=\,biomedical) with recency (pre-cutoff 2021--2023; post-cutoff: in-domain\,=\,the 2024--25 temporal supplement, out-domain\,=\,the released 2026 biomedical entries). $\Delta_{\mathrm{dom}}$ is the pure domain effect (out-pre $-$ in-pre, recency held fixed). Two readings: (i)~the out-domain post-cutoff column is ${\approx}1.0$ for every model, reflecting the post-cutoff date heuristic (89\% of those false positives cite the future date) rather than domain transfer; (ii)~once recency is matched, $\Delta_{\mathrm{dom}}$ is negative or small for calibrated verifiers and large only for the recall-aggressive Qwen3-235B. Sorted by out-domain pre-cutoff \gls{fpr}; DeepSeek-R1 omitted (see text). $^{\S}$Sonnet~4.6 and Opus~4.7 use the May collection in the in-domain columns, not the September re-run of \cref{tab:results}; their out-domain runs are from July.}
\label{tab:crossdomain_llm}
\centering
\small
\begin{tabular}{lccccc}
\toprule
& \multicolumn{2}{c}{\textbf{pre-cutoff}} & & \multicolumn{2}{c}{\textbf{post-cutoff}} \\
\cmidrule(lr){2-3}\cmidrule(lr){5-6}
\textbf{Model} & in-dom. & out-dom. & $\Delta_{\mathrm{dom}}$ & in-dom. & out-dom. \\
\midrule
Gemini 2.5 Flash  & 0.100 & 0.013 & $-0.09$ & 0.595 & 0.991 \\
Gemini 2.5 Pro    & 0.050 & 0.020 & $-0.03$ & 0.250 & 1.000 \\
Llama 4 Maverick  & 0.146 & 0.059 & $-0.09$ & 0.763 & 1.000 \\
Claude Sonnet 4.6$^{\S}$ & 0.127 & 0.099 & $-0.03$ & 0.120 & 0.957 \\
GPT-5.1           & 0.411 & 0.105 & $-0.31$ & 0.759 & 1.000 \\
GPT-5.4           & 0.228 & 0.112 & $-0.12$ & 0.413 & 1.000 \\
Claude Opus 4.7$^{\S}$   & 0.072 & 0.125 & $+0.05$ & 0.073 & 0.948 \\
Mistral Large     & 0.250 & 0.178 & $-0.07$ & 0.793 & 1.000 \\
DeepSeek-V3.2     & 0.702 & 0.408 & $-0.29$ & 0.759 & 1.000 \\
Qwen3-VL-235B     & 0.551 & 0.553 & $+0.00$ & 0.887 & 1.000 \\
Qwen3-235B        & 0.533 & \cellcolor{cellbad}0.974 & $+0.44$ & 0.809 & 1.000 \\
\bottomrule
\end{tabular}
\end{table}

\paragraph{Provider content filtering on biomedical citations.}
Both Anthropic models we query through OpenRouter (Claude Opus~4.7 and Sonnet~4.6) return an empty, content-filtered response (\texttt{finish\_reason:\ content\_filter}) on the same 5 of 152 valid biomedical entries (3.3\%) in the recency-matched split: all virology and immunology papers on SARS-CoV-2 inhibitor resistance, ACE2 receptor-binding and antibody-escape mutations, and HIV envelope neutralization, i.e.\ dual-use--adjacent titles the provider filter blocks.
The refusal is deterministic: it reproduces at completion budgets from 512 to 4096 tokens, so it is not output truncation, and a trivial control prompt to the same endpoint returns normally, so it is not an outage or a credit fault.
These entries fall to the verifier's error path and score as UNCERTAIN, i.e.\ committed-VALID under our convention, so they do not inflate false-positive rates. They do depress detection: the same error path catches 10 hallucinated entries for Opus and 9 for Sonnet, which then score as missed detections, so the cost is one of availability for the \gls{fpr} column and of accuracy for the \gls{dr} column. The released prediction dumps record every one of these records as \texttt{[Error fallback] Parse error: None}, indistinguishable there from a generic API failure; the \texttt{content\_filter} finish reason was observed at run time and is not recoverable from the release.
The filter never fires on the ML-venue core, so the failure mode is invisible in-domain and surfaces only out of domain: a deployment consideration specific to safety-filtered providers, orthogonal to the verdict-quality question the rest of this section studies.

\begin{takeaway}
\textbf{Takeaway.} The cross-domain \gls{fpr} rise is mostly a recency artifact, not a domain effect. For the \glspl{llm}, 89\% of the released biomedical false positives are post-cutoff date heuristics that fall to 2\% once the entries are pre-cutoff; the residual domain effect is negative or small for calibrated verifiers and large only for a model that already over-flags in its own regime. For \texttt{bibtex-updater}, canonical metadata shows the released 0.375 was inflated by scrape noise: its true out-of-domain behavior is a coverage drop---it abstains on 71\% of valid biomedical citations---at an in-domain-level \gls{fpr} (0.112). A Stage-2 Sonnet diagnoser recovers that coverage ($0.29\to1.00$ on valid entries) at \gls{fpr} 0.189: the out-of-domain abstention is a recoverable gap. Precision is a property of the tool--regime pair, and the pair that matters is domain \emph{and} recency together.
\end{takeaway}

\paragraph{Tool-augmented LLM baseline.}
\label{app:tool_augmented}
To test whether combining \gls{llm} reasoning with API-backed verification is greater than the sum of its parts, we augment GPT-5.1 with \texttt{bibtex-updater}'s structured output.
For each entry, we first run \texttt{bibtex-check} to obtain the verification status, mismatched fields, confidence score, and APIs consulted, then inject this evidence into an augmented prompt that instructs the model to use the tool findings as evidence while applying its own judgment.
\cref{tab:tool_augmented} compares the augmented model against its components.

\begin{table}[t]
\caption{\textbf{Tool-augmented GPT-5.1 vs.\ standalone components on \texttt{dev\_public}.} The augmented model improves precision and calibration over GPT-5.1 alone but does not match bibtex-updater's detection rate.}
\label{tab:tool_augmented}
\centering
\small
\begin{tabular}{lcccccc}
\toprule
\textbf{Baseline} & \textbf{DR $\uparrow$} & \textbf{FPR $\downarrow$} & \textbf{F1 $\uparrow$} & \textbf{MCC $\uparrow$} & \textbf{TW-F1 $\uparrow$} & \textbf{ECE $\downarrow$} \\
\midrule
GPT-5.1 (standalone)        & 0.837 & 0.411 & 0.766 & 0.442 & 0.822 & 0.190 \\
bibtex-updater (standalone, v1.2.0)  & 0.820 & 0.051 & 0.880 & 0.768 & 0.892 & 0.395 \\
GPT-5.1 + bibtex-updater     & 0.843 & 0.144 & 0.856 & 0.698 & 0.872 & \textbf{0.078} \\
\bottomrule
\end{tabular}
\end{table}

The augmented model improves over standalone GPT-5.1 on every headline metric: detection rate rises slightly (84.3\% vs.\ 83.7\%) while \gls{fpr} drops sharply (0.144 vs.\ 0.411) and \gls{ece} improves to 0.078, the best calibration of any baseline.
However, it still falls short of \texttt{bibtex-updater}'s standalone \gls{fpr} (0.144 vs.\ 0.051).

\cref{tab:tool_augmented_pertype} reveals why.
On the two types where bibtex-updater excels and GPT-5.1 struggles---\texttt{swapped\_authors} and \texttt{near\_miss\_title}---the augmented model tracks the \gls{llm} rather than the tool: \texttt{swapped\_authors} reaches 47.8\% against bibtex-updater's 98.5\% (standalone GPT-5.1: 44.8\%), and \texttt{near\_miss\_title} falls from the \gls{llm}'s own 57.7\% to 50.0\% against the tool's 88.5\%.
The \gls{llm} overrides tool-detected metadata mismatches, treating them as potential API artifacts rather than genuine hallucination signals.
The one type where evidence injection buys anything is \texttt{chimeric\_title}, which rises to 97.9\% from standalone GPT-5.1's 89.4\%: the tool evidence helps the \gls{llm} \emph{confirm} an existing suspicion rather than revise a judgment. On three types---\texttt{fabricated\_doi}, \texttt{near\_miss\_title} and \texttt{plausible\_fabrication}---the augmented model lands below both of its components at once.

\begin{table}[t]
\caption{\textbf{Per-type detection rates for tool-augmented GPT-5.1 vs.\ standalone components.} The augmented model fails to transfer \texttt{bibtex-updater}'s strength on metadata-based types. All three columns are scored on the released labels; the \textbf{GPT-5.1} column reproduces the \texttt{per\_type\_metrics} block of \texttt{llm\_openai\_dev\_public.json}. Red marks the three cells where evidence injection falls below both standalone components.}
\label{tab:tool_augmented_pertype}
\centering
\small
\setlength{\tabcolsep}{3pt}
\begin{tabular}{llccc}
\toprule
\textbf{Tier} & \textbf{Type} & \textbf{GPT-5.1} & \textbf{\shortstack{\texttt{bibtex-}\\\texttt{updater}}} & \textbf{Augmented} \\
\midrule
\multirow{4}{*}{1}
& \texttt{fabricated\_doi}        & 1.000 & 1.000 & \cellcolor{cellbad}0.923 \\
& \texttt{nonexistent\_venue}     & 1.000 & 0.436 & 0.795 \\
& \texttt{placeholder\_authors}   & 1.000 & 1.000 & 1.000 \\
& \texttt{future\_date}           & 1.000 & 1.000 & 1.000 \\
\midrule
\multirow{5}{*}{2}
& \texttt{chimeric\_title}        & 0.894 & 1.000 & 0.979 \\
& \texttt{wrong\_venue}           & 0.851 & 0.447 & 0.681 \\
& \texttt{swapped\_authors}       & 0.448 & 0.985 & 0.478 \\
& \texttt{preprint\_as\_pub.}     & 0.742 & 0.677 & 0.677 \\
& \texttt{hybrid\_fabrication}    & 1.000 & 1.000 & 1.000 \\
\midrule
\multirow{2}{*}{3}
& \texttt{near\_miss\_title}      & 0.577 & 0.885 & \cellcolor{cellbad}0.500 \\
& \texttt{plausible\_fabrication} & 0.934 & 1.000 & \cellcolor{cellbad}0.882 \\
\bottomrule
\end{tabular}
\end{table}

These results demonstrate that na\"ive evidence injection---presenting tool output as \gls{llm} context---is insufficient to realize the complementarity between parametric and retrieval-based verification.
The \gls{llm} treats tool evidence as \emph{advisory} rather than \emph{authoritative}, defaulting to its own judgment when conflicts arise.
More structured integration strategies---such as forcing acceptance of tool-detected field mismatches, weighted ensembling, or agentic tool-use where the \gls{llm} iteratively queries APIs---may better combine the complementary strengths.

\section{Robustness ablations}
\label{app:ablations}

We report four ablations that probe whether the precision-ceiling finding is an artifact of a single design choice---a prompt, a threshold, an input field, or a single annotator---alongside the selective-prediction analysis of \cref{app:coverage}.
We read all four as \emph{robustness evidence for the ranking}, not as tuning for best numbers: the tool ranking is what transfers across the regimes these ablations vary (\cref{sec:crosssplit_robustness})---it does \emph{not} survive a change in what counts as a hallucination, which \cref{app:taxonomy_fold} measures separately---so we test its invariance to phrasing (\cref{app:ablation_prompt}), operating point (\cref{app:ablation_threshold}), input format (\cref{app:ablation_format}), and rater (\cref{app:ablation_kappa}).

\subsection{What counts as a hallucination changes the ranking}
\label{app:taxonomy_fold}
Four of the fourteen types label entries whose cited work exists. \texttt{wrong\_venue} assigns a real paper to the wrong conference, \texttt{preprint\_as\_published} cites a real preprint as though it appeared at a venue, \texttt{partial\_author\_list} truncates a real author list without marking it, and \texttt{arxiv\_version\_mismatch} cites a real preprint with the wrong venue and a shifted year. In each case the reference points at a work a reader can find; what is wrong is the description. The other ten types cover works that do not exist. A verifier that flags all fourteen is doing two jobs, and the benchmark scores them as one.

The distinction is not marginal. On \texttt{test\_public} the four modes are 139 of 519 hallucinated entries ($26.8\%$), and on \texttt{dev\_public} 158 of 606 ($26.1\%$). They are not confined to \texttt{stress\_test}, and the stress split is $61.5\%$ real-paper modes in its own right.

We re-score all 20 tool configurations three ways (\cref{tab:taxonomy_fold}): as shipped; with the four modes folded out of the evaluation set entirely; and with them scored as false accusations, so a flag on one counts against the tool. The ranking is stable to neither. Folding them out moves 12 of 20 configurations in rank; scoring them as false accusations moves 16 of 20 and changes the leader. The cascade of \cref{tab:results} leads as shipped at MCC $0.941$ ($0.938$ conservative) and falls to $0.668$ ($0.665$) as false accusations; \texttt{bibtex-updater} falls less, from $0.738$ to $0.688$, and takes first place under that reading. Folded out, the cascade keeps its lead ($0.936$ against $0.845$). The reversal has a plain cause, visible in the tools' detection rates on the four modes alone: the cascade catches nearly all of them ($0.993$) while \texttt{bibtex-updater} catches $0.597$ and \gls{doi}-only $0.065$. More of the cascade's headline credit comes from flagging papers that exist.

A second control bounds how much verification those detections require. A classifier with access only to surface features---which fields are present, how long the strings are, how many authors there are, whether the venue string says arXiv, the entry type, the year---and no network access and no semantics reaches \gls{dr} $0.468$ at \gls{fpr} $0.045$ on \texttt{dev\_public} and \gls{dr} $0.388$ at \gls{fpr} $0.096$ on \texttt{test\_public}, against a majority-class baseline of $0.000$. Of the four modes' positives $80\%$ are perturbations of real entries, while $95\%$ of the valid entries they are scored against were scraped from real bibliographies, so roughly $40\%$ of the separation is a property of how the entries were made rather than of what they claim.

We report this rather than relabel. Which of the four modes a deployed verifier should flag depends on what the verifier is for: a venue chair screening for fabricated citations wants the ten fabrication modes, while a copy-editing pass wants all fourteen. The benchmark as released serves the second reading, and \cref{tab:results} should be read as a ranking under it. Readers with the first use in mind should take the folded-out column, where the cascade still leads \texttt{bibtex-updater} ($0.936$ against $0.845$). Public venue guidance does not agree on this line. ICML 2026's area-chair instructions count a reference as hallucinated only when no document with a title anywhere near it can be found, and exclude a real venue that is incorrect for the reference, a swapped author list, and a few missing authors~\citep{icml2026acinstructions}: \texttt{wrong\_venue}, \texttt{preprint\_as\_published} and \texttt{arxiv\_version\_mismatch} fall on the excluded side, and \texttt{partial\_author\_list} does when the truncation is small. ICLR 2026's review retrospective instead describes flagged references as ones that ``referred to documents that didn't exist and/or contained egregiously incorrect bibliographic information''~\citep{iclr2026retrospective}, merging the two classes, and ACL 2026 lists hallucinated references as grounds for desk rejection without defining them~\citep{acl2026statement}. We therefore report both readings rather than choosing one.

\begin{table}[!htbp]
\caption{\textbf{What counts as a hallucination changes the ranking (\texttt{test\_public}, 20 configurations).} Real-mode \gls{dr} is the detection rate on the four types whose cited work exists (\texttt{wrong\_venue}, \texttt{preprint\_as\_published}, \texttt{partial\_author\_list}, \texttt{arxiv\_version\_mismatch}); MCC is reported as shipped, with those four types folded out of the evaluation set, and with them scored as false accusations. Rows are named by their released result file; the cascade rows are the Sonnet~4.6 cascade on the shared Stage~1, and \texttt{bibtexupdater} is the rescored run.}
\label{tab:taxonomy_fold}
\centering
\footnotesize
\setlength{\tabcolsep}{4pt}
\fitwidth{%
\begin{tabular}{lcccc}
\toprule
\textbf{Configuration} & \textbf{Real-mode DR} & \textbf{MCC as shipped} & \textbf{MCC folded out} & \textbf{MCC as false accusations} \\
\midrule
\texttt{cascade\_db\_diagnosis} & 0.992 & 0.938 & 0.932 & 0.665 \\
\texttt{cascade\_db\_diagnosis\_aggressive} & 0.993 & 0.941 & 0.936 & 0.668 \\
\texttt{llm\_openrouter\_claude\_sonnet\_4\_6} & 0.863 & 0.743 & 0.767 & 0.569 \\
\texttt{bibtexupdater} & 0.597 & 0.738 & 0.845 & 0.688 \\
\texttt{llm\_agentic\_btu\_sonnet\_4\_6} & 0.993 & 0.721 & 0.701 & 0.512 \\
\texttt{llm\_openrouter\_claude\_opus\_4\_7} & 0.791 & 0.700 & 0.748 & 0.573 \\
\texttt{llm\_agentic\_btu\_openai} & 0.921 & 0.661 & 0.668 & 0.500 \\
\texttt{llm\_tool\_augmented} & 0.755 & 0.601 & 0.647 & 0.497 \\
\texttt{llm\_openai\_gpt54} & 0.655 & 0.544 & 0.603 & 0.472 \\
\texttt{llm\_openrouter\_deepseek\_r1} & 0.725 & 0.489 & 0.530 & 0.407 \\
\texttt{llm\_agentic\_openai} & 0.899 & 0.464 & 0.479 & 0.366 \\
\texttt{llm\_openrouter\_llama\_4\_maverick} & 0.450 & 0.451 & 0.530 & 0.444 \\
\texttt{llm\_openrouter\_gemini\_pro} & 0.252 & 0.427 & 0.517 & 0.459 \\
\texttt{llm\_openrouter\_gemini\_flash} & 0.324 & 0.404 & 0.482 & 0.414 \\
\texttt{llm\_openai} & 0.799 & 0.397 & 0.422 & 0.322 \\
\texttt{llm\_openrouter\_mistral} & 0.475 & 0.394 & 0.484 & 0.424 \\
\texttt{llm\_openrouter\_qwen\_max} & 0.914 & 0.372 & 0.374 & 0.276 \\
\texttt{llm\_openrouter\_qwen} & 0.899 & 0.355 & 0.356 & 0.261 \\
\texttt{llm\_openrouter\_deepseek\_v3} & 0.863 & 0.244 & 0.272 & 0.218 \\
\texttt{doi\_only} & 0.065 & 0.211 & 0.273 & 0.274 \\
\bottomrule
\end{tabular}%
}
\end{table}

The temporal-mechanism probes (the recall and late-cutoff controls, cutoff-aware prompting, and the thinking-budget boundary) live with failure mode~(iii) in \cref{app:temporal}; the bootstrap CIs are in \cref{app:bootstrap}.
The prompt-variant, field-\gls{loo}, and rater runs use a fresh dated OpenRouter snapshot (2026-05-31) that does not reproduce the main-run absolute aggregates; we therefore read each as a \emph{within-run} difference, robust to endpoint drift (\cref{sec:limitations}).

\subsection{Prompt-sensitivity sweep}
\label{app:ablation_prompt}

We sweep four prompt variants---\texttt{default} (the paper's prompt), \texttt{notaxo} (taxonomy removed), \texttt{uncertain} (abstention explicitly encouraged), and \texttt{terse} (compressed instructions)---over four models on a stratified $n{=}150$ \texttt{dev\_public} sample (81 hallucinated / 69 valid), at temperature 0 and seed 42 (\cref{tab:ablation_prompt}); GPT-5.1 runs through its OpenAI-direct endpoint, the other three through OpenRouter.

The finding is two-part. First, \emph{the model ranking is prompt-invariant}: mean pairwise Spearman $\rho=0.90$ for \gls{f1}, \gls{dr}, and \gls{fpr} across the four variants (Sonnet~4.6 $>$ GPT-5.1 $>$ DeepSeek-V3.2 $>$ Gemini~2.5~Flash on \gls{f1} at \texttt{default}), and the sole departure from $\rho=1.0$ is a $0.001$ \gls{f1} near-tie between GPT-5.1 ($0.794$) and Sonnet~4.6 ($0.793$) under the \texttt{uncertain} variant: a tie, not a reordering. This is the result that defends the single-prompt design of \cref{sec:baselines}.
Second, \emph{the absolute \gls{fpr} is wording-sensitive}: GPT-5.1 shows the largest swing: \Gls{fpr} $0.580$ at \texttt{default} down to $0.212$ under \texttt{terse} ($-36.8$\,pp), with its UNCERTAIN rate reaching $19.3\%$ under \texttt{uncertain} (from $0\%$). The abstention-encouraging \texttt{uncertain} variant drops Sonnet~4.6's \gls{fpr} from $0.121$ to $0.015$ ($-10.6$\,pp) and DeepSeek-V3.2's from $0.899$ to $0.600$ ($-29.9$\,pp), while Gemini~2.5~Flash moves only $-4.3$\,pp. Sonnet's UNCERTAIN rate rises from $2.0\%$ (\texttt{notaxo}/\texttt{terse}) to $8.7\%$ (\texttt{uncertain}), a ${\sim}7$\,pp coverage swing from wording alone. Pooled over all models and non-default variants, the verdict-flip rate against the default is $17.4\%$ (all entries) and $13.6\%$ when UNCERTAIN-involving flips are excluded, with GPT-5.1 the most prompt-sensitive model ($25.8\%$ mean flip).

The practical consequence is a scope bound on the post-cutoff \gls{fpr} \emph{magnitude}: because a single wording change can move \gls{fpr} by $10$--$37$\,pp, we treat the absolute post-cutoff false-alarm rate (\cref{sec:temporal_robustness}) as prompt-conditional and rest the temporal claim on the cross-regime \emph{ranking}, which this sweep shows is prompt-invariant.

\begin{table}[t]
\caption{\textbf{Prompt-sensitivity sweep} ($n{=}150$ \texttt{dev\_public}; temperature 0; seed 42; OpenRouter models on a fresh 2026-05-31 snapshot, GPT-5.1 via the OpenAI-direct endpoint). DR, FPR, F1, ECE, and UNCERTAIN-rate per model and prompt variant. Model F1/DR ranking is near-invariant to the variant (mean pairwise Spearman $\rho=0.90$); the absolute FPR is not (the \texttt{uncertain} and \texttt{terse} variants drop it sharply).}
\label{tab:ablation_prompt}
\centering
\small
\setlength{\tabcolsep}{4pt}
\begin{tabular}{llccccc}
\toprule
\textbf{Model} & \textbf{Variant} & \textbf{DR $\uparrow$} & \textbf{FPR $\downarrow$} & \textbf{F1 $\uparrow$} & \textbf{ECE $\downarrow$} & \textbf{UNC.} \\
\midrule
\multirow{4}{*}{Sonnet~4.6}
 & default   & 0.909 & 0.121 & 0.903 & 0.087 & 4.7\% \\
 & notaxo    & 0.744 & 0.101 & 0.811 & 0.098 & 2.0\% \\
 & uncertain & 0.667 & 0.015 & 0.793 & 0.083 & 8.7\% \\
 & terse     & 0.759 & 0.074 & 0.833 & 0.113 & 2.0\% \\
\midrule
\multirow{4}{*}{GPT-5.1}
 & default   & 0.914 & 0.580 & 0.759 & 0.233 & 0.0\% \\
 & notaxo    & 0.802 & 0.420 & 0.743 & 0.225 & 0.0\% \\
 & uncertain & 0.800 & 0.250 & 0.794 & 0.145 & 19.3\% \\
 & terse     & 0.720 & 0.212 & 0.755 & 0.170 & 6.0\% \\
\midrule
\multirow{4}{*}{DeepSeek-V3.2}
 & default   & 0.975 & 0.899 & 0.712 & 0.372 & 0.0\% \\
 & notaxo    & 0.938 & 0.768 & 0.724 & 0.335 & 0.0\% \\
 & uncertain & 0.872 & 0.600 & 0.735 & 0.282 & 4.7\% \\
 & terse     & 0.889 & 0.667 & 0.724 & 0.315 & 0.0\% \\
\midrule
\multirow{4}{*}{Gemini~2.5~Flash}
 & default   & 0.469 & 0.130 & 0.594 & 0.316 & 0.0\% \\
 & notaxo    & 0.370 & 0.087 & 0.513 & 0.324 & 0.0\% \\
 & uncertain & 0.383 & 0.087 & 0.525 & 0.314 & 0.0\% \\
 & terse     & 0.321 & 0.101 & 0.456 & 0.414 & 0.0\% \\
\midrule
\multicolumn{2}{l}{Mean pairwise Spearman $\rho$} & \multicolumn{2}{l}{F1/DR: $0.90$} & \multicolumn{3}{l}{FPR: $0.90$} \\
\bottomrule
\end{tabular}
\end{table}

\begin{takeaway}
\textbf{Takeaway.} Prompt wording moves the absolute numbers and leaves the ranking alone: a single variant shifts \gls{fpr} by $10$--$37$\,pp (GPT-5.1: $0.580\to0.212$ under \texttt{terse}) and Sonnet's coverage by ${\sim}7$\,pp, whereas the model ordering holds at mean pairwise Spearman $\rho=0.90$, with the sole departure a $0.001$ \gls{f1} near-tie. This is why the temporal claims rest on cross-regime rankings and every absolute post-cutoff \gls{fpr} reads as prompt-conditional (\cref{sec:temporal_robustness}).
\end{takeaway}

\subsection{\acrshort{gepa} prompt optimization}
\label{app:prompt_opt}

The robustness check above bounds how far wording moves the metrics; it does not say how far a \emph{deliberate} search through wording can move them. That question is sharpened by failure mode~(iii): the cutoff experiments show post-cutoff over-flagging (\cref{sec:temporal_robustness}), the cutoff-aware check shows part of it answers to instructions (\cref{app:cutoff-aware}), and the sweep prices wording high enough to cover the \gls{fpr} gap between a small model and the frontier (\cref{app:ablation_prompt}). If a prompt is worth that much, the precision gap we report might be a prompting effect rather than a capability difference. We test that directly, and the answer is no.

We optimize the paper's verification prompt with \gls{gepa}, a reflective evolutionary search that mutates a prompt from natural-language feedback on its own failures~\citep{agrawal2026gepareflectivepromptevolution}. The seed is the paper's default prompt (\cref{app:prompt-template}) and the model is Claude Haiku~4.5. The objective is balanced binary accuracy on a $200$-entry \texttt{dev\_public} validation set and under-flagging is penalized as hard as over-flagging; \textsc{uncertain} scores $0$. \textbf{Optimization touched \texttt{dev\_public} only; every number below is \texttt{test\_public}}, which the optimizer never saw.

\begin{table}[t]
\caption{\textbf{\acrshort{gepa} prompt optimization on Claude Haiku~4.5.} Both rows are the same model under the same decoding settings; only the prompt differs. Bal.\ is balanced accuracy, $(\text{DR}+1-\text{FPR})/2$, on \texttt{test\_public}; Val.\ objective is the same metric on the 200-entry \texttt{dev\_public} validation set the optimizer tuned on. Abstentions score 0 in both.}

\label{tab:prompt_opt}
\centering
\small
\begin{tabular}{lcccc}
\toprule
 & \textbf{Val.\ objective} & \multicolumn{3}{c}{\textbf{Held out: \texttt{test\_public} $(n{=}831)$}} \\
\cmidrule(lr){2-2}\cmidrule(lr){3-5}
\textbf{Prompt} & \texttt{dev\_public} $(n{=}200)$ & \textbf{DR $\uparrow$} & \textbf{FPR $\downarrow$} & \textbf{Bal.\ $\uparrow$} \\
\midrule
Seed (paper default) & 0.700 & 0.871 & 0.468 & 0.702 \\
\acrshort{gepa}-optimized & 0.800 & 0.746 & 0.224 & 0.761 \\
\midrule
$\Delta$ & $+0.100$ & $-0.125$ & $-0.244$ & $+0.059$ \\
\bottomrule
\end{tabular}
\end{table}

Optimization works, but does not close the gap. Claude Opus~4.7 on the untuned default prompt  catches far more hallucinations than the optimized small model at a similar false-positive rate (\gls{dr} $0.900$ vs.\ $0.746$, \gls{fpr} $0.205$ vs.\ $0.224$) on the same split (\cref{tab:test_public_full}). \emph{Prompt optimization does not move a small model to the frontier's operating point.} We state the claim no more broadly than that: only Haiku was optimized, and it is compared against frontier models running the shared default prompt, so this is an optimized-vs-default comparison.

A note on protocol, since prompt optimizers leak by default. The reflection step sees per-entry feedback on failures, which is an easy channel for the optimizer to write dataset specifics into the prompt. We therefore constrain reflection to describe failure patterns in general terms and forbid it from copying author names, titles, citation keys, or split statistics.
The optimized prompt is released as \path{results/gepa_haiku/guarded/best_prompt.txt}.

\subsection{Threshold and aggregation ablation}
\label{app:ablation_threshold}

We re-score the stored per-entry confidences on \texttt{dev\_public} (no API calls) to test two operating-point choices: the decision threshold and the field-aggregation rule (\cref{tab:ablation_threshold}).

\paragraph{Threshold.} Confidences are effectively quantized, so threshold tuning buys almost nothing: the gap between the default $0.5$ threshold and the best-F1 threshold is below $0.35$\,pp for seven of eight tools ( GPT-5.4 $0.34$\,pp, Opus~4.7 $0.20$\,pp, Sonnet~4.6 $0.09$\,pp, the rest ${\leq}0.07$\,pp). The lone exception is Gemini~2.5~Flash ($8.22$\,pp), and it buys F1 only by trading FPR up to $0.343$. AUROC orders the verifiers as expected (Sonnet $0.927$, Opus $0.903$, GPT-5.4 $0.834$, then the recall-aggressive open-weight models $0.61$--$0.74$). For the seven quantized-confidence tools the fixed-$0.5$ operating point of \cref{tab:results} is therefore near-optimal---best-F1 is chosen in-sample on \texttt{dev\_public}, so this lower-bounds the true headroom---while Gemini~2.5~Flash is the exception noted above. The ranking is threshold-robust regardless, in contrast to peer benchmarks that hard-code an unjustified match threshold.

\paragraph{Aggregation.} The benchmark's detector flags an entry when \emph{any one} of the cross-database sub-tests fails (the ``any-miss'' rule---the benchmark's internal detector, distinct from the agentic harness's \emph{any-no-match} rule in \cref{app:aggregation}). Sweeping stricter quorums trades DR for FPR monotonically: on the structured sub-tests, any-miss gives DR $0.950$ / FPR $0.000$, while requiring two, three, or all four misses reduces DR to $0.269$, $0.048$, $0.002$; on the field-level resolver the same sweep drops DR from $0.759$ (any-miss) to $0.128$ (unanimous). A noisy-voter ensemble makes the same point with real verifiers: treating the twelve zero-shot \glspl{llm} with stored per-entry predictions (\cref{tab:llm_agreement}) as twelve independent voters, flagging whenever any single voter flags reproduces the any-no-match profile (DR $0.995$ / FPR $0.881$), simple majority ($\geq 7/12$) gives DR $0.779$ / FPR $0.191$ / F1 $0.803$, and supermajority and unanimity buy lower FPR at steep DR cost. The benchmark deploys the any-miss rule (F1 $0.975$); requiring cross-database agreement instead nudges F1 to $0.985$ at a small false-positive cost (FPR $0.000\to0.035$), so any-miss sits within $1$\,pp of the family's F1 maximum while holding FPR at zero.

\begin{table}[t]
\caption{\textbf{Threshold and aggregation ablation on \texttt{dev\_public}} (offline re-score of stored confidences; no API calls). \emph{Top:} per-tool AUROC, default-$0.5$ F1, and the F1 gap to the best threshold---near zero for all but Gemini~2.5~Flash, confirming the fixed operating point is near-optimal. \emph{Bottom:} the field-aggregation sweep; the benchmark uses the any-miss rule, which sits within $1$\,pp of the family's F1 maximum while holding FPR at zero, and stricter quorums trade DR for FPR monotonically.}
\label{tab:ablation_threshold}
\centering
\small
\setlength{\tabcolsep}{4pt}
\begin{tabular}{lcccc}
\toprule
\multicolumn{5}{l}{\emph{(a) Threshold: default $0.5$ vs.\ best-F1 (8 tools)}} \\
\textbf{Tool} & \textbf{AUROC} & \textbf{F1@0.5} & \textbf{best-F1} & \textbf{gap (pp)} \\
\midrule
Claude Sonnet~4.6 & 0.927 & 0.893 & 0.894 & 0.09 \\
Claude Opus~4.7   & 0.903 & 0.890 & 0.892 & 0.20 \\
GPT-5.4           & 0.834 & 0.783 & 0.786 & 0.34 \\
Mistral Large     & 0.744 & 0.742 & 0.742 & 0.00 \\
DeepSeek-R1       & 0.741 & 0.739 & 0.739 & 0.07 \\
Gemini~2.5~Flash  & 0.739 & 0.631 & 0.713 & 8.22 \\
Qwen3-235B        & 0.695 & 0.744 & 0.744 & 0.00 \\
DeepSeek-V3.2     & 0.609 & 0.727 & 0.727 & 0.00 \\
\midrule
\multicolumn{5}{l}{\emph{(b) Aggregation rule (cross-DB sub-tests / noisy 12-voter ensemble)}} \\
\textbf{Rule} & \textbf{DR} & \textbf{FPR} & \textbf{F1} & \textbf{MCC} \\
\midrule
structured: any-miss ($k{\geq}1$/4)      & 0.950 & 0.000 & 0.975 & 0.948 \\
structured: cross-DB-agreement only      & 1.000 & 0.035 & 0.985 & 0.968 \\
structured: majority ($k{\geq}3$/4)      & 0.048 & 0.000 & 0.091 & 0.150 \\
field-level: any-miss ($k{\geq}1$/4)     & 0.759 & 0.139 & 0.809 & 0.619 \\
field-level: two-fields ($k{\geq}2$/4)   & 0.365 & 0.031 & 0.525 & 0.407 \\
ensemble: majority ($\geq 7/12$)       & 0.779 & 0.191 & 0.803 & 0.586 \\
ensemble: supermajority ($\geq 11/12$) & 0.495 & 0.047 & 0.645 & 0.493 \\
\bottomrule
\end{tabular}
\end{table}

\subsection{Input-format and field leave-one-out}
\label{app:ablation_format}

We test whether detection rests on a surface format tell or on field content by varying the input representation (full BibTeX vs.\ structured fields) and dropping one field at a time (title / authors / venue / year / \gls{doi}), on the same $n{=}150$ stratified sample with two models (\cref{tab:ablation_format}). This is the dynamic counterpart to the static format-tell audit (\cref{app:shortcuts}).

Two results. \emph{Title carries most of the signal.} For Gemini~2.5~Flash, dropping the title raises \gls{fpr} by $+35.5$\,pp over the structured baseline ($0.145\to0.500$) and dropping authors by $+18.8$\,pp, while venue, year, and \gls{doi} move \gls{fpr} little; DeepSeek-V3.2---already near-saturated at \gls{fpr} $0.81$ structured---shows the same direction at smaller magnitude ($+12.9$\,pp for title). \emph{Format matters less than content}: switching between full BibTeX and structured fields shifts \gls{dr}/\gls{fpr} by at most $7.4$/$14.5$\,pp. Detection therefore leans on title and author semantics rather than a surface artifact, reinforcing the construct-validity argument that the result is not driven by a single co-designed field or by formatting (\cref{app:codesign,app:shortcuts}).

\begin{table}[t]
\caption{\textbf{Input-format and field leave-one-out on \texttt{dev\_public}} ($n{=}150$; temperature 0; seed 42; fresh 2026-05-31 snapshot). $\Delta$FPR is measured against each model's structured-field baseline. Dropping the title or authors spikes FPR; dropping venue/year/DOI does not: detection leans on title/author content, not a format tell. (DOI LOO affects only the $71$ DOI-bearing entries.)}
\label{tab:ablation_format}
\centering
\small
\setlength{\tabcolsep}{4pt}
\begin{tabular}{llcccc}
\toprule
\textbf{Model} & \textbf{Condition} & \textbf{DR $\uparrow$} & \textbf{FPR $\downarrow$} & \textbf{F1 $\uparrow$} & \textbf{$\Delta$FPR} \\
\midrule
\multirow{7}{*}{Gemini~2.5~Flash}
 & full        & 0.543 & 0.101 & 0.667 & --- \\
 & structured  & 0.617 & 0.145 & 0.709 & (base) \\
 & $-$title    & 0.789 & 0.500 & 0.709 & $+0.355$ \\
 & $-$authors  & 0.709 & 0.333 & 0.709 & $+0.188$ \\
 & $-$venue    & 0.519 & 0.101 & 0.646 & $-0.043$ \\
 & $-$year     & 0.593 & 0.275 & 0.649 & $+0.130$ \\
 & $-$DOI      & 0.469 & 0.145 & 0.589 & $\phantom{+}0.000$ \\
\midrule
\multirow{7}{*}{DeepSeek-V3.2}
 & full        & 1.000 & 0.957 & 0.711 & --- \\
 & structured  & 0.975 & 0.812 & 0.731 & (base) \\
 & $-$title    & 0.988 & 0.940 & 0.712 & $+0.129$ \\
 & $-$authors  & 0.987 & 0.896 & 0.719 & $+0.084$ \\
 & $-$venue    & 0.926 & 0.826 & 0.704 & $+0.014$ \\
 & $-$year     & 0.926 & 0.754 & 0.721 & $-0.058$ \\
 & $-$DOI      & 0.963 & 0.841 & 0.719 & $+0.029$ \\
\bottomrule
\end{tabular}
\end{table}

\subsection{Multi-rater reliability proxy}
\label{app:ablation_kappa}

\emph{This is an automated multi-rater reliability proxy, not human inter-annotator agreement.}
Three independent \gls{llm} raters (Sonnet~4.6, DeepSeek-V3.2, Gemini~2.5~Pro) label 132 blinded entries---80 real-world-incident hallucinations and the 52 relabel-recovered real papers---and we report their agreement (UNCERTAIN mapped to HALLUCINATED, binary; \cref{tab:ablation_kappa}). It does not substitute for human \gls{iaa}, which remains future work (\cref{sec:limitations}); it serves to test whether independent raters reproduce the relabel.

Overall agreement is only fair: Fleiss' $\kappa=0.24$~\citep{fleiss1971kappa} on the ``fair'' band of the Landis--Koch scale~\citep{landis1977kappa}, with pairwise Cohen's $\kappa$ ranging from $0.45$ (Sonnet--DeepSeek, moderate) down to $0.03$ (DeepSeek--Gemini~Pro, slight). Against the corrected ground-truth labels---the database-backed labels after the relabel audit (\cref{app:dataset})---DeepSeek-V3.2 agrees best ($\kappa=0.716$, accuracy $0.871$) and Gemini~2.5~Pro worst ($\kappa=0.026$); the majority vote reaches accuracy $0.720$ ($\kappa=0.340$).

The diagnostic split is the point. Majority-vote accuracy is $0.975$ on the 80 real-world hallucinations but only $0.327$ on the 52 relabel-recovered real papers: the \gls{llm} raters confidently agree on genuine fabrications yet systematically \emph{over-flag} the recovered real papers, the same over-flagging failure mode \textsc{Hallmark} is built to measure. Independent raters would have repeated the original labeling error, which both corroborates that the relabel audit corrected a real bias and motivates the precision-via-abstention contribution (\cref{app:coverage}).

\begin{table}[t]
\caption{\textbf{Multi-rater reliability proxy (not human IAA)} on 132 blinded \texttt{dev\_public} entries (80 real-world-incident hallucinations, 52 relabel-recovered real papers), 3 independent LLM raters, binary (UNCERTAIN${\to}$HALLUCINATED). The diagnostic split does the work: the raters agree on genuine fabrications but over-flag the recovered real papers, exactly the failure mode the benchmark measures; the red-shaded cell marks that diagnostic.}
\label{tab:ablation_kappa}
\centering
\small
\setlength{\tabcolsep}{5pt}
\begin{tabular}{lcc}
\toprule
\textbf{Agreement statistic} & \multicolumn{2}{c}{\textbf{Value}} \\
\midrule
Fleiss' $\kappa$ (3 raters, binary) & \multicolumn{2}{c}{$0.238$ (fair)} \\
Cohen's $\kappa$ Sonnet--DeepSeek   & \multicolumn{2}{c}{$0.454$ (moderate)} \\
Cohen's $\kappa$ Sonnet--Gemini~Pro & \multicolumn{2}{c}{$0.266$ (fair)} \\
Cohen's $\kappa$ DeepSeek--Gemini~Pro & \multicolumn{2}{c}{$0.029$ (slight)} \\
\midrule
\textbf{Rater} & \textbf{$\kappa$ vs.\ ground truth} & \textbf{acc.\ vs.\ ground truth} \\
DeepSeek-V3.2     & 0.716 & 0.871 \\
Sonnet~4.6        & 0.360 & 0.727 \\
Gemini~2.5~Pro    & 0.026 & 0.576 \\
Majority vote     & 0.340 & 0.720 \\
\midrule
\textbf{Majority-vote accuracy by pool} & & \\
real-world hallucinations ($n{=}80$)    & \multicolumn{2}{c}{0.975} \\
relabel-recovered real papers ($n{=}52$) & \multicolumn{2}{c}{\cellcolor{cellbad}0.327} \\
\bottomrule
\end{tabular}
\end{table}

\begin{takeaway}
\textbf{Takeaway.} Independent raters reproduce the failure mode the benchmark measures: majority-vote accuracy is $0.975$ on the 80 real-world hallucinations yet $0.327$ on the 52 relabel-recovered real papers, so three independent \gls{llm} raters would have repeated the labeling error the ground-truth audit corrected. With Fleiss' $\kappa=0.24$ overall, \gls{llm} raters are no substitute for database-backed ground truth, nor for the human \gls{iaa} that remains future work (\cref{sec:limitations}).
\end{takeaway}

\section{Extended limitations}
\label{app:limitations_full}
The seven paragraphs below are the full-length limitations discussion that \cref{sec:limitations} ranks and compresses.

\paragraph{Dataset scale and coverage.}
The benchmark is small relative to the full diversity of citation hallucinations.
At 2{,}526 entries, most types on \texttt{dev\_public} and \texttt{test\_public} clear $\geq30$ instances---enough for meaningful per-type comparison---but a few fall below it, and \texttt{test\_hidden} does not reach the floor at all (\cref{app:statistics}).
Prevalence differs across splits (54.2--62.5\%), so we emphasize the prevalence-independent metrics (\gls{dr}, \gls{fpr}).
We cover English-language BibTeX from five ML venues (2021--2023); the taxonomy and sub-tests are designed to be domain-agnostic but validated mainly in that regime.
A 500-entry cross-domain split (PubMed/bioRxiv plus non-ML CS venues), together with a recency-matched, canonically-resolved rebuild, separates domain from post-cutoff recency out of regime (\cref{app:codesign}). The separation reframes the result: the released split's apparent cross-domain \gls{fpr} rise is mostly a recency artifact---the \glspl{llm} flag 2026 dates as impossible---and once entries are pre-cutoff the residual domain effect is negative or small for the calibrated verifiers, large only for a single recall-aggressive model that already over-flags in its own regime (\cref{tab:crossdomain_llm}). For \texttt{bibtex-updater}, canonical metadata leaves the \gls{fpr} at its in-domain level while coverage falls: out of domain the cost is coverage, not precision. Two failure modes stay specific to biomedical data---provider safety filters block ${\sim}3\%$ of valid biomedical citations for the Anthropic models, distinct from any verdict error---while humanities and non-English settings remain future work.
 Just over half the valid entries lack a \gls{doi} ($53.0\%$ on \texttt{dev\_public}, $55.4\%$ on \texttt{test\_public})---many ML papers are never formally published, and DBLP records for ICML, ICLR and NeurIPS carry no \gls{doi} at all---which raises the \gls{fpr} of verifiers that treat a missing identifier as suspicious.

\paragraph{Synthetic--real gap.}
Most hallucinated entries are perturbation-generated, and the real-world anchor is thin.
The benchmark holds 108 real-world entries (plus 280 \gls{llm}-generated), type-skewed (55\% \texttt{plausible\_fabrication}), with 15 compound-failure entries that involved judgment calls without \emph{human} inter-annotator agreement.
We report an automated three-rater reliability proxy (Fleiss' $\kappa=0.24$, fair inter-rater agreement; \cref{app:ablation_kappa}) that corroborates the relabel audit but does not substitute for human \gls{iaa}, which remains future work.
Expanding real-world coverage from retraction databases is a priority.

\paragraph{Temporal fragility and the calibration question.}
\Gls{llm} verifiers carry an implicit training-cutoff bias that we can only partly disentangle.
\cref{sec:temporal_robustness} measures the post-cutoff \gls{fpr} rise and the later-cutoff resistance; a recall probe and a non-Anthropic late-cutoff control point away from pure memorization for Opus~4.7 (and only partly for Sonnet~4.6), but the recall measure is model-derived and residual confounds---system-prompt handling, default temperature, native-vs-OpenRouter routing---remain.
We therefore read the resistance as a descriptive signal, not a causal attribution.

\paragraph{Controls and residual gaps.}
Three controls probe the calibration-versus-contamination question, none decisive (\cref{app:reviewer_experiments}).
(i)~A recall probe on valid 2024--2025 papers is consistent with calibration over memorization for Opus~4.7 (it accepts $84\%$ of papers it cannot recall), while Sonnet~4.6's lower \gls{fpr} is partly recall-driven.
(ii)~A third-provider late-cutoff control (DeepSeek-V4-Pro) also holds its post-cutoff \gls{fpr} low relative to the over-flagging cohort ($0.36$ vs.\ $0.93$ for GPT-5.1 on the same $n{=}300$ subsample, at full coverage), though its cutoff matches rather than exceeds the Anthropic pair's (\cref{app:reviewer_experiments}).
(iii)~GPT-5.1's run-to-run variance is small (\gls{f1} std $0.002$ over three runs), so single-run rankings seem to be stable to sampling noise, at least for some models.
The open gaps: the cross-provider control is a single third-party model, a native-API replication of the Anthropic runs is outstanding, and contamination is reduced but not fully excluded even for Opus~4.7.

\paragraph{Compute and token budget.}
Compute caps the coverage of reasoning-mode and extended-thinking variants.
Reasoning models cost far more per entry (DeepSeek-R1 at a median ${\sim}19$\,s vs.\ $1$--$10$\,s, \cref{tab:cost_latency}), so we did not run extended-thinking variants on the full benchmark (\cref{fig:thinking-budget} maps where their JSON-output contract holds), the agentic harness is capped at five tool calls per entry, and HaRC/verify-citations are excluded because neither produces its own verdicts at benchmark scale (\cref{app:harc_disclaimer}).

\paragraph{Regime conditionality and co-design.}
The ``\gls{fpr} decides'' framing is calibrated to the reviewer-bound venue-audit regime: low prevalence, finite reviewer attention, a false alarm cheap to dismiss.
The \gls{ppv} \emph{ranking} is itself prevalence-invariant---sweeping prevalence over $1$--$5\%$ does not reorder the tools (\cref{tab:ppv_sweep})---so what changes across regimes is whether \emph{any} verifier reaches a usable absolute precision and the cost asymmetry $c_{\text{FN}}/c_{\text{FP}}$; under high prevalence or a high cost of a missed fabrication, recall and human triage regain primacy.
Separately, \texttt{bibtex-updater} and the taxonomy were developed in parallel, so the taxonomy reflects fundamental verification steps but the construct-overfitting risk is non-zero; we report it as a co-designed reference (\cref{app:codesign}).
 A second co-design channel runs through the valid-entry inclusion rule, and it is arguably stronger than the taxonomy: \cref{sec:dataset} admits a valid entry only after confirming \gls{doi} resolution, title existence in at least two databases, and author--venue consistency, so every valid entry is database-confirmable by construction, and \gls{fpr}---the axis we rank on---is computed on exactly that half.
The rule therefore favors database-lookup architectures structurally, independently of how the taxonomy was derived.
What bounds the effect is that it does not decide the outcome: an independent parametric verifier matches it: Gemini~2.5~Pro's \gls{fpr} ($.050$) is level with the co-designed tool's ($.051$).
Whether rankings on synthetic hallucinations predict performance on real errors remains open~\citep{hardt2025benchmarks}; the 108 real-world entries are an initial signal, and the codebase tracks the number of times the dev split has been evaluated against during development, an adaptive-data-analysis hygiene measure that bounds the overfitting introduced by repeatedly reusing the same held-out set~\citep{dwork2015}.

\paragraph{Source availability is a second measurement condition.}
Every number a retrieval-dependent verifier produces is conditional on whether the databases answered, and we report none of that conditioning per row.
It is not a small effect. Holding tool and split fixed and correcting only the source conditions---keys, credentials and a 20-per-minute pace---moved the fraction of entries with at least one lookup that never completed from $53.7\%$ to $10.0\%$, and the run got \emph{faster} rather than slower, $2.4$\,s per entry against $5.7$: a starved source turns into a retry storm that costs more than the wait it avoids.
On a 119-entry probe holding model and prompt fixed, $12\%$ of flags cleared with arXiv starved against $24\%$ with it answering: the same verifier on the same entries, half the corrections.
Nothing in this paper controls for it: our rows were collected under whatever the sources were doing that day, which bounds how exactly any of them can be reproduced, and we flag it as an evaluation condition that citation-verification benchmarks, ours included, do not yet report.

\paragraph{Endpoint drift and reproducibility.}
Several ablations and the two Anthropic risk--coverage curves were collected against the OpenRouter API on a dated snapshot that does not reproduce the main-run aggregates.
The Anthropic endpoints drift over time, and for these two models the drift is large enough to dominate several of the effects we measure.
Two independently collected later runs agree on its direction and rough size: an OpenRouter snapshot puts Opus~4.7 at $0.072{\to}0.162$ and Sonnet~4.6 at $0.127{\to}0.165$, and the 2026-09-02 re-run lands at $0.154$ and $0.161$ (\cref{tab:endpoint_drift}). Both roughly double the May \gls{fpr} under a fixed model ID, prompt, split, temperature and seed.
That re-run also persists per-entry predictions, which the original main run did not, so the Sonnet/Opus coverage cells and dev confidence intervals are now recoverable and are reported rather than marked ``n/a''.
Our convention follows from the size of the drift rather than from a preference for either collection. Cross-sectional comparisons take the latest run, so \cref{tab:results} reports 2026-09-02 for these two rows. Any comparison whose quantity is a \emph{difference} stays inside one collection, because a $13$\,pp endpoint shift swamps the differences at stake: the post-cutoff shift in \cref{tab:temporal_supplement} is under $1$\,pp for both models, so those rows remain wholly on May, and the agentic-inflation ratio of failure mode~(i) is likewise quoted within May.
Mixing the two would not merely add noise, it would report drift as a finding---pairing a September 21--23 \gls{fpr} with a May 24--25 one yields $-4.1$ and $-8.1$\,pp, i.e.\ apparent \emph{improvement} on post-cutoff papers.
Rankings are the more robust reading in any case, since drift shifts absolute \gls{fpr} in the same direction across models while largely preserving their order (cross-regime ranking at Spearman $\rho{=}0.90$ under prompt perturbation, \cref{app:ablation_prompt}).
Within-run deltas (prompt-variant, field-\gls{loo}, rater-agreement) difference conditions measured against the same endpoint on the same day and so survive the drift.
The released corpus is pinned at tag \texttt{v1.2.0}, carrying the ground-truth audit of \cref{app:dataset}.\footnote{Later data releases (\texttt{v1.2.1}, \texttt{v1.2.2}) replace stale generation-pipeline explanation strings with diagnoses verified against CrossRef/arXiv/DBLP records (per-entry fix log in the repository); labels, hallucination types, tiers, sub-tests, and split membership are byte-identical to \texttt{v1.2.0}. Tag \texttt{v1.2.3} additionally retypes four entries to \texttt{plausible\_fabrication}/Tier~3 (\texttt{ffe715a15b16} from \texttt{fabricated\_doi} and \texttt{ec01d96455e0} from \texttt{preprint\_as\_published} on \texttt{dev\_public}; \texttt{ab9c13051a56} and \texttt{cad998c28243} from \texttt{placeholder\_authors} on \texttt{test\_public}) and moves the baseline results to \texttt{data/v1.2/}. Split membership and \textsc{valid}/\textsc{hallucinated} labels are unchanged, so aggregate \gls{dr}, \gls{fpr} and \gls{f1} are unaffected; the per-type and per-tier counts in this paper are those of \texttt{v1.2.3}.}

\paragraph{Diagnostic resolution is lower than six.}
Three of the six sub-tests carry little discriminative load on this corpus.
Cross-database agreement is expected-fail for all 14 hallucination types and expected-pass for every valid entry, so it separates the two classes perfectly and localizes nothing; \texttt{fields\_complete} is expected-fail for no type and varies only for \texttt{plausible\_fabrication}; and \texttt{doi\_resolves} is not applicable, or varies by entry, for seven of the fourteen (\cref{tab:taxonomy}).
The effective number of sub-tests that discriminate \emph{between} hallucination types is therefore nearer three than six, which bounds how finely a failure signature can attribute an error and is a target for the next taxonomy revision.

\section{Broader impact and deployment safeguards}
\label{sec:broader_impact}

\textsc{Hallmark} supports scientific integrity by enabling systematic comparison of citation-verification tools, informing venue deployment as \gls{llm}-assisted writing becomes standard.
A taxonomy of hallucination types could theoretically aid adversaries, but the defensive value of understanding these types outweighs the risk; the taxonomy is derived from publicly documented incidents, not novel attack research.
The benchmark is open-source (MIT) and GPU-free, running on a single CPU regardless of institutional resources.

\paragraph{Intended use, and two uses we advise against.}
\textsc{Hallmark} is a diagnostic for tool builders and venue operators: it measures which verifier fails on which citation error, and at what precision cost.
It is not built for autonomous rejection of submissions, nor for screening authors for research misconduct, and we advise against both.
Both inferences require establishing intent, which no verifier in this evaluation attempts and no metadata check can support: a flag means the metadata did not resolve against the databases queried, not that an author fabricated anything.
Our own measurements are the argument against automating the decision.
At a venue-realistic ${\sim}2\%$ base rate the best verifiers in the cohort catch one true hallucination per 6--10 flags and the most aggressive fewer than one in 35 (\cref{tab:ppv}), so no operating point we measured would make an automated gate defensible.

\paragraph{Deployment safeguards.}
Three safeguards follow from those measurements rather than from principle.
A human decides: a verifier's output routes an entry to a reviewer for checking and never stands as a verdict on the citation.
A flag arrives with its evidence---which databases were queried and what each returned---so that an author can rebut it without reverse-engineering the tool.
And an appeal route exists and is cheap: an author who supplies a resolving identifier or a canonical database record clears the flag without adjudication, which is the correct outcome for the recency and indexing failures below, where the tool is wrong and the citation is fine.

\paragraph{Who a false positive lands on.}
False alarms are not distributed evenly across authors, and the unevenness is measurable rather than speculative.
Averaged over the eighteen verifiers with released per-entry predictions, \gls{fpr} orders the five venues identically on both public splits---ICML $0.44$, ICLR $0.36$, NeurIPS $0.32$, AAAI $0.13$, CVPR $0.09$---and the mean pairwise Spearman correlation between verifiers' per-venue rates is $+0.56$, so the burden is a property of the corpus rather than of any one tool.
What the ordering tracks is bibliographic completeness, not venue standing: no valid ICML, ICLR or NeurIPS entry in the public splits carries a \gls{doi}, against every entry at CVPR and AAAI.
Per-verifier spread across those venues runs from $1.3$\,pp for Gemini~2.5~Pro and $6.3$\,pp for \texttt{bibtex-updater} to $69.5$\,pp for GPT-5.1 and $93.9$\,pp for the agentic harness, so a low-\gls{fpr} verifier also buys uniformity, while an aggressive operating point concentrates its errors on whoever is least well indexed.
Publication recency behaves as a cliff rather than a gradient: within 2021--2023 no verifier's year-to-year spread exceeds $19$\,pp, but on the 35 real February-2026 arXiv preprints in the corpus---post-cutoff for every model evaluated---all eighteen verifiers over-flag and fifteen reach \gls{fpr} $1.000$, and on the 448-entry 2024--2025 supplement all twelve \glspl{llm} are worse on 2025 than on 2024 (\cref{sec:temporal_robustness,app:temporal_supplement}).
The authors most exposed are therefore those publishing recent work and work whose venue records carry no resolvable identifier, which is close to the inverse of who can most easily absorb the cost.
A venue that deploys a verifier without first measuring its own subgroup false-positive rates will impose this burden without observing it.

\paragraph{Scope of the evidence.}
The conclusions here are established on English-language BibTeX from five ML venues (2021--2023), extended by a 500-entry cross-domain split covering biomedical and non-ML computer-science sources and by an authentic-\gls{llm}-output supplement spanning 42 multidisciplinary topics (\cref{app:codesign,app:walters_wilder}).
They are not established for non-English citation practices, for humanities conventions where monographs and archival sources dominate, or for fields whose canonical identifiers differ from the \gls{doi}/arXiv pair these tools query.
Deploying a verifier in those settings requires re-measuring its false-positive rate there first, and this benchmark does not license the transfer.

\printacronyms[title={Abbreviations}, toctitle={Abbreviations}]

\ifneuripschecklist
  \newpage
\fi
\newpage
\ifdraftnotes\listoftodos[]\fi
\end{document}